\documentclass[10ptt]{mdpi} 

\usepackage{slashed,epsfig,amssymb,amsmath,mathrsfs,amsthm,graphicx,float,color}
\usepackage[matrix,frame,arrow]{xy}
\usepackage[matrix,frame,arrow]{xy}
\usepackage{amsmath}

\newcommand{\qw}[1][-1]{\ar @{-} [0,#1]}

\newcommand{\measureD}[1]{*{\xy*+=+<.5em>{\vphantom{\rule{0em}{.1em}#1}}*\cir{r_l};p\save*!R{#1} \restore\save+UC;+UC-<.5em,0em>*!R{\hphantom{#1}}+L **\dir{-} \restore\save+DC;+DC-<.5em,0em>*!R{\hphantom{#1}}+L **\dir{-} \restore\POS+UC-<.5em,0em>*!R{\hphantom{#1}}+L;+DC-<.5em,0em>*!R{\hphantom{#1}}+L **\dir{-} \endxy} \qw}

\newcommand{\multigate}[2]{*+<1em,.9em>{\hphantom{#2}} \qw \POS[0,0].[#1,0];p !C *{#2},p \save+LU;+RU **\dir{-}\restore\save+RU;+RD **\dir{-}\restore\save+RD;+LD **\dir{-}\restore\save+LD;+LU **\dir{-}\restore}
\newcommand{\ghost}[1]{*+<1em,.9em>{\hphantom{#1}} \qw}

\newcommand{\Qcircuit}[1][0em]{\xymatrix @*[o] @*=<#1>}  
 \renewcommand{\Qcircuit}[1][0em]{\xymatrix @*=<#1>}

\newcommand{\pureghost}[1]{*+<1em,.9em>{\hphantom{#1}}}
\newcommand{\multiprepareC}[2]{*+<1em,.9em>{\hphantom{#2}}\save[0,0].[#1,0];p\save !C
  *{#2},p+RU+<0em,0em>;+LU+<+.8em,0em> **\dir{-}\restore\save +RD;+RU **\dir{-}\restore\save
  +RD;+LD+<.8em,0em> **\dir{-} \restore\save +LD+<0em,.8em>;+LU-<0em,.8em> **\dir{-} \restore \POS
  !UL*!UL{\cir<.9em>{u_r}};!DL*!DL{\cir<.9em>{l_u}}\restore}
\newcommand{\prepareC}[1]{*{\xy*+=+<.5em>{\vphantom{#1\rule{0em}{.1em}}}*\cir{l^r};p\save*!L{#1} \restore\save+UC;+UC+<.5em,0em>*!L{\hphantom{#1}}+R **\dir{-} \restore\save+DC;+DC+<.5em,0em>*!L{\hphantom{#1}}+R **\dir{-} \restore\POS+UC+<.5em,0em>*!L{\hphantom{#1}}+R;+DC+<.5em,0em>*!L{\hphantom{#1}}+R **\dir{-} \endxy}}
\newcommand{\poloFantasmaCn}[1]{{{}^{#1}_{\phantom{#1}}}}

\usepackage{graphicx,subfigure}
\preto{\abstractkeywords}{\nolinenumbers}
 
\usepackage{subfigure}
\newcommand{\Z}{\mathbb{Z}}
\newcommand{\R}{\mathbb{R}}
\newcommand{\CC}{\mathbb{C}}

\newcommand{\set}[1]{\mathsf{#1}}
\newcommand{\spc}[1]{\mathcal{#1}}

\newcommand{\grp}[1]{\mathsf{#1}}
\newcommand{\Irr}{\set{Irr}}

\newcommand{\Span}{{\mathsf{Span}}}
\newcommand{\Lin}{\mathsf{Lin}}

\newcommand{\Rank}{\mathsf{Rank}}
\def\>{\rangle}
\def\<{\langle}

\newcommand{\bs}[1]{\boldsymbol{#1}}     

\newcommand{\map}[1]{\mathcal{#1}}

\newcommand{\St}{{\mathsf{St}}}

\newcommand{\Transf}{{\mathsf{Transf}}}
\newcommand{\CChan}{{\mathsf{Chan}}}
\newcommand{\QSt}{{\mathsf{QSt}}}

\newcommand{\Act}{{\mathsf{Act}}}

\newcommand{\Deg}{{\mathsf{Deg}}}

\newtheorem{theo}{Theorem}

\newtheorem{prop}{Proposition}
\newtheorem{cor}{Corollary}
\newtheorem{defi}{Definition}
\newtheorem{rem}{Remark}
\newtheorem{ru}{Rule}

\newtheorem{eg}{Example}
 
\newcommand{\Proof}{{\bf Proof. \,}}

\def\Tr{\operatorname{Tr}}

\Title{Agents, subsystems, and the conservation of information}

\Author{Giulio Chiribella $^{1,2,3,4\star}$}

\address{%
$^{1}$ Department of Computer Science, University of Oxford, Parks Road, Oxford, UK\\
$^2$ Canadian Institute for Advanced Research, CIFAR Program in Quantum Information Science\\
$^3$  Department  of  Computer  Science,  The  University  of  Hong  Kong,  Pokfulam  Road,  Hong
Kong\\
$^4$ HKU Shenzhen Institute of Research and Innovation, 
Kejizhong 2nd Road, Shenzhen, China.}

\corres{Correspondence: giulio.chiribella@cs.ox.ac.uk}

\abstract{Dividing the world into subsystems is an important component of the scientific method. The choice of subsystems, however,  is not defined {\em a priori}. Typically, it is dictated by experimental capabilities,  which may be different for  different agents.  Here we propose a way to define subsystems in general physical theories, including theories beyond quantum and classical mechanics.   Our construction  associates every agent $A$ with a subsystem  $S_A$, equipped with its set of states and its set of transformations. In quantum theory, this construction accommodates   the notion of subsystems as factors of a tensor product Hilbert space, as well as  the notion of subsystems associated to a subalgebra of operators. Classical systems can be interpreted as subsystems  of  quantum systems in different ways, by applying our construction to agents who have access to different sets of operations, including  multiphase covariant channels and certain sets of free operations arising in the resource theory of quantum coherence.      After illustrating the basic definitions, we restrict our attention to closed systems, that is, systems where all physical transformations act invertibly and where all states can be generated from a fixed initial state. 
For closed systems,  we propose a dynamical definition of pure states, and show that all the states of all subsystems admit a canonical purification. This result extends the purification principle to a broader setting, in which coherent superpositions can be interpreted as purifications of incoherent mixtures.  }


\begin{document}
\maketitle

The composition of systems and operations is a fundamental primitive in our modelling of the world.  
 It has been investigated in depth in quantum information theory \cite{nielsen2000quantum,kitaev2002classical}, and  
in the foundations of quantum mechanics, where composition has played a key role from the early days of Einstein-Podolski-Rosen \cite{einstein1935can} and Schroedinger \cite{schrodinger1935discussion}. At the level of frameworks, the most recent developments are the compositional frameworks of   general probabilistic theories \cite{hardy2001quantum,barnum2007generalized,barrett2007information,chiribella2010probabilistic,barnum2011information,hardy2011foliable,hardy2013formalism,chiribella2014dilation,chiribella2016quantum,hardy2016reconstructing,dariano2017quantum} and     categorical quantum mechanics \cite{abramsky,kinder,picturalism,cqm,coecke2017picturing}.  

The mathematical structure underpinning  compositional approaches is the structure of monoidal category \cite{picturalism,selingersurvey}. Informally, a monoidal category describes circuits,   in which wires represent systems and boxes represent operations, as in the following picture  
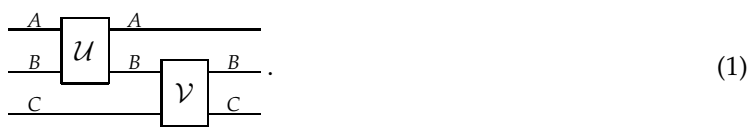
\begin{align}
\begin{aligned}
\Qcircuit @C=1em @R=.7em @! R {    
&  \poloFantasmaCn{A} \qw & \multigate{1}{ \map  U  }   &\poloFantasmaCn{A}\qw  &   \qw  & \qw &\qw  \\
&  \poloFantasmaCn{B} \qw & \ghost{ \map  U  }   &\poloFantasmaCn{B}\qw  &   \multigate{1}{\map V}  & \poloFantasmaCn{B}\qw  &   \qw        \\ 
&  \poloFantasmaCn{C} \qw & \qw   & \qw  &   \ghost{\map V}  & \poloFantasmaCn{C}\qw  &   \qw        }  
   \end{aligned} ~ .\end{align}
The composition of systems is described by a binary operation denoted by $\otimes$, and referred to as the ``tensor product'' \footnote{Note that $\otimes$ is not necessarily a tensor product of vector spaces.}.   The system $A \otimes B$ is interpreted as the composite system made of subsystems $A$ and $B$.   Larger systems are built in a bottom-up fashion, by combining subsystems together.  
 For example, a quantum system of dimension $d=  2^n$ arises as the composite of  $n$ single qubits. 

In some situations,   having a rigid decomposition into subsystems is neither the most convenient nor the most natural approach.   For example, in algebraic quantum field theory \cite{haag2012local} it is natural to start from  a single system---the field---and then to identify subsystems, e.g. spatial or temporal modes. The construction of the subsystems is rather flexible, as there is no privileged decomposition of the field into modes.    Another example of flexible decomposition into subsystems arises in quantum  information,  where it is crucial to identify degrees of freedom that can be treated as ``qubits". Viola,  Knill, and Laflamme \cite{viola2001constructing} and Zanardi, Lidar, and Lloyd \cite{zanardi2004quantum} proposed that the partition of a system into subsystems should  depend on which operations are  experimentally accessible. 
 This flexible definition of subsystem has been exploited in quantum error correction, where  decoherence free subsystems are used to construct logical qubits that are untouched by noise  \cite{palma1996quantum,zanardi1997noiseless,lidar1998decoherence,knill2000theory,zanardi2000stabilizing,kempe2001theory}.  
 The logical qubits are described by ``virtual subsystems"  of the total Hilbert space \cite{zanardi2001virtual}, and in general such subsystems are spread over many physical qubits.  In all these examples, the subsystems are constructed through an algebraic procedure, whereby the subsystems are associated to algebras   of observables \cite{bratteli1987operator}.  For general physical theories, however, the notion of ``algebra of observables" is less appealing, because generally the multiplication of  two observables may not be defined.  For example, in the framework of general probabilistic theories \cite{hardy2001quantum,barnum2007generalized,barrett2007information,chiribella2010probabilistic,barnum2011information,hardy2011foliable,hardy2013formalism,chiribella2014dilation,chiribella2016quantum,hardy2016reconstructing,dariano2017quantum} observables represent measurement procedures, and there is no notion of ``multiplication of two measurement procedures''.

 In this paper we propose a  construction of subsystems that  can be applied to general physical theories, even in scenarios where observables and measurements are not included in  the framework.       The core of our construction is to associate subsystems to sets of {\em operations}, rather than observables.  To fix ideas, it is helpful to think that the operations can be performed by some {\em agent}.  Given a  set of operations, the construction extracts the degrees of freedom that are acted upon {\em only} by those operations, identifying  a ``private space'' that only the agent  can access.    Such a private space then becomes the subsystem, equipped with its own set of states and its own set of operations. 
    This construction is  closely related to an approach proposed  by Kr\"amer and del Rio,  in which the states of a subsystem are identified with equivalence classes of states of the  global system under sets of operations \cite{kraemer2017operational}.  In this paper, we  extend the equivalence relation  to transformations, providing a complete description of the subsystems.   We illustrate the construction in a several examples, including  
 \begin{enumerate}
 \item  quantum subsystems  associated to the tensor product of two Hilbert spaces 
 \item  subsystems associated to an subalgebra of self-adjoint operators on a given Hilbert space
 \item  classical systems of quantum systems 
 \item  subsystems associated to the action of a group representation acting on a given Hilbert space.   
 \end{enumerate}
 In particular, the example of classical channels   has interesting implications for the resource theory of coherence \cite{aberg2006quantifying,baumgratz2014quantifying,levi2014quantitative,winter2016operational,chitambar2016critical,chitambar2016comparison,marvian2016quantify,yadin2016quantum}. 
 Our construction implies that many different types of agents, corresponding to  different choices of free operations in the resource theory of coherence, are associated to the same subsystem, namely the classical subsystem of the given quantum system.       Specifically, the classical system  arises from strictly incoherent operations \cite{yadin2016quantum},   physically incoherent operations  dephasing covariant operations  \cite{chitambar2016critical,chitambar2016comparison},  phase covariant operations \cite{marvian2016quantify}, and  multiphase covariant operations \footnote{To the best of our knowledge, multiphase covariant operations have not been considered so far in the resource theory of coherence.}. Notably, we do not obtain classical subsystems from the maximally incoherent operations  \cite{aberg2006quantifying},  incoherent operations \cite{baumgratz2014quantifying,levi2014quantitative}, which are the first two sets of free operations proposed in the resource theory of coherence. For these two types of operations, we find that the associated subsystem is the whole quantum system.

 After examining the above examples, we explore the general features of our construction.  An interesting feature  is that certain properties, such as the impossibility of instantaneous signalling between two distinct subsystems, arise {\em by fiat}, rather then being postulated as physical requirements.   
   This fact is potentially useful  for the project of finding new axiomatizations of quantum theory    \cite{chiribella2011informational,hardybig,masanes2011derivation,dakic2011quantum,masanes2012existence,wilce2012conjugates,barnum2014higher}, because it suggests that some of the axioms assumed in the usual (compositional) framework may turn out to be consequences of the very  definition of subsystem. Leveraging on this fact, one could hope to find axiomatizations with a smaller number of axioms that pinpoint exactly the distinctive features of quantum theory.      In addition, our construction suggests a {\em desideratum} that every truly fundamental axiom  should  arguably satisfy:  {\em an axiom for quantum theory should hold for all possible subsystems of quantum systems}.  We call this requirement {\em Consistency Across Subsystems.}   If one accepts our broad definition of subsystems, then Consistency Across Subsystems is a  very non-trivial requirement, which is not easily satisfied.  For example, the Subspace Axiom \cite{hardy2001quantum}, stating that all systems with the same number of distinguishable states are equivalent, does not satisfy Consistency Across Subsystems, because classical subsystems are not equivalent to the corresponding quantum systems, even if the dimensions are the same.   
   
   In general,  proving that Consistence Across Subsystems is satisfied may require  great effort. Rather than inspecting the existing axioms and checking whether or not they are consistent across subsystems, one can try to formulate the axioms in a way that guarantees the validity of this property.  We illustrate this idea in the case of the Purification Principle \cite{chiribella2010probabilistic,chiribella2012quantum,chiribella2013quantumtheory,chiribella2014dilation,chiribella2016quantum,chiribella2015conservation,dariano2017quantum}, which is the key ingredient in the quantum axiomatization of Refs.  \cite{chiribella2011informational,chiribella2016quantum,dariano2017quantum} and  plays a central role in the axiomatic foundation of quantum thermodynamics \cite{chiribella2015entanglement,chiribella2017microcanonical,chiribella2016entanglement} and  quantum information protocols   \cite{chiribella2010probabilistic,dariano2017quantum,lee2016generalised,lee2016deriving,lee2017oracles}.  Specifically, we show that the Purification Principle holds for  {\em closed systems}, defined as systems where all transformations are invertible, and where every state can be generated from a fixed initial state by the action of a suitable transformation.   Closed systems  satisfy the  Conservation of Information \cite{susskind2008black}, {\em i.e.} the requirement that physical dynamics should send distinct states to distinct states.   Moreover, the states of the closed systems can be interpreted as ``pure''.    In this setting,  the general notion of subsystem captures the idea of purification, and extends it to a broader setting, allowing us to regard coherent superpositions as the ``purifications" of classical probability distributions.

The paper is structured as follows.  In Section \ref{sec:related} we outline related works.   In Section \ref{sec:framework} we present the main framework and the construction of subsystems. The framework is illustrated with five concrete examples in Section \ref{sec:examples}.   In Section \ref{sec:keystructures}  we discuss the key structures arising from our construction, such as the notion of partial trace and  the validity of the no-signalling property.    In Section \ref{sec:FirstAxioms} we put identify two requirements, concerning the existence of agents with non-overlapping sets of operations, and the ability to all states from a given initial state. We also highlight the relation between the second requirement and the notion of causality.    We then move to systems satisfying the Conservation of Information   (Section \ref{sec:conservation}) and we formalize an abstract  notion of closed systems (Section \ref{sec:closed}). For such systems, we provide a dynamical notion of pure states, and we  prove that every subsystem satisfies   the Purification Principle  (Section \ref{sec:purification}).  A macro-example, dealing with  group representations in quantum theory is provided in Section \ref{sec:groups}, where we interpret the duality between representation spaces and multiplicity spaces as a partition of the system into subsystems.  Finally, the conclusions are drawn in Section \ref{sec:conclusions}.

\section{Related works}\label{sec:related}

In quantum theory, the canonical route to the definition of subsystems is to consider commuting algebras of observables, associated to independent subsystems.  
The idea of defining independence in terms of commutation has a long tradition in quantum field theory and, more recently, quantum information theory.  In algebraic quantum field theory \cite{haag2012local}, the  local subsystems associated to causally disconnected regions of spacetime are described by commuting $C$*-algebras.    A closely related approach is to associate quantum systems to von Neumann algebras, which can be characterized as  double commutants  \cite{takesaki1979theory}.    In quantum error correction, decoherence free subsystems are associated to the commutant of the noise operators \cite{knill2000theory,zanardi2000stabilizing,zanardi2001virtual}.    In this context, Viola,  Knill, and Laflamme \cite{viola2001constructing} and Zanardi, Lidar, and Lloyd \cite{zanardi2004quantum} made the point that subsystems should be defined operationally, in terms of sets of experimentally accessible operations.   
 The canonical approach of associating subsystems to subalgebras  was further generalized by Barnum, Knill,  Ortiz,  and Viola \cite{barnum2004subsystem,barnum2003generalizations}, who  proposed the notion of { generalized entanglement}, i.e. entanglement relative to a subspace of operators. Later, Barnum,  Ortiz, Somma, and Viola explored this notion in the context of general probabilistic  theories \cite{barnum2005generalization}. 
  
The above works provided a concrete model  of subsystems that inspired the present work. An important difference, however,  is that here we will not use the notions of observable and expectation value. In fact, we will not use any probabilistic notion, making our construction usable also in frameworks where no notion of measurement is present. 
This makes the construction appealingly simple, although the downside is that more work will have  to be done in order to recover the probabilistic features that are built-in in other frameworks.  

More recently,  del Rio, Kr\"amer, and Renner \cite{del2015resource}  proposed a general framework for representing the knowledge of agents in general  theories (see also the  PhD theses of del Rio \cite{delrio2015resources} and Kr\"amer \cite{kraemer2016restricted}).  
  Kr\"amer and del Rio further developed the framework to address a number of questions related to locality,  associating agents to monoids of operations, and introducing a relation, called {\em convergence through a monoid}, among states of a global system \cite{kraemer2017operational}. 
  Here we will  extend this relation to transformations, and use it to propose a general definition of subsystem, equipped with its set of states and its set of transformations.

Another related work is the work of Brassard and Raymond-Robichaud on no-signalling and local realism \cite{brassard2017equivalence}.   There, the authors adopt an equivalence relation on  operations, whereby two transformations are equivalent iff they can be transformed into one another through composition with a local reversible transformation.  Such a relation is  related to the equivalence relation on transformations considered in this paper, in the case of systems satisfying  the Conservation of Information.  It is interesting to observe that, notwithstanding the different scopes of Ref. \cite{brassard2017equivalence} and this paper,   the Conservation of Information  plays an important role in both.  Ref.   \cite{brassard2017equivalence}, along with  discussions  with Gilles Brassard during QIP 2017 in Seattle, provided inspiration for the present paper.

\section{Constructing subsystems}\label{sec:framework}

Here we outline the basic definitions and the construction of subsystems.  

\subsection{A pre-operational  framework}

Our starting point is to consider a single system $S$, with a given set of states and a given set of transformations. One could think $S$ to be the whole universe, or, more modestly, our ``universe of discourse'', representing a fragment of the world of which we have made a mathematical model.  
We denote by $\St (S)$ the set of  states of the system (sometimes called the ``state space''), and  by 
$\Transf (S)$ be the set of transformations the system can undergo.  
We assume that $\Transf(S)$ is  equipped with a composition operation $\circ$, which maps a pair of transformations $\map A$ and $\map B$ into the transformation $\map B\circ \map A$. The transformation $\map B\circ \map A$ is physically  interpreted as the transformation occurring when  $\map B$  happens right before  $\map A$.    We also assume that there exists an identity operation $\map I_S$, satisfying the condition $\map A \circ \map I_S  =  \map I_S\circ \map A  =  \map A$ for every transformation $\map A \in \Transf (A)$.  In short, we assume that the physical  transformations form a  {\em monoid}.

We do not assume any structure on the state space $\St (S)$: in particular, we do not assume that $\St (S)$ is convex.  
We do assume, however, is that there is an action of the monoid $\Transf (S)$  on the set $\St (S)$: given an input state $\psi \in \St(S)$ and a transformation $\map T \in \Transf (S)$, the action of the transformation produces the output state  $\map T  \psi\in \St (S)$.

\begin{eg}[Closed quantum systems]
Let us illustrate the basic framework with a textbook example, involving a closed quantum system evolving under unitary dynamics. Here,   $S$ is a quantum system of dimension $d$, and the state space $\St (S)$ is the set of pure quantum states, represented as  rays on the complex vector space $\CC^d$, or equivalently, as rank-one projectors.   With this choice, we have 
\begin{align}
\St (S)   =  \Big\{  |\psi\>\<\psi|     \, : \quad   |\psi\>  \in  \CC^d  \, ,  \quad \< \psi|\psi\>  = 1 \Big\} \, .
\end{align}  

The physical transformations are represented by   unitary channels, {\em i.e.} by maps of the form $|\psi\>\<\psi|   \mapsto  U  |\psi\>\<\psi|U^\dag$, where $U  \in  M_d (\CC)$ is a unitary   $d$-by-$d$  matrix over the complex field.  In short, we have 
\begin{align}
\Transf (S)   =  \Big\{  U \cdot U^\dag       \, : \quad   U  \in   M_d ( \CC)  \, ,  \quad   U^\dag U    =  U^\dag U  =   I  \Big\} \, ,
\end{align}  
where $I$ is the identity matrix.  The physical transformations form a monoid, with the composition operation induced by the  matrix multiplication $(U\cdot U^\dag)\circ  (V \cdot V^\dag)  :=  (UV)  \cdot  (UV)^\dag$.   

\end{eg}

\begin{eg}[Open quantum systems]
Generally, a quantum system can be in a mixed state and can undergo an irreversible evolution.     To account for this scenario, we must take the state space $\St (S)$ to  be the set of all density matrices. For a system  of dimension $d$, this means that the state space is
\begin{align}
\St (S)   =  \Big\{ \rho   \in  M_d(\CC) \, : \quad  \rho  \ge 0  \, \quad  \Tr[\rho]=1   \Big\} \, ,
\end{align}  
where  $\Tr[\rho]   =  \sum_{n=1}^d  \<n|  \rho|n\>$ denotes the matrix trace, and $\rho\ge 0$ means that  the matrix  $\rho$ is positive semidefinite.  
$\Transf(S)$ is the set of all   quantum channels \cite{holevo2003statistical}, {\em i.e.} the set of all linear, completely positive, and trace-preserving maps from $M_d (\CC)$ to itself. The action of the quantum channel $\map T$ on a generic state $\rho$ can be specified through the Kraus representation \cite{kraus1983states}  
\begin{align}
\map T (\rho)  =  \sum_{i=1}^r   T_i \rho  T_i^\dag \, ,
\end{align}
where $\{T_i\}_{i=1}^r  \subseteq M_d (\CC)$ is a set of matrices satisfying the condition $\sum_{i=1}^r  T_i^\dag T_i  =  I$.   The composition of two transformations   $\map T$ and $S$ is given by the composition of the corresponding linear maps.  
\end{eg}
Note that, at this stage, there is no notion of measurement in the framework.   The sets $\St (S)$ and $\Transf(S)$ are meant as a model of system $S$ irrespectively of anybody's ability to measure it, or even to operate on it. 
  For this reason, we call this layer of the framework {\em pre-operational}.    One can think of the pre-operational framework as the arena in which agents will act.  
   Of course, the physical description of such an arena might have been suggested by experiments done earlier on by other agents, but this fact is inessential for the scope of this paper.  
 
 \subsection{Agents} 
 
Let us introduce agents into the picture. 
In our framework an agent  $A$ is identified a  set of actions, 
denoted as $\Act (A; S)$ and interpreted as the possible actions of $A$ on $S$.    Since the actions must  be allowed  physical processes,  the inclusion   $\Act (A; S)  \subseteq \Transf (S)$ must hold.    It is natural, but not strictly necessary,  to assume that the concatenation of two actions is a valid action, and that the identity transformation is a valid action.   When these assumptions are made,  $\Act (A;S)$ is a monoid. Still, the construction presented in the following will   hold not only for monoids, but also  for generic sets $\Act (A;S)$. 
Hence, we adopt the following minimal definition
\begin{defi}[Agents]
An agent $A$ is identified by a subset $\Act (A;S)  \subseteq \Transf (S)$. 
\end{defi}
Note that this definition captures only one aspect of agency.  Other aspects---such as the ability to gather information, make decisions, and interact with other agents---are important too,   but  not necessary for the scope of this paper.  

We also stress that the interpretation of the set  $\Act (A;S)$ as the set of {\em actions of an agent} is  not  strictly necessary for the validity of our results.  Nevertheless, the notion of ``agent'' here is a useful personification, because helps explaining  the rationale of our construction  \footnote{The role of the agent   is somehow similar to the role of a ``probe charge'' in classical electromagnetism. The probe charge need not  exist in reality, but helps---as a conceptual tool---to give  operational meaning to the magnitude and direction of the electric field.}.

In general, the set of actions  available to agent $A$ may be smaller than the set of all physical transformations on $S$.  Also, there may be other agents that  act  on system $S$ independently of agent $A$.   We define the independence of actions in the following way: 
\begin{defi}
Agents $A$ and $B$ {\em act independently} if the order in which they act is irrelevant, namely
\begin{align}\label{independence}
\map A\circ \map B   =  \map B\circ \map A  \, ,\qquad \forall \map A  \in  \Act (A;S) \, , \map B \in  \Act (B;S) \, .
\end{align} 
\end{defi}
In a very primitive sense, the above relation expresses the fact that $A$ and $B$ act on ``different degrees of freedom'' of the system.

\begin{rem}[{\bf Commutation of transformations vs commutation of observables}]  Commutation conditions similar to Eq. (\ref{independence}) are of fundamental importance in  quantum field theory, where they are known under the names of  ``Einstein causality'' \cite{haag1962postulates}  and ``Microcausality'' \cite{haag1964algebraic}.  However, the similarity should not mislead the reader. The field theoretic conditions are expressed in terms of operator algebras. The condition is  that the operators associated to independent systems  commute. For example, a system localized in a certain region could be associated to the operator algebra $\sf A$, and another system localized in another region could be associated to the operator algebra $\sf B$.   In this situation, the  commutation condition  reads 
\begin{align}   CD = DC \qquad \forall C\in{\sf A}, \quad \forall D \in {\sf B} \,.
\end{align}  
In contrast, Eq. (\ref{independence}) is a condition {\em on the transformations},  and not on the observables, which are not even described by our framework. 
 In the field theoretic example, the transformations  of the systems  are described by linear, completely positive maps.    Eq. (\ref{independence}) is a condition on the completely positive maps, and not to the elements of the algebras $\sf A$ and $\sf B$.   In Section \ref{sec:examples} we will     bridge the gap between our framework and the usual algebraic framework, focussing on the scenario where $\sf A$ and $\sf B$ are finite dimensional von Neumann algebras.  
\end{rem}


\subsection{Adversaries and degradation}
From the point of view of  agent  $A$,   it is important to identify the degrees of freedom that no other agent $B$ can affect.    In an adversarial setting,  agent   $B$ can be viewed as an adversary that tries to control as much of the system  as possible. 
\begin{defi}[Adversary]
Let $A$ be an agent and let $\Act (A;S)$ be her set of operations.  An {\em adversary} of $A$ is an agent $B$ that acts independently of $A$, {\em i.e.} an agent $B$ whose set of actions satisfies
\begin{align}
\Act (B;S)   \subseteq \Act (A;S)'    :   =  \Big\{  \map B  \in \Transf(S) \, :    \quad  \map B \circ \map A  =  \map A\circ \map B \, , \forall \map A \in  \Act(A;S)\Big\}\, .
\end{align}
\end{defi} 
Like the agent, the adversary is a conceptual tool, which will be used to illustrate our notion of subsystem. The adversary need not be a real physical entity, localized outside the agent's laboratory, and trying to counteract the agent's actions.  Mathematically, the adversary is just a subset of the commutant of $\Act(A;S)$.  Again, the interpretation of $B$ as an ``adversary'' is a way to ``give life to to the mathematics'', and to illustrate the rationale of our construction.   

When $B$ is interpreted as an adversary,  we can think of his actions as a ``degradation'', which compromises  states and transformations. 
We denote the degradation relation as  $\succeq_B$, and write 
\begin{align}
\phi  \succeq_B  \psi  \qquad {\rm iff}  \qquad &  \exists \map B  \in  \Act(B; S)  :   \psi =  \map B  \, \phi  \, ,  \\    
\map S  \succeq_B  \map T  \qquad {\rm iff}  \qquad &  \exists \map B_1,\map B_2  \in  \Act(B; S)  :  \, \map T  = \map B_1  \circ  \map S  \circ \map B_2 
\end{align}
for $\phi,\psi  \in  \St(S)$ or $\map S, \map T \in \Transf (S)$.

The states that can be obtained by degrading $\psi$ will be denoted as  
\begin{align}\Deg _B(\psi)  :=  \Big\{     \map B \psi  \, : \quad \map B\in\Act (B;S) \Big\}
\end{align} 
and the transformations that can be obtained by degrading $\map T$ will be denoted as 
\begin{align}
\Deg_B (\map T):=  \Big\{   \map B_1\circ  \map T\circ \map B_2  \, : \quad \map B_1,\map B_2 \in\Act (B;S)    \Big\}
\end{align}

The more operations $B$ can perform, the more powerful $B$ will be as an adversary.   
The most powerful adversary compatible with the independence condition (\ref{independence}) is the adversary that can implement all physical transformations in the commutant of $\Act(A;S)$:  
\begin{defi} The {\em maximal adversary} of agent $A$ is the agent $A'$ that can perform actions  $\Act \big(A' ;  S\big)  :  = \Act (A; S)'$. 
\end{defi}
Note that the actions of the maximal adversary are automatically a monoid, even if the set $\Act(A;S)$ is not. Indeed,   
\begin{itemize}
\item the identity map $\map I_S$ commutes with all operations  in $\Act(A;S)$, and
\item if $\map B$ and $\map B'$ commute with every operation in $\Act(A;S)$, then also their composition  $\map B \circ \map B'$ will commute with all the operations in $\Act (A;S)$.  
\end{itemize}
In the following we will use the  maximal adversary to define the subsystem associated to agent $A$.  

\subsection{The states of the  subsystem} 

Given an agent $A$, we think of  the  subsystem $S_A$ to be the collection of all degrees of freedom that are unaffected by the action of the maximal adversary $A'$.   Consistently with this intuitive picture,  we partition the states of $S$ into disjoint subsets, with the interpretation that two  states are in the same subset   if and only if they correspond to the same state of   subsystem $S_A$.   We  denote by $\Lambda_\psi$ the subset  of $\St (S)$ containing the state $\psi$. 

To construct the state space of the subsystem, we adopt the following rule: 
\begin{ru}\label{rule:sameclass}
If the state $\psi$ is obtained from the state $\phi$ through degradation, {\em i.e.} if $ \psi  \in  \Deg_{A'} (\phi)$, then $\psi$ and $\phi$  must correspond to the same state of subsystem $S_A$, {\em i.e.} one must have $\Lambda_\psi  = \Lambda_{\phi}$. 
\end{ru}

Rule \ref{rule:sameclass} imposes that all  states in the set  $\Deg_{A'} (\psi)$ must be contained in the set $\Lambda_\psi$. 
 Furthermore, we have the following simple fact:  
 \begin{prop}\label{prop:inter}
 If the sets $\Deg_{A'} (\phi)$ and $\Deg_{A'} (\psi)$  have non-trivial intersection,   then  $\Lambda_\phi  =  \Lambda_{\psi}$
 \end{prop}
 \Proof By Rule {rule:sameclass}, every element of    $\Deg_{A'} (\phi)$ is contained in  $\Lambda_\phi$. Similarly,  every element of 
      $\Deg_{A'} (\psi)$ is contained in  $\Lambda_\psi$.  Hence, if   $\Deg_{A'} (\phi)$ and $\Deg_{A'} (\psi)$  have non-trivial intersection, then also $\Lambda_\phi$ and $\Lambda_\psi$ have non-trivial intersection.  Since the sets  $\Lambda_\phi$ and $\Lambda_\psi$ belong to a disjoint partition, we conclude that $\Lambda_\phi =\Lambda_\psi$.  \qed 
   
      \medskip 
   
          Generalizing the above argument, it is clear that two states $\phi$ and $\psi$ must be in the same subset $\Lambda_\phi  =  \Lambda_{\psi}$  if there exists a finite sequence $(\psi_1,\psi_2,\dots,  \psi_n) \subseteq  \St (S)$ such that 
\begin{align}
\psi_1 =\phi  \, ,  \qquad \psi_n  =  \psi  \, ,\qquad    {\rm and} \qquad  \Deg_{A'} (\psi_i)  \cap  \Deg_{A'} (\psi_{i+1})\not  =  \emptyset   \quad \forall  i\in  \{1,2,\dots,  n-1\} \, .   
\end{align}
 When this is the case, we write $\phi   \simeq_{A }\psi $.  Note that the relation $\phi   \simeq_{A} \psi $ is  an equivalence relation.    When the relation $\phi   \simeq_{A}\psi $ holds, we say that $\phi$
 and $\psi$ are {\em equivalent for agent $A$}.   We denote the equivalence class of the state $\psi$  by $[\psi]_A$. 
 
  By Rule \ref{rule:sameclass},  the whole equivalence class $[\psi]_A$ must be contained in the set $\Lambda_\psi$, meaning that all states in the equivalence class   must correspond to the same state of subsystem $S_A$. 
 Since we are not constrained by any other condition,   we make the minimal choice
 \begin{align}
 \Lambda_\psi   :=  [\psi]_{A} \,     .
 \end{align}   
 In summary, the state space of system $S_A$ is 
 \begin{align}\label{sta}
 \St (S_A)    :  =  \Big\{  [\psi]_{A}  :    ~      \psi \in\St (S) \Big\} \, . 
 \end{align}

  \subsection{The transformations of a subsystem} 
  
The   transformations of system  $S_A$ can also be constructed through  equivalence classes.    But before taking equivalence classes,  we need a candidate set of transformations that we can interpret as  acting {exclusively on the  degrees of freedom associated to agent $A$}.   The largest candidate  set is the  set of all transformations that commute with the actions  of the maximal adversary $A'$, namely  
   \begin{align}
\Act (A';S)'     =  \Act (A;S)^{\prime \prime}   \, . 
   \end{align}
In general $\Act  (A;S)^{\prime \prime}$ could be larger than $\Act (A;S)$, in agreement with the fact  the set of physical transformations of system $S_A$    could be larger than the set of operations that agent $A$ can perform. For example, agent $A$ could have access only to a subset of noisy operations on her subsystem, while another, more technologically advanced agent could perform more accurate operations on the same subsystem. 

For two transformations  $\map S$ and $\map T$ in $\Act  (A;S)^{\prime \prime}$, the degradation relation  $\succeq_{A'}$ takes the simple form  
\begin{align}
\map S  \succeq_{A'}  \map T  \qquad {\rm iff}  \qquad &  \, \map T  = \map B  \circ  \map S \quad {\rm for ~some~}  \map B \in  \Act(A'; S)    \, . 
\end{align}

As we did for the set of states,  we now partition the set $\Act(A;S)^{\prime \prime}$ into disjoint subsets, with the interpretation that two transformations act in the same way on the subsystem $S_A$ if and only if they belong to the same subset.    Let us denote by $\Theta_\map A$ the subset  containing the transformation $\map A$.   
To find the appropriate partition of  $\Act(A;S)^{\prime \prime}$ into disjoint subsets, we adopt the following rule
\begin{ru}\label{rule:sameclasstransf}
If the transformation $\map T \in  \Act(A;S)^{\prime \prime}$ is obtained from the transformation $\map S  \in  \Act(A;S)^{\prime \prime}$ through degradation, {\em i.e.} if $ \map T  \in  \Deg_{A'} (\map S)$, then $\map T$ and $\map S$  must act in the same way on the subsystem $S_A$, {\em i.e.} they must satisfy  $\Theta_{\map T}  = \Theta_{\map S}$. 
\end{ru}
Intuitively, the motivation for the above rule is that $\map T$ and $\map S$ differ only by an operation performed by the adversary, and system $S_A$ is {\em defined} as the system that is not affected by the action of the adversary.

    Rule \ref{rule:sameclasstransf} implies that all the transformations in $\Deg_{A'} (\map T)$ must be contained in $\Theta_{\map T}$.    Moreover, we have the following 
    \begin{prop}
 If the sets $\Deg_{A'} (\map S)$ and $\Deg_{A'} (\map T)$  have non-trivial intersection,   then  $\Theta_{\map S}  =  \Theta_{\map T}$.  
 \end{prop}
 \Proof By Rule \ref{rule:sameclasstransf}, every element of    $\Deg_{A'} (\map S)$ is contained in  $\Theta_{\map S} $. Similarly,  every element of  $\Deg_{A'} (\map T)$ is contained in  $\Theta_{\map T} $.  Hence, if   $\Deg_{A'} (\map S)$ and $\Deg_{A'} (\map T)$  have non-trivial intersection, then also $\Theta_{\map S} $ and $\Theta_{\map T} $ have non-trivial intersection.  Since the sets  $\Lambda_{\map S}$ and $\Lambda_{\map T}$ belong to a disjoint partition, we conclude that $\Lambda_{\map S} =\Lambda_{\map T}$.  \qed 
   
   \medskip 
   
   Using the above proposition, we obtain that  the equality $\Theta_{\map T}  = \Theta_{\map S}$ whenever  there exists a finite sequence $(\map A_1,\map A_2,\dots, \map A_n)  \subseteq \Act (A;S)^{\prime \prime}$ such that  
\begin{align}
\map A_1 =\map S  \, ,  \qquad \map A_n  =  \map T  \, ,\qquad    {\rm and} \qquad  \Deg_{A'} (\map A_i)  \cap  \Deg_{A'} (\map A_{i+1})  \not = \emptyset  \quad \forall  i\in  \{1,2,\dots,  n-1\} \, .   
\end{align}
When the above relation is satisfied, we  write  $\map S  \simeq_{A}  \map T$. Note  that   $\simeq_{A}$   is an equivalence relation.  When the relation  $\map S  \simeq_{A}  \map T$ holds, we say that {\em $\map S$ and $\map T$ are equivalent for agent $A$}.  

 By Rule \ref{rule:sameclasstransf},  all the elements of  $[\map T]_{A}$ must  be contained in the set $\Theta_{\map T}$, {\em i.e.} they should correspond to the same transformation on $S_A$.  Again, we make the minimal choice: we  stipulate that the set $\Theta_{\map T}$ coincides exactly with the equivalence class $[\map T]_{A}$. Hence, the transformations  of subsystem $S_A$ are 
   \begin{align}
 \Transf (S_A)  :  =  \Big \{  [\map T]_{A}  : ~  \map T  \in  \Act (A; S)^{\prime \prime}  \Big\} \, .
 \end{align}

The composition of two transformations $[\map T_1]_{A}$ and $[\map T_2]_{A}$ is defined in the obvious way, namely 
\begin{align}\label{def:composition}
[\map T_1]_{A}  \circ [\map T_2]_{A}   :  =  [\map T_1 \circ \map T_2]_{A} \, .
\end{align}
 Similarly, the action of the transformations on the states is as
  \begin{align}\label{def:action}
[ \map T]_{A}   \,    [ \psi]_{A}  : =  [\map T \, \psi]_{A} \, .
 \end{align}
In Appendix \ref{app:wellposed} we show that  definitions  (\ref{def:composition}) and (\ref{def:action}) are well-posed, in the sense that their right hand sides are independent of the choice of representatives within the equivalence classes.

 \bigskip

{\bf Remark.}  It is important not to confuse the transformation $\map T\in  \Act (A;S)^{\prime \prime}$ with the equivalence class $[\map T]_{A}$: the former is a transformation {\em on the whole system} $S$, while the latter is  a transformation {\em only on subsystem} $S_A$.  
To keep track of the distinction,  we define the {\em restriction}  of the action  $\map T \in\Act (A;S)^{\prime \prime}$ to the subsystem $S_A$ via the map 
\begin{align}
\pi_A   (\map T)   :  =  [\map T]_{A}   \qquad \forall \map T \in  \Act (A;S)^{\prime \prime}  \, . 
\end{align}

\begin{prop}
The restriction map $\pi_A:   \Act (A;S)^{\prime \prime}  \to  \Transf (S_A)$ is a monoid homomorphism, namely $\pi_A  (\map I_S)  =  \map I_{S_A}$ and $\pi_A  (\map S  \circ \map T)    =  \pi_A  (\map S)  \circ  \pi_A  (\map T)$ for every pair of transformations $\map S,\map   T\in\Act (A;S)^{\prime \prime}$.  
\end{prop}
\Proof Immediate from the definition (\ref{def:composition}). \qed 

\medskip 

\section{Examples of agents, adversaries, and subsystems}\label{sec:examples}  

In this section we illustrate the construction of subsystems  in five concrete examples.  

\subsection{Tensor product of two quantum systems}\label{subsec:product}

Let us start from the obvious example, which will serve as a sanity check for the soundness of  our construction.  
Let $S$ be  a  quantum system with Hilbert space of the tensor product form $\spc H_{S}   =  \spc H_A\otimes \spc H_B$.     The states of $S$ are all the density operators on the Hilbert space $\spc H_S$.  The space of all linear operators from $\spc H_S$ to itself will be denoted as $\Lin (\spc H_S)$,  so that 
\begin{align}
\St  (S)    =  \Big\{     \rho  \in  \Lin (\spc H_S)\,  : \quad  \rho \ge 0  ,  \quad  \Tr [\rho]  =1 \Big\} \, .
\end{align}  
The transformations are all the quantum channels (linear, CP, and trace-preserving linear maps) from $\Lin  (\spc H_S)$ to itself. We will denote the set of all channels on system $S$ as $\CChan (S)$.      Similarly, we will use the notation $\Lin  (\spc H_A)$   [$\Lin (\spc H_B)$]
 for the spaces of linear operators from $\spc H_A$ [$\spc H_B$] to itself,  and the notation $\CChan  (A)$  [$\CChan(  B)$] for the quantum channels from $\Lin (\spc H_A)$  [$\Lin (\spc H_B)$] to itself.

We can now define an agent  $A$ whose actions are all quantum channels acting locally on system $A$, namely 
\begin{align}
\Act (A;S)    :=  
\Big \{    \map A  \otimes \map I_B  \, :   \quad  \map A  \in  \CChan (A) \Big\} \, ,   
\end{align} 
where    $\map I_B$ denotes the identity map on $\Lin (\spc H_B)$.  
It is relatively easy to see that the commutant of $\Act (A;S)$ is  
\begin{align}
\Act (A;S)'   =   \Big \{    \map I_A  \otimes \map B  \, : \quad   \map B  \in  \CChan (B) \Big\}    
\end{align}
(see Appendix \ref{app:commutant} for the proof).  
Hence, the maximal adversary of agent $A$ is the adversary $A'  =  B$ that has full control on the Hilbert space  $\spc H_B$.  Note also  that one has $\Act(A;S)^{\prime \prime}   =   \Act (A;S)$.

Now, the following fact holds: 
\begin{prop} 
Two states $\rho,\, \sigma \in \St (S)$ are equivalent for agent $A$  if and only if $\Tr_B  [\rho]  =  \Tr_B [\sigma]$, where $\Tr_B$ denotes the partial trace over the Hilbert space $\spc H_B$.   
\end{prop}  
\Proof  Suppose that the equivalence $\rho  \simeq_{A} \sigma$ holds.  By definition, this means that there exists  a finite sequence $(\rho_1,\rho_2,\dots,  \rho_n)$ such that 
\begin{align}
\rho_1 =\rho  \, ,  \qquad \rho_n  =  \sigma  \, ,\qquad    {\rm and} \qquad  \Deg_B (\rho_i)  \cap  \Deg_B (\rho_{i+1})\not  =  \emptyset   \quad \forall  i\in  \{1,2,\dots,  n-1\} \, .   
\end{align} 
In turn, the condition of non-trivial intersection implies that, for every $i \in  \{1,2,\dots, n-1\}$,  one has 
\begin{align}\label{poiu}
(\map I_A \otimes \map B_i)  \, (\rho_i)     =   (\map I_A\otimes \widetilde{\map B}_i) \,  (\rho_{i+1})  \, ,
\end{align}
where  $\map B_i$ and  $\widetilde{ \map B}_i$ are two quantum channels in  $\CChan (B)$.   Since $\map B_i$ and $\widetilde {\map B_i}$ are trace-preserving, the above condition implies  $\Tr_B  [\rho_i]  =  \Tr_B [\rho_{i+1}]$, as one can see by taking   the partial trace on $\spc H_B$  on both sides of Eq. (\ref{poiu}).   In conclusion, we obtained the equality  $\Tr_B [\rho]   \equiv  \Tr_B  [\rho_1]   =  \Tr_B[\rho_2]  =  \dots  =  \Tr_B [\rho_n]  \equiv  \Tr_B [\sigma]$.   

Conversely, suppose that the condition  $\Tr_B  [\rho]  =  \Tr_B [\sigma]$ holds.   Then, one has 
\begin{align}\label{above}
(\map I_A \otimes \map B_0)  \, (\rho)     =   (\map I_A\otimes  {\map B}_0) \,  (\sigma)  \, ,
\end{align} where $\map B_0 \in  \CChan (B)$ is the erasure channel defined as $\map B_0   (\cdot)   =   \beta_0 \,  \Tr_B  [\cdot] $,   $\beta_0$ being an arbitrary density matrix in $\Lin (\spc H_B)$.        Since $\map I_A\otimes \spc B_0$ is an element of $\Act (B;S)$,   Eq. (\ref{above}) shows that the intersection between $\Deg_B (\rho)$ and $\Deg_B (\sigma)$ is non-empty.  Hence, $\rho$ and $\sigma$ correspond to the same state of system $S_A$. \qed   

\medskip 

We have seen that two global states $\rho, \sigma \in \St (S)$ are equivalent for agent  $A$ if and only if they have the same partial trace over $B$.   Hence, the  state space of the subsystem $S_A$ is
\begin{align}
\St (S_A)   =   \Big \{   \Tr_B[\rho]\, :   \qquad   \rho  \in  \St (S) \Big \}  \, ,
\end{align}  
 consistently with the standard prescription of quantum mechanics. 

\medskip 

Now, le us consider the transformations.   It is not hard to show that two transformations $\map T,\map S  \in  \Act( A;S)^{\prime \prime}$  are equivalent if and only if $\Tr_B  \circ \map T =  \Tr_B \circ \map S$  (see Appendix \ref{app:commutant} for the details).  Recalling that the transformations in $\Act(A;S)^{\prime \prime}$ are of the form $\map A\otimes \map I_B$, for some $\map A  \in  \CChan (A)$, we obtain that the set of transformations of $S_A$ is 
\begin{align}
\Transf (S_A)  =    \CChan (A)  \,.
\end{align}
 In summary, our construction correctly identifies the  quantum subsystem associated to the Hilbert space $\spc H_A$, with the right set of states and the right set of physical transformations. 
 

\subsection{Subsystems associated to finite dimensional von Neumann algebras}\label{subsec:vN}  

In this example we show that our notion of subsystem is equivalent to the traditional notion of subsystem based on an algebra of observables.     For simplicity, we restrict our attention to a quantum system $S$ with finite dimensional Hilbert space $\spc H_S  \simeq \CC^d$, $d<\infty$.  
 With this choice, the  state space  $\St (S)$ is the set of all density matrices in $M_d (\CC)$ and the transformation monoid $\Transf(S)$ is the set of all quantum channels (linear, completely positive, trace-preserving maps) from $M_d (\CC)$ to itself.

We now define an agent $A$  associated with a  von Neumann algebra  $\sf A  \subseteq M_d (\CC)$.   In the finite dimensional setting, a von Neumann algebra is just a matrix algebra that contains the identity operator and is closed under the matrix adjoint.  
Every such algebra  can be decomposed in a block diagonal form.  Explicitly, one can decompose  the Hilbert space $\spc H_S$ as 
  \begin{align}\label{wedderburn}
  \spc H_S  =  \bigoplus_k  \,   \left ( \spc H_{A_k}  \otimes \spc H_{B_k} \right) \, ,  
 \end{align}
  for appropriate Hilbert spaces $\spc H_{A_k}$  and $\spc H_{B_k}$.   Relative to this decomposition, the elements of the algebra $\sf A$   are characterized as follows:
  \begin{align}\label{blockA}
 C  \in {\sf A} \qquad \Longleftrightarrow  \qquad C  =  \bigoplus_k  \left(   C_k\otimes I_{B_k}\right)  \, , 
  \end{align} 
where $C_k$ is an operator in $\Lin (\spc H_{A_k})$, and $I_{B_k}$ is the identity on $\spc H_{B_k}$.   The elements of the commutant  algebra $\sf A'$  are characterized as  follows:
\begin{align}\label{blockA'}
  D \in {\sf A'}  \qquad \Longleftrightarrow  \qquad  D  =  \bigoplus_k  \left(   I_{A_k}  \otimes   D_k\right)   \, , 
  \end{align}   
  where $I_{A_k}$ is the identity  on $\spc H_{A_k}$ and $D_k$ is an operator in $\Lin (\spc  H_{B_k})$. 
  
 We grant  agent $A$ the ability to implement all  quantum channels with Kraus operators in the algebra $\sf A$, {\em i.e.} all quantum channels in the set  
 \begin{align}
\CChan  (\sf A)  &:=  \Big\{  \map C  \in  \CChan (S) \, :      \quad \map C (\cdot)   =  \sum_{i=1}^r    C_i  \cdot C_i^\dag \,, \quad    C_i   \in  \sf A   ~  \forall i\in  \{1,\dots, r\}\Big\} \, . 
\end{align}

The maximal adversary of agent $A$ is the agent $B$ who can implement all the quantum channels that commute with the channels in $\CChan (\sf A)$, namely 
\begin{align}
\Act (B;S)     =  \CChan  (\sf A)' \, .
\end{align} 
In Appendix \ref{app:algebras}, we prove that  $\CChan (\sf A)'$ coincides with the set of  quantum channels with Kraus operators in the commutant of the algebra  $\sf A$:       in formula, 
\begin{align} 
\CChan  (\sf A)'    =  \CChan  (\sf A')  \, .  
\end{align} 

As in the previous example,  the states of subsystem $S_A$  can be characterized as ``partial traces'' of the states in $S$, provided that one adopts the right  definition of ``partial trace''.    Denoting the commutant of the algebra $\sf A$ by $\sf B  :  =  \sf A'$, one can define the ``partial trace over the algebra $\sf B$'' as the  channel    $\Tr_{\sf B}:  \Lin  (\spc H_S)  \to \bigoplus_k  \Lin (\spc H_{A_k})$ specified by the relation 
\begin{align}\label{partialtrace}
\Tr_{\sf B} (\rho)  :  =  \bigoplus_k \,        \Tr_{B_k}\Big [\Pi_k  \, \rho  \Pi_k \Big] \, , 
\end{align}  
where $\Pi_k$ is the projector on the subspace $\spc H_{A_k}\otimes \spc H_{B_k} \subseteq \spc H_S$,  and $\Tr_{B_k}$ denotes the partial trace over the space $\spc H_{B_k}$.  With  definition (\ref{partialtrace}),  is not hard to see that two states are equivalent for $A$ if and only if they have the same partial trace over $\sf B$: 
\begin{prop}\label{prop:algebrastates} 
Two states $\rho,\, \sigma \in \St (S)$ are equivalent for $A$ if and only if $\Tr_{\sf B}  [\rho]  =  \Tr_{\sf B} [\sigma]$. 
\end{prop}  
The proof is provided in Appendix \ref{app:algebras}.   In summary, the states of system $\St (S_A)$ are obtained from the states of $S$ via partial trace over $\sf B$, namely
\begin{align}
\St (S_A)    =  \Big  \{  \Tr_{\sf B} (\rho)  \, :  \quad  \rho  \in   \St (S)  \Big\}   \, .
\end{align}
Our construction is consistent with the standard algebraic construction, where the states of system $S_A$ are defined as restrictions of the global states to the subalgebra $\sf A$:  indeed, for every element $C  \in  \sf A$, we have the relation
\begin{align}
\Tr [    C  \,  \rho]  &  =    \Tr \left[ \left(  \bigoplus_k  \,   C_k\otimes I_{B_k}  \right)  \,   \rho  \right]  \nonumber\\
  &  = \sum_k  \,  \Tr[  (C_k\otimes I_{B_k} )\,  \Pi_k  \rho \Pi_k]  \nonumber \\
  &  = \sum_k  \Tr  \Big\{  C_k  \,  \Tr_{B_k}  [\Pi_k  \rho \Pi_k]  \Big\} \nonumber \\
  &   =  \Tr\Big\{   \check C\Tr_{\sf B}  [\rho] \Big\}  \, , \qquad  \check C   :  =  \bigoplus_k  C_k  \, ,
\end{align} 
meaning that the restriction of the state $\rho$ to the subalgebra $\sf A$ is in one-to-one correspondence with the state $\Tr_{\sf B} [\rho]$.  

Alternatively, the states of the subsystem $S_A$  can be characterized as the density matrices of the block diagonal form 
\begin{align}
\sigma   = \bigoplus_k\,  p_k  \,  \sigma_k \, ,
\end{align}
where $(p_k)$ is a probability distribution, and each $\sigma_k$ is a density matrix in $\Lin  (\spc H_{A_k})$.  In    Appendix \ref{app:algebras} we characterize 
 the transformations of the subsystem $S_A$ as   quantum channels   $\map A$ of the form  
\begin{align}
\map A      =  \bigoplus_k  \,  \map A_k  \, ,  
\end{align}
where $\map A_k:   \Lin  (\spc H_{A_k}) \to \Lin (\spc H_{A_k})$ is a linear, CP, and trace-preserving map. 
 In summary, the subsystem $S_A$ is a direct sum of quantum systems, corresponding to the Hilbert spaces $\spc H_{A_k}$.

\subsection{Coherent superpositions vs incoherent mixtures  in  closed-system quantum theory}\label{subsec:coherence}  

We now analyze an example involving only pure states and reversible transformations.  Let $S$ be a single quantum system with  Hilbert space  $\spc H_S   =  \CC^d \, , d< \infty$, equipped with a distinguished orthonormal basis $\{  |n\>  \}_{n=1}^d$.  As the state space,  we consider  the set of   pure quantum states:  in formula,
\begin{align}
\St (S)   =  \Big \{ |\psi\>\<\psi| \, : \quad |\psi\>  \in \CC^d \, , \quad  \<\psi|\psi\>  = 1       \Big\} \, .
\end{align}   
As the  set of transformations, we consider  the set of unitary   channels: in formula,   
\begin{align}
\Transf (S)   =  \Big\{  U \cdot U^\dag       \, : \quad   U  \in   M_d ( \CC)  \, ,  \quad   U^\dag U    =  U^\dag U  =   I  \Big\} \, .
\end{align}

To agent $A$, we grant the ability to implement  all unitary channels corresponding to diagonal unitary matrices, {\em i.e.}  matrices of the form 
\begin{align}\label{multiphaseU}
U_{\bs \theta}   =  \sum_k\,   e^{i\theta_k} \,   |k\>\<k|   \,  , \qquad \bs \theta =   (\theta_1,\dots, \theta_d)  \in  [0,2\pi)^{\times d} \,,  
\end{align} 
where each phase $\theta_n$ can vary independently of the other phases.   In formula, the set of actions of agent $A$ is 
\begin{align}
\Act  (A;S)   =  \Big \{     U_{\bs \theta}  \cdot  U_{\bs \theta}^\dag \, :  \quad  U_{\bs \theta} \in   \Lin (\spc H_S) \, ,\quad  U_{\bs \theta} ~{\rm as~in~Eq.~(\ref{multiphaseU})}     \Big\}
\end{align}

The peculiarity of this example is that the actions of the maximal  adversary $A'$ are exactly the same as the actions of $A$. It is immediate to see that $\Act (A;S)$ is included in $\Act (A';S)$,  because all  operations of agent $A$ commute.  With a bit of extra work one can see  that, in fact,  $\Act (A;S)$ and $\Act (A';S)$ coincide.

Let us look at the subsystem associated to agent $A$. 
   The equivalence relation among states takes a simple form:  
   \begin{prop}\label{prop:diagU}  
   The pure states corresponding to the unit vectors  $|\phi\>,  |\psi\>  \in \spc H_S$ are equivalent for $A$ if and only if $|\psi\>  =  U|\phi\>$ for some diagonal unitary  matrix $U$.  
   \end{prop} 
   \Proof Suppose that there exists a finite sequence $(|\psi_1\>,  |\psi_2\>,\dots,  |\psi_n\>)$ such that 
\begin{align*}
|\psi_1\> =|\phi\>  \, ,  \qquad |\psi_n\>  =  |\psi \>  \, ,\qquad    {\rm and} \qquad  \Deg_{A'} (|\psi_i\>\<\psi_i|)  \cap  \Deg_{A'} (|\psi_{i+1}\>\<\psi_{i+1}|)\not  =  \emptyset   \quad \forall  i\in  \{1,2,\dots,  n-1\} \, .   
\end{align*}   
 This means that, for every $i\in  \{1,\dots,  n-1\}$, there exist two diagonal unitary matrices $U_i$ and $\widetilde U_i$ such that  $U_i  |\psi_i\>    =   \widetilde U_i  |\psi_{i+1}\>$, or equivalently, 
 \begin{align}
 |\psi_{i+1}\>   =    \widetilde U_i^\dag  \, U_i  |\psi_{i}\>  \, . 
 \end{align}  
 Using the above relation for all values of $i$,  we obtain $|\psi\>  =   U  |\phi\>$ with $U  :=   \widetilde U_{n-1}^\dag  \, U_{n-1}     \cdots \widetilde U_2^\dag  \, U_2    \widetilde U_1^\dag  \, U_1$.  
 
Conversely, suppose that the condition $|\psi\>  =  U|\phi\>$  holds for some diagonal unitary matrix $U$.  Then, the intersection $\Deg_{A'} (  |\phi\>\<\phi|) \cap  \Deg_{A'}  (|\psi\>\<\psi|)$ is non-empty, which implies  that $|\phi\>\<\phi|$ and  $|\psi\>\<\psi|$ are in the same equivalence class.  \qed

\medskip     Using Proposition \ref{prop:diagU}  it is immediate  to see that the equivalence class $[|\psi\>\<\psi|]_{A'}$ is uniquely identified by the diagonal density matrix  $\rho  =  \sum_k \,  |\psi_k|^2 \,  |k\>\<k|$.    Hence, the state space of system $S_A$ is the set of diagonal density matrices 
 \begin{align}
 \St (S_A )  =  \Big \{   \rho  =  \sum_k  \,   p_k  \,  |k\>\<k|  \, :   \quad p_k  \ge 0 ~ \forall k \, , \quad \sum_k   p_k  =  1 \Big\} \, .
 \end{align}
The set of transformations of system $S_A$ is trivial, because the actions of $A$ coincide with the actions of the adversary $A'$, and therefore they are all  in the equivalence class of the identity transformation. In formula, one has  
\begin{align}
\Transf  (S_A)  =  \Big\{  \map I_{S_A}\Big\} \, .
\end{align}  

\subsection{Classical subsystems in open-system quantum theory}\label{subsec:multiphase}

This example is of the same flavour as the previous one, but  is more elaborate and more interesting. 
Again, we  consider a  quantum system  $S$ with Hilbert space $\spc H  =  \CC^d$. Now, we take $\St (S)$ to be the whole set of density matrices in  $M_d(\CC)$ and $\Transf (S)$ to be the whole set of quantum channels from $M_d (\CC)$ to itself. 

We grant to agent $A$ the ability to perform every multiphase covariant channel, that is, every quantum channel $\map M$ satisfying the condition  
\begin{align}  
\map U_{\bs \theta} \circ \map M  =  \map M \circ \map U_{\bs \theta} \qquad \forall \bs \theta   = (\theta_1,\theta_2,\dots,  \theta_d) \in  [0,2\pi)^{\times d} \, , 
\end{align}
where $\map U_{\bs \theta}  =  U_{\bs \theta}  \cdot U_{\bs \theta}^\dag$  is the unitary channel corresponding to the diagonal unitary $U_{\bs \theta }   =  \sum_k  \,  e^{i\theta_k} \,  |k\>\<k|$.   Physically, we can interpret the restriction to multiphase covariant channels as the lack of a reference for the definition of the phases in the basis $\{|k\>  \, ,  k=1,\dots, d\}$.    

 It turns out that the maximal  adversary of agent $A$ is the agent  $A'$ that can perform  every {\em basis-preserving channel}   $\map B$,  that is, every channel satisfying the condition  
\begin{align}
\map B(|k\>\<k|)   =     |k\>\<k|  \qquad \forall k\in\{1,\dots, d\} \, .
\end{align}  
Indeed, we have the following 
\begin{theo}\label{theo:dualmultiphase}
The monoid of multiphase covariant channels and the monoid of basis-preserving channels are  the commutant of one another.  
\end{theo}
The proof, presented in Appendix \ref{app:multi-bpres}, is based on the characterization of the basis-preserving  channels provided in \cite{buscemi2005inverting,buscemi2007quantum}. 


We now show that states of system $S_A$ can be characterized as classical probability distributions. 
\begin{prop}\label{prop:diagrho}
For every pair of states  $\rho ,\sigma  \in \St (S)$, the following are equivalent 
\begin{enumerate}
\item $\rho$ and $\sigma$ are equivalent for agent $A$
\item  $  \map D ( \rho )    = \map D(  \sigma  ) $, where $\map D$ is the completely dephasing channel $\map D  (\cdot) :=   \sum_k    \,   |k\>\< k|  \cdot  |k\>\<k|$.
\end{enumerate}
\end{prop}   
\Proof  Suppose that Condition 1 holds, meaning that  there exists a sequence $(\rho_1, \rho_2,\dots, \rho_n)$ such that  
\begin{align}
\rho_1 = \rho  \, , \qquad \rho_n=  \sigma \, , \qquad  \forall i \in  \{1,\dots, n-1\} \,  \exists   \map B_i  \, ,  \widetilde {\map B_i} \in  \Act(B;S)  :  ~  \map B_i  (\rho_i)  =  \widetilde {\map B_i}  (\rho_{i+1}) \, ,
\end{align} 
where  $\map B_i$ and $\widetilde {\map B_i}$ are basis-preserving channels. The above equation  implies 
\begin{align}\label{anima}
 \< k|  \map B_i  (\rho_i)  |k\>  =   \<  k|  \widetilde{\map B_i}  (\rho_{i+1})  |k\>  \, .
 \end{align}
Now, we use the relation $  \< k|  \map B  (\rho)  |k\> =  \<  k|   \rho  |k\> $, valid for every basis-preserving channel \cite{buscemi2005inverting}.   Applying this relation on  both sides of Eq. (\ref{anima}) we obtain the condition  
\begin{align}
 \< k|  \rho_i  |k\>  =   \<  k|  \rho_{i+1}  |k\>  \, ,
 \end{align}
valid for every $k\in  \{1,\dots, d\}$. Hence, all the density matrices $(\rho_1,\rho_2,\dots ,\rho_n)$ must have the same diagonal entries, and, in particular, Condition 2 must hold.  

Conversely, suppose that Condition 2 holds.   Since the dephasing channel $\map D$ is obviously basis-preserving,  we obtained the condition $\Deg_{A'} (\rho)  \cap \Deg_{A'} (\sigma)  \not  =  \emptyset$, which implies that $\rho$ and $\sigma$ are equivalent for  agent $A$.   In conclusion, Condition 1 holds.  \qed 

\medskip  
Proposition \ref{prop:diagrho} guarantees that  the states of system $S_A$ is in one-to-one correspondence with diagonal density matrices, and therefore, with classical probability distributions: 
  in formula,
\begin{align}
\St (S_A)    =  \Big \{   (p_k)_{k=1}^d  \, :  \quad  p_k\ge 0  ~\forall k   \, , \quad \sum_k p_k=1  \,     \Big\}\, .
\end{align}  

The transformations of system $S_A$ can be characterized as \emph{transition matrices}, namely  
   \begin{align}\label{classicalchan}
    \Transf (S_A)   =  \Big\{   [ P_{jk} ]_{j\le d ,\,  k\le d   }  \,: \quad      P_{jk}  \ge 0 ~\forall j,k \in \{1,\dots, d\} \, , ~ \sum_{j}\,  P_{jk}  =  1   ~\forall k\in \{1,\dots, d\}  \Big\} \, .
    \end{align}   
   
The proof of Eq. (\ref{classicalchan})  is provided in Appendix \ref{app:classical}.

   In summary,  agent $A$ has control on a classical system,  whose states are  diagonal density matrices  and whose transformations are classical transition matrices.

\subsection{Classical systems from free operations in the resource theory of coherence}
In the previous example we have seen that classical systems arise from  agents who have access to the monoid of multiphase covariant channels.  In fact, classical systems can arise  in many other  ways,  corresponding to agents who have access to different monoids of operations.  In particular, we find that several types of free operations in the  resource theory of coherence \cite{aberg2006quantifying,baumgratz2014quantifying,levi2014quantitative,winter2016operational,chitambar2016critical,chitambar2016comparison,marvian2016quantify,yadin2016quantum} identify classical systems.   Specifically,   consider  the monoids  of 
\begin{enumerate}
\item {\em Strictly incoherent operations \cite{yadin2016quantum}, i.e.}  quantum channels $\map T$ with the property that, for every Kraus operator $T_i$, the map $\map T_i  (\cdot)  =  T_i \cdot T_i  $ satisfies the condition $\map D  \circ \map T_i  = \map T_i  \circ \map D$, where $\map D$ is the completely dephasing channel.
\item {\em Dephasing covariant  operations \cite{chitambar2016critical,chitambar2016comparison,marvian2016quantify}, i.e.}  quantum channels $\map T$ satisfying the condition  $\map D  \circ \map T  = \map T  \circ \map D$.
\item {\em Phase covariant channels  \cite{marvian2016quantify}, i.e.} quantum channels   $\map T$ satisfying the condition $\map T\circ \map U_\varphi  =  \map U_\varphi  \circ \map T$, $\forall  \varphi  \in  [0,2\pi)$, where $\map U_\varphi $ is the unitary channel associated to the unitary matrix $U_\varphi  =   \sum_k    \,   e^{i  k\varphi}\,  |k\>\<k|$.
\item {\em Physically incoherent  operations \cite{chitambar2016critical,chitambar2016comparison}, i.e.}  quantum channels that are convex combinations of channels $\map T$ admitting a Kraus representation where each Kraus operator   $T_i$ is of the form   
\begin{align}\label{physinc}
T_i   =    U_{\pi_i}   \,  U_{{\bs \theta}_i} \,  P_i \,,
\end{align} where $U_{\pi_i}$ is a unitary that permutes the elements of the computational basis,  $U_{{\bs \theta}_i}$ is a diagonal unitary, and $P_i$ is a projector on a subspace spanned by a subset of vectors in the computational basis.  
\end{enumerate}

For each of the monoids 1-4, our construction yields the classical subsystem consisting of diagonal density matrices.  The transformations of the subsystem are just the classical channels.    The proof is presented in Appendix \ref{app:classicalsi}.  

Notably, other choices of free operations, such as the {\em maximally incoherent operations \cite{aberg2006quantifying}} and the  {\em incoherent operations} \cite{baumgratz2014quantifying}, do {\em not} identify classical subsystems.  The maximally incoherent operations are the  quantum channels $\map T$ that map diagonal density matrices to diagonal density matrices, namely $\map T  \circ \map D   =   \map D \circ \map T \circ \map D$, where $\map D$ is the completely dephasing channel.   The incoherent operations are the quantum channels $\map T$ with the property that, for every Kraus operator $T_i$, the map $\map T_i  (\cdot)  =  T_i \cdot T_i  $ sends diagonal matrices to diagonal matrices, namely  $\map T_i  \circ \map D   =   \map D \circ \map T_i \circ \map D$. 

In Appendix \ref{app:classicalno} we show that incoherent and maximally incoherent operations do not identify classical subsystems: the subsystem associated to these operations is the whole quantum system.  This result can be understood from the analogy between these    operations and  non-entangling operations in the resource theory of entanglement \cite{chitambar2016critical,chitambar2016comparison}. Non-entangling operations do not generate entanglement, but nevertheless they cannot (in general) be implemented with local operations and classical communication. Similarly, incoherent and maximally incoherent operations do not generate coherence, but they cannot (in general) be implemented with incoherent states and coherence non-generating unitary gates. An agent that performs these operations must have access to more degrees of freedom than just a classical subsystem.  Heuristically, this observation justifies the fact that the subsystem associated to these operations is not the classical subsystem, but the whole quantum system.  

At the mathematical level, the problem is that the incoherent and maximally incoherent operations do not necessarily commute with the dephasing channel  $\map D$.  In our construction, commutation with the dephasing channel is essential for retrieving classical subsystems.     In general, we have the following theorem:  
\begin{theo}
Every set of operations that 
\begin{enumerate}
\item contains the set of classical channels, and
\item  commutes with the dephasing channel 
\end{enumerate}
 identifies a $d$-dimensional classical subsystem of the original $d$-dimensional quantum system. 
\end{theo}
The proof is provided in Appendix \ref{app:classicalsi}.

\section{Key structures: partial trace and no signalling}\label{sec:keystructures} 

In this section we go back to the general construction of subsystems, and we analyse the main structures arising from it. First, we observe that the definition of subsystem guarantees {\em by  fiat} the validity of the no-signalling principle, stating that  operations performed on one subsystem cannot affect the  state of  an independent  subsystem.  Then, we show that our construction of subsystems allows one to build a category, where the objects are subsystems and the morphisms are physical transformations among them. 

 \subsection{The partial trace and the no signalling property}  

We defined the states of system $S_A$ as  equivalence classes.  In more physical terms, we can regard the map $\psi  \mapsto  [ \psi]_{A}$ as an operation of discarding,  which takes system $S$ and throws away the degrees of freedom reachable by  the maximal adversary $A'$. In our adversarial picture, ``throwing away some degrees of freedom'' means leaving them under the control of the adversary, and considering only the part of the system that remains under the control of the agent.     
\begin{defi}
The {\em partial trace over $A'$}  is the function $\Tr_{A'}:  \St (S) \to \St (S_A)$, defined by $\Tr_{A'}  (\psi)  =  [\psi]_{A}$ for a generic $\psi \in\St (S)$. 
\end{defi}
The reason for the notation $\Tr_{A'}$ is that in quantum theory  the operation $\Tr_{A'}$ coincides with the partial trace of matrices, as shown in the example of Subsection \ref{subsec:product}.    For subsystems associated to von Neumann algebras, the partial trace is the ``partial trace over the algebra'' defined in Subsection \ref{subsec:vN}. For subsystems associated to multiphase covariant channels or dephasing covariant operations, the partial trace is the completely dephasing channel, which     ``traces out'' the off-diagonal elements of the density matrix.  

With the partial trace notation, the states of system $S_A$ can be succinctly written as 
\begin{align}
\St (S_A)  =  \Big\{  \rho  =   \Tr_{A'} (\psi)  \, :     \quad \psi  \in  \St (S)   \Big\} \, .
\end{align}  
Denoting $B  :  =  A'$,  we  have the important relation  
\begin{align}
\label{nosig} \Tr_B   \circ  \map B  =  \Tr_B  \qquad      &\forall \map B\in\Act (B;S)  \,. 
\end{align}

Eq. (\ref{nosig}) can be regarded as  the {\em no signalling property}:  the actions of agent $B$ cannot lead to any change on the system of agent $A$.   Of course, here the no signalling  property holds {\em by fiat}, precisely because of the way the subsystems are defined!   

The construction of subsystems has the merit to clarify the status of the no-signalling principle.   No-signalling is often associated to space-like separation, and heuristically justified through the idea that physical influences should propagate within the light cones.  However, locality is only {\em  a sufficient condition} for the no signalling property. Spatial separation implies no signalling, but the converse is not necessarily true: every pair of distinct quantum systems satisfies the no-signalling condition, even if the two systems are spatially contiguous.  In fact, the no-signalling condition holds even for virtual subsystems of a {\em single}, spatially localized system.    Think for example of a quantum particle localized in the  $xy$ plane. 
 The particle can be regarded as a composite system, made of two virtual subsystems: a particle localized on the $x$ axis, and another particle  localized on the $y$ axis.  The no-signalling property  holds for these two subsystems, even if they are not separated in space.  
  As Eq. (\ref{nosig}) suggests, the validity of the no-signalling property has more to do with the way subsystems are constructed, rather than the way the subsystems are distributed in space.     

 
   \subsection{A baby category}  

Our construction of subsystems defines a category, consisting of three objects, $S, S_A$, and $S_B$, where $S_B$ is the subsystem associated to the agent $B  =  A'$.  The sets $\Transf (S)$, $\Transf(S_A)$, and $\Transf(S_B)$ are the endomorphisms from $S$ to $S$,  $S_A$ to $S_A$, and $S_B$ to $S_B$, respectively.  
  The morphisms from $S$ to $S_A$ and from $S$ to $S_B$ are defined as 
  \begin{align}
  \Transf(S  \to  S_A)   =  \Big\{ \Tr_B \circ   \map T \, : \quad \map T\in\Transf (S)  \Big\}   
  \end{align}
and 
  \begin{align}
  \Transf(S  \to  S_B)   =  \Big\{ \Tr_A \circ   \map T \, : \quad \map T\in\Transf (S)  \Big\}  \, , 
  \end{align}
  respectively.    
  
Morphisms from $S_A$ to $S$, from $S_B$ to $S$, from $S_A$ to $S_B$, or from $S_B$ to $S_A$, are not naturally defined.  In Appendix \ref{app:category}, we provide a mathematical construction that enlarges the sets of transformations, making all sets non-empty. Such a construction allows us to reproduce a categorical structure known as  a   {\em splitting of idempotents} \cite{selinger2008idempotents,coecke2018two}

 \section{Non-Overlapping Agents, Causality, and the Initialization Requirement}\label{sec:FirstAxioms}  
 
In the previous sections, we developed a general framework, applicable to arbitrary physical systems.    In this section and in the remainder of the paper, we identify some desirable properties that the global systems may enjoy. 

\subsection{Dual pairs of agents}

 So far, we have taken the perspective of agent $A$.   Let us now take the perspective of the maximal adversary $A'$.  We  consider $A'$ as the agent, and denote his maximal  adversary as $A^{\prime \prime}$.     By definition, $A^{\prime \prime}$ can perform every action in the commutant of $\Act(A';S)$, namely
\begin{align}
\Act (A^{\prime \prime};S)   =  \Act  (A';S)'  =  \Act  (A;S)^{\prime \prime} \, .
\end{align} 
Obviously, the set of actions allowed to agent $A^{\prime \prime}$ includes the set of actions allowed to agent $A$.  At this point, one could continue the construction and consider the maximal adversary of agent $A^{\prime \prime}$.     However, no new agent would appear at this point: the maximal adversary of agent $A^{\prime \prime}$ is agent $A^{\prime}$ again.     When two agents have this property, we call them a {\em dual pair:}  
\begin{defi}  
Two agents $A$ and $B$ form a dual pair iff  $\Act  (A;S)   =  \Act  (B;S)'$ and $\Act (B;S)  =  \Act  (A;S)'$. 
\end{defi}
All the examples in Section \ref{sec:examples} are examples of dual pairs of agents.  

It is easy to see that an agent $A$ is part of a dual pair if and only if the set  $\Act  (A;S)$ coincides with its double commutant $\Act (A; S)^{\prime \prime}$.    

 \subsection{Non-Overlapping  Agents}  
 
Suppose that agents $A$ and $B$ form a dual pair. In general, the actions in  $\Act (A;S)$ may have a non-trivial intersection with the actions in $\Act (B;S)$.  This situation does indeed  happen, as we have seen in Subsections \ref{subsec:coherence} and \ref{subsec:multiphase}. 
Still, it is important to examine the special case where the actions of $A$ and $B$ have only  trivial intersection, corresponding to the identity action $\map I_S$.  When this is the case, we say that the agents $A$ and $B$ are {\em non-overlapping:}
\begin{defi}
Two agents $A$ and $B$ are non-overlapping iff $\Act (A;S)   \cap  \Act (B;S)   =  \{\map I_S\}$. 
\end{defi}              

Dual pairs of non-overlapping agents  are characterized by the fact that the actions monoids have trivial center:  
\begin{prop}
Let $A$ and $B$ be a dual pair of agents. Then, the following are equivalent: 
\begin{enumerate}
\item $A$ and $B$ are  non-overlapping 
\item $\Act (A;S)$ has trivial center
\item $\Act (B;S)$ has trivial center.
\end{enumerate}
\end{prop}

\Proof Since agents  $A$ and $B$ are dual to each other, we have $\Act  (B;S)   =  \Act  (A;S)'$ and $\Act (A;S)  =  \Act  (B;S)'$.   Hence, the intersection  $\Act (A;S)   \cap  \Act (B;S) $ coincides with the center of $\Act (A;S)$, and with  the center of $\Act (B;S)$.  The non-overlap condition holds if and only if the center is trivial.    \qed

\medskip 

Note that the existence of non-overlapping dual pairs is a condition on the transformations of the whole system $S$:  
\begin{prop}\label{prop:centertriv}
The following are equivalent:  
\begin{enumerate}
\item system $S$ admits a dual pair of non-overlapping agents
\item  the monoid $\Transf (S)$ has trivial center. 
\end{enumerate}
\end{prop}
\Proof Assume that Condition 1 holds for a pair of agents $A$ and $B$.  Let $\grp C (S)$ be the center of $\Transf (S)$.   By definition,  $\grp C  (S)$ is contained into $\Act(B;S)$, because $\Act(B;S)$ contains all the transformations that commute with those in $\Act (A;S)$.   Moreover, the elements of $\grp C(S)$ commute with all elements of $\Act (B;S)$, and therefore they are in the center of $\Act (B;S)$.  Since $A$ and $B$ are a non-overlapping dual pair, the center of $\Act (B;S)$ must be trivial, and therefore $\grp C (S)$ must be trivial.  Hence, Condition 2 holds. 

Conversely, suppose that Condition 2 holds. In that case, it is enough to take $A$ to be the {\em maximal agent}, {\em i.e.} the agent $A_{\max}$ with $\Act \left(A_{\max};S\right)  =  \Transf (S)$. Then, the  maximal adversary of $A_{\max}$ is the agent $B=  A_{\max}'$ with $\Act (B;S)   =  \Act  \left(A_{\max};S\right)'   =    \grp C(S)   =   \{  \map I_S\}$.  By definition, the two agents form a non-overlapping dual pair.  Hence, Condition 1 holds.  \qed

\medskip 

The existence of dual pairs of non-overlapping agents is a desirable property, which may be used to characterize  ``good systems'': 
\begin{defi}[Non-Overlapping Agents]\label{ax:nonoverlap}
We say that system $S$  satisfies the {\em Non-Overlapping Agents Requirement} if there exists  at least one dual pair of non-overlapping agents acting on $S$. 
\end{defi}

The Non-Overlapping Agents Requirement guarantees that the total system $S$ can be regarded as a subsystem: if $A_{\max}$ is the {\em maximal agent} ({\em i.e.} the agent who has access to all transformations on $S$),  then  the subsystem $S_{A_{\max}}$   is  the whole system $S$.   A more formal statement of this fact is provided in Appendix \ref{app:SaS}.

\subsection{Causality}

The Non-Overlapping Agents Requirement guarantees that the subsystem associated to a maximal agent ({\em i.e.} an agent who has access to all possible transformations) is the whole system $S$.   On the other hand, it is natural to expect that a minimal agent, who has no access to any transformation, should be associated to the trivial system,~{\em i.e.} the system with a single state and a single transformation.
The fact that the minimal agent is associated to the trivial system is important, because it equivalent to a property of causality \cite{chiribella2010probabilistic,coecke2013causal,chiribella2016quantum,coecke2016terminality}.    Precisely, we have the following 
\begin{prop}\label{prop:causality}
Let $A_{\min}$ be the minimal agent and let $A_{\max}$ be its maximal adversary, coinciding with the maximal agent. Then, the following conditions are equivalent
\begin{enumerate}
\item  $S_{A_{\min}}$ is the trivial system
\item  one has $\Tr_{A_{\max}}  [\rho]   =  \Tr_{A_{\max}} [\sigma] $ for every pair of states $\rho,\sigma\in\St (S)$. 
\end{enumerate}
\end{prop}
\Proof By definition, the state space of $S_{A_{\min}}$ consists of states of the form $\Tr_{A_{\max}}[\rho]$, $\rho \in \St (S)$.     Hence, the state space contains only one state if and only if Condition 2 holds \footnote{The fact that $S_{A_{\min}}$ has only one transformation is true by definition:  since the adversary of $A_{\min}$ is the maximal agent, one has $\map T  \in  \Deg_{A_{\max}}  (\map I_S)$ for every transformation $\map T\in\Transf (S)$. Hence, every transformation is in the equivalence class of the identity.}. \qed  

\medskip 

With a little abuse of notation, we may denote the trace over  $A_{\max}$ as $\Tr_S$, because $A_{\max}$ has access to all transformations on system $S$.    With this notation, the causality condition reads 
\begin{align}\label{causality}
\Tr_S  [\rho]   =    \Tr_S  [\sigma]  \qquad \forall  \rho,\sigma\in \St (S) \, .
\end{align}
It is interesting to note that, unlike no signalling, causality does not necessarily hold in the framework of this paper.  This is because the trace $\Tr_S$ is defined as the quotient with respect to all possible transformations, and having a single equivalence class is a non-trivial property.   
One possibility is to demand the validity of this property, and to call a system {\em proper}, only if it satisfies the causality condition (\ref{causality}).  In the following subsection we will see a requirement that guarantees the validity of the causality condition.

\subsection{The Initialization Requirement}  
The ability to prepare states from a fixed initial state is important in the circuit model of quantum computation, where qubits are initialized to the state $|0\>$, and more general states are generated  by applying quantum gates.  More broadly, the ability to  initialize the system in a given state and to  generate  other  states from it  is important for applications in quantum control and adiabatic quantum computing.   Motivated by these considerations, we formulate  the following definition:

 \begin{defi}
A system $S$ satisfies the {\em Initialization Requirement} if there exists at least a {\em cyclic state}   $\psi_0 \in  \St (S)$ from which any other state can be generated, meaning that, for every other state $\psi  \in \St (S)$ there exists a transformation $\map T \in  \Transf (S)$  such that $\psi  =  \map T  \psi_0$. 
\end{defi}

The Initialization Requirement is satisfied  in quantum theory, both at the pure state level and at  the mixed state level.  At the pure state level,  every unit vector $|\psi\>  \in \spc H_S$ can be generated from a fixed unit vector $|\psi_0\> \in  \spc H_S$ via a unitary transformation $U$.  At the mixed state level, every density matrix $\rho$ can be generated from a fixed density matrix $\rho_0$ via the erasure channel $\map C_\rho (\cdot)  =  \rho  \,  \Tr[\cdot] $.    By the same argument, the initialization requirement is also satisfied when $S$ is a system in an operational-probabilistic theory \cite{chiribella2010probabilistic,hardy2011foliable,hardy2013formalism,chiribella2014dilation,chiribella2016quantum} and when $S$ is a system in a causal  process theory \cite{coecke2013causal,coecke2016terminality}.

The Initialization Requirement guarantees that minimal  agents  are associated with trivial systems: 
\begin{prop}
Let  $S$ be a system satisfying the Initialization Requirement, and let $A_{\min}$ be the {\em minimal agent}, {\em i.e.} the agent that can only perform the identity transformation.  Then, the subsystem $S_{A_{\min}}$ is trivial: $\St \left(S_{A_{\min}}\right)$ contains only one state and $\Transf \left(S_{A_{\min}}\right)$ contains only one transformation.    
\end{prop}  
\Proof By definition, the maximal adversary of $A_{\min}$ is the maximal agent $A_{\max}$, who has access to all physical transformations.   Let $\psi_0$ be the cyclic state. By the Initialization Requirement, the set  $\Deg_{A_{\max}}(\psi_0)$ is the whole state space $\St (S)$. Hence, every state is equivalent to the state $\psi_0$  modulo $A_{\max}$.  In other words, $\St \left(S_{A_{\min}}\right)$ contains only one state.    \qed 

\medskip  

The Initialization Requirement guarantees the validity of causality, thanks to Proposition \ref{prop:causality}.   
In addition, the Initialization Requirement is important also independently of the causality property. For example,  later in the paper we will use it to formulate an abstract notion of {\em closed  system}. 

\section{The Conservation of Information}\label{sec:conservation}  

In this section we consider systems where all transformations are {\em  invertible}.   In such systems, every transformation can be thought as the result of some deterministic dynamical law.    The different transformations in $\Transf(S)$ can be interpreted as different dynamics, associated to different values of  physical parameters, such as coupling constants or external control parameters. 
 
\subsection{Logically invertible vs physically invertible}

\begin{defi}A transformation $\map T  \in  \Transf (S)$ is  {\em logically invertible} iff the map 
\begin{align}\widehat{\map T}:  \quad \St (S) \to \St (S) \, , \quad     \psi  \mapsto  \map T  \psi 
\end{align} 
is injective. 
\end{defi} 
Logically invertible transformations can be interpreted as  evolutions of the system that preserve the distictness of states.    At the fundamental level, one may require that all physical evolutions be logically invertible, a requirement that is sometimes called  the  {\em Conservation of Information} \cite{susskind2008black}. In the following we will explore the consequences of such requirement: 
\begin{defi}[Logical Conservation of Information]  System $S$ satisfies the Logical Conservation of Information if all transformations in $\Transf (S)$ are logically invertible. 
\end{defi} 
The requirement is well-posed, because the invertible transformations form a monoid. Indeed, the identity transformation is invertible,  and that the composition of two invertible transformations is invertible.

A special case of logical invertibility is physical invertibility, defined as follows:
\begin{defi}A transformation $\map T  \in  \Transf (S)$ is  {\em physically invertible} iff there exists another transformation $\map T'\in  \Transf (S)$ such that  $\map T'\circ \map T  =  \map I_S $. 
\end{defi} 
Physical invertibility is more than just mathematical invertibility:  not only should the map   $\map T$  be invertible  as a function on the state space, but also its inverse should be a member of the  monoid of physical transformations.  
In light of this observation, we state a stronger version of the Conservation of Information, requiring physical invertibility:  
\begin{defi}[Physical Conservation of Information]  
System $S$ satisfies the Physical Conservation of Information if all transformations in $\Transf (S)$ are physically invertible. 
\end{defi}

The difference between Logical and Physical Conservation of Information is highlighted by the following example: 

\begin{eg}[Conservation of Information in closed-system quantum theory]  Let $S$ be a closed quantum system  described by a separable, infinite-dimensional Hilbert space  $\spc H_S$, and let $\St (S)$ be the set of pure states, represented as rank-one density matrices 
\begin{align}
\St (S)  =  \Big\{   |\psi\>\<  \psi|  \, :    \quad |\psi\>   \in  \spc H_S  \, , \quad \<\psi|\psi\>  = 1 \Big\}  \, .
\end{align} 
One possible choice of transformations is the monoid of isometric channels 
\begin{align}\label{isophys}\Transf (S)  =  \Big \{   V\cdot V^\dag \, :   \quad  V  \in  \Lin(S) \, ,  \quad  V^\dag V  = I\Big\} \,.
\end{align}    This choice of transformations satisfies the Logical Conservation of Information, but violates the Physical Conservation of Information, because in general the map $V^\dag \cdot  V$ fails to be trace-preserving, and therefore fails to be an isometric channel.   For example, consider the shift operator  
  \begin{align}\label{shift}
  V =  \sum_{n=0}^{\infty}  |n+1\>\<n|  \, .
  \end{align}
The operator $V$  is an isometry but its inverse $V^\dag$ is not. As a result, the channel $V^\dag \cdot V$ is not an allowed physical  transformation according to Eq.  (\ref{isophys}). 

An alternative choice of physical  transformations is  the set of unitary channels
\begin{align}\Transf (S)  =  \Big \{   V\cdot V^\dag \, :  \quad   V  \in  \Lin(S) \, , \quad  V^\dag V  =  VV^\dag  =   I\Big\} \,. 
\end{align}   
With this choice, the Physical Conservation of Information is satisfied: every physical transformation is invertible and the inverse is a physical transformation.  
\end{eg}

\subsection{Systems satisfying the Physical Conservation of Information}  

In a system satisfying the Physical Conservation of Information, the transformations are not only physically invertible, but also physically {\em reversible}, in the following sense: 

\begin{defi}
A transformation $\map T \in  \Transf(S)$ is {\em physically reversible}    iff there exists another transformation $\map T'  \in  \Transf(S)$   such that $\map T' \circ \map T  =  \map T\circ\map T'  =  \map I_S$. 
\end{defi}

With the above definition, we have the following:
\begin{prop}
If system $S$ satisfies the Physical Conservation of Information, then every physical transformation is physically reversible.  The monoid $\Transf(S)$ is a group, hereafer denoted as $\grp G(S)$. 
\end{prop}

\Proof  Since $\map T$ is physically invertible, there exists a transformation $\map T'$ such that $\map T'\circ \map T= \map I_S$.   Since the Physical Conservation of Information holds, $\map T'$ must be physically invertible, meaning that there exists a transformation $\map T^{\prime \prime}$ such that $\map T^{\prime \prime}  \circ \map T'  = \map I_S$. Hence, we have 
\begin{align}
\map T^{\prime \prime}   = \map T^{\prime \prime} \circ  (\map T' \circ \map T)  =  (\map T^{\prime \prime}  \circ \map T'  ) \circ \map T  =  \map T  \,  .
\end{align}  
Since $\map T^{\prime \prime}  = \map T$, the invertibility condition $\map T^{\prime \prime} \circ \map T'  =  \map I_S$ becomes  $\map T \circ \map T'=  \map I_S$. Hence, $\map T$ is reversible and $\Transf (S)$ is a group.  \qed 

\medskip


\subsection{Subsystems of systems satisfying the Physical Conservation of Information}

Imagine that an agent $A$ acts on a  system  $S$ satisfying the Physical Conservation of Information.  We assume that the actions of agent $A$  form a subgroup of $\grp G(S)$, denoted as $\grp G_A$.   The maximal  adversary of $A$ is the adversary  $B  =  A'$, who  has access to all transformations in the set
\begin{align}
\grp G_B  :  =  \grp G_{A}'  = \Big\{    \map U_B  \in \grp G (S)  \,  :  \quad      \map U_B \circ \map U_A  =  \map U_A \circ \map U_B  \, , \quad \forall \map U_A \in  \grp G(A)  \Big\} 
 \, .
\end{align}
It is immediate to see that the set $\grp G_B$ is a group.  We call it the {\em adversarial group}.      

The equivalence relations used to define subsystems can be greatly simplified.  Indeed, it is easy to see that two states $\psi, \psi'  \in \St (S)$ are equivalent for  $A$ if and only if there exists a  transformation $\map U_B  \in  \grp G_B$ such that
 \begin{align}
    \psi'  =  \map U_B  \psi  \,. 
\end{align}  
Hence, the states of the subsystem $S_A$ are  orbits of the group $\grp G_B$: for every $\psi\in\St (S)$, we have 
\begin{align}\label{states=orbits}
\Tr_B [\psi]   :  =  \Big\{  \map U_B \psi \,:  \quad \map U_B \in  \grp G_B  \Big\} \, .
\end{align}
Similarly, the degradation of a transformation $\map U  \in  \grp G (S)$  yields the orbit  
\begin{align}
\Deg_{B}(  \map U)   =  \Big\{   \map U_{B,1}  \circ \map U  \circ  \map U_{B,2}    \, : \qquad  \map U_{B,1}  ,  \map U_{B,2}  \in  \grp G_B \Big\} \, .
\end{align} 
 It is easy to show that  the transformations of the subsystem $S_A$ are the orbits of the group $\grp G_B$:  
  \begin{align}
  \Transf (S_A)   = \Big\{ \pi_A  (\map U)  \, : \quad \map U\in\grp G_A^{\prime \prime}  \Big\} \, ,\qquad  \pi_A  (\map U)  :  =  \Big\{ \map U_B\circ \map U  \, : \quad \map U_B \in \grp G_B   \Big\} \, .
  \end{align}

 \section{Closed systems}\label{sec:closed}  

Here we define an abstract notion of ``closed systems'', which captures the essential features of what is traditionally called a closed system in quantum theory. Intuitively, the idea is that all the states of the closed system are ``pure'' and all the evolutions are reversible.

An obvious problem in defining closed system is that in our framework we do not have a notion of ``pure state''.
  To circumvent the problem we define the closed systems in the following way: 
\begin{defi}
System $S$ is {\em closed} iff it satisfies the Logical Conservation of Information and the Initialiation Requirement, that is, iff  
\begin{enumerate}
\item every transformation is logically invertible
\item there exists a state $\psi_0  \in  \St (S)$ such that, for every other state $\psi \in \St (S)$, one has $\psi  =  \map V  \psi_0$ for some suitable transformation $\map V  \in  \Transf (S)$. 
\end{enumerate}    
\end{defi}

For a closed system, we nominally say that all the states in $\St (S)$ are ``pure'', or, more precisely, ``dynamically pure''.   
This definition is generally different  from the usual definition of pure states as extreme points of convex sets, or from the compositional definition of pure states as states with only product extensions \cite{chiribella2014distinguishability}.      First of all, dynamically pure states are {\em not a subset} of the state space:  provided that the right conditions are met, they are {\em all} the states.  
Other  differences between the usual notion of pure states and the notion of dynamically pure states  are highlighted by the following example:  

\begin{eg}
Let $S$ be a  system in which all  states are of the form $U  \rho_0   U^\dag$, where $U$ is a generic 2-by-2 unitary matrix, and $\rho_0  \in  M_2 (\CC)$ is a fixed 2-by-2 density matrix.   For the transformations, we allow all unitary channels  $U\cdot U^\dag$.  By construction, system $S$ satisfies the initialization Requirement, as one can generate every state from the initial state $\rho_0$.   Moreover, all the transformations of system $S$ are unitary and therefore the Conservation of Information is satisfied, both at the physical and the logical level.  Therefore, the states of system $S$ are dynamically pure.     Of course, the  states  $U\rho_0  U^\dag$    need not be extreme points of the convex set of all density matrices, {\em i.e.} they need not be rank-one projectors.  They are so only when the cyclic state $\rho_0$ is rank-one.  
  
On the other hand, consider a similar example,  where 
\begin{itemize}
\item system $S$ is a qubit
\item the states are pure states, of the form $|\psi\>\<\psi|$ for a generic unit vector $|\psi\>  \in  \CC^2$
\item the transformations are unitary channels $V  \cdot V^\dag$, where the unitary matrix $V$ has {\em real} entries 
\end{itemize}
Using the Bloch sphere picture,  the physical transformations are rotations around the $y$ axis.  Clearly, the Initialization Requirement is not satisfied, because there is no way to generate arbitrary points on the sphere using only rotations around the $y$ axis.  In this case, the states of $S$ are pure in the convex set sense, but not dynamically pure.  
\end{eg}

 For closed systems satisfying the Physical Conservation of Information, every pair of pure states are interconvertible: 
 \begin{prop}[Transitive action on the pure states] If system $S$ is closed and satisfies the Physical Conservation of Information,  then for every pair of states $\psi, \psi'  \in\St (S)$  there exists a reversible transformation $\map U\in \grp G(S)$ such that $\psi'  =  \map U\psi$.  
\end{prop}
\Proof  By the Initialization Requirement, one has $\psi  =  \map V\psi_0$ and $\psi'  = \map V' \psi_0$ for suitable $\map V,\map V'\in\Transf (S)$.  By the Physical Conservation of Information, all the tranformations in $\Transf(S)$ are physically reversible.  Hence, $\psi' =   \map V'  \circ \map V^{-1}  \psi   = \map U  \psi$, having defined $\map U  = \map V'\circ \map V^{-1}$. \qed

\medskip  The requirement that all pure states be connected  by reversible transformations has featured  in many axiomatizations of quantum theory, either directly \cite{hardy2001quantum,dakic2011quantum,masanes2011derivation,masanes2012existence}, or indirectly as a special case of other axioms \cite{chiribella2011informational,barnum2014higher}. Comparing our framework with the framework of general probabilistic theories, we can see that the dynamical definition of pure states refers to a rather specific situation, in which all pure states are connected, either to each other (in the case of physical reversibility) or with to a fixed cyclic state (in the case of logical reversibility).

\section{Purification}\label{sec:purification}  

Here we show that closed systems  satisfying the  Physical Conservation of Information also satisfy the purification property \cite{chiribella2010probabilistic,chiribella2012quantum,chiribella2013quantumtheory,chiribella2014dilation,chiribella2016quantum,chiribella2015conservation,dariano2017quantum}, namely  the property that  every mixed state can be modelled as a pure state of a larger system in a canonical way.     Under a certain regularity assumption, the same holds for closed systems satisfying only the Logical Conservation of Information.

 \subsection{Purification in systems satisfying the Physical Conservation of Information} 
 
\begin{prop}[Purification]
Let $S$ be a closed system satisfying the Physical Conservation of Information.  Let $A$ be an agent in $S$, and let $B=  A'$ be its maximal adversary. Then,  for every state $\rho \in \St (S_A)$, there exists a pure state $\psi\in\St (S)$, called the {\em purification of $\rho$},  such that  $\rho   =  \Tr_B  [\psi]$.  
Moreover,  the purification of $\rho$  is {\em essentially unique}:  if $\psi'\in\St (S)$ is another  pure state with $\Tr_B  [\psi]  =  \rho$, then there exists a reversible  transformation $\map U_B\in\grp G_B$ such that $\psi'   =   \map U_B  \psi$.  
\end{prop}

\Proof By construction, the states of system $S_A$ are orbits of states of system $S$ under the adversarial group $\grp G_B$.   By Equation \ref{states=orbits}, every two states $\psi,\psi'\in \St (S)$  in the same orbit are connected by an element of $\grp G_B$. \qed  

\medskip 

Note that the notion of purification used here is  more general  than the usual notion  of purification in quantum information and quantum foundations.  The most important difference is that system $S_A$ need not be a factor in a tensor product.    Consider  the example of the coherent superpositions vs classical mixtures (subsection \ref{subsec:coherence}).   There, systems $S_A$ and $S_B$ coincide, their states are classical probability distributions, and the  purifications are coherent superpositions.   Two purifications of the same classical state ${\bf p}  =  (p_1,p_2,\dots,  p_d)$ are  two rank-one projectors $|\psi\>\<\psi|$ and $|\psi'\>\<\psi'|$ corresponding to unit vectors of the form  
\begin{align}
|\psi\>  =  \sum_n  \,   \sqrt{p_n}  \, e^{i\theta_n} \,  |n\>   \qquad {\rm and} \qquad |\psi'\>  =   \sum_n  \,   \sqrt{p_n}  \, e^{i\theta'_n} \,  |n\> \, .
\end{align} 
One purification can be obtained from the other by applying a diagonal unitary operator, corresponding to a diagonal unitary matrix.  Specifically,  one has 
\begin{align}
|\psi'\>   =  U_B |\psi\>   \qquad {\rm with~}\quad U_B  =  \sum_n e^{i  (\theta_n'-\theta_n)}\, |n\>\<n| \, . 
\end{align}

For finite dimensional quantum systems, the notion of purification  proposed here encompasses   both the notion of entanglement and the notion of coherent  superposition.   The case of  infinite dimensional systems will be discussed in the next subsection.  

\subsection{Purification in systems satisfying the Logical Conservation of Information}

For infinite dimensional quantum systems, every density matrix can be purified, but not all purifications are connected by reversible transformations.  Consider for example  the unit vectors 
\begin{align}
|\psi\>_{AB}   = \sqrt {1-x^2}  \,   \sum_{n=0}^{\infty}   \,  x^n  \,  |n\>_A\otimes |n\>_B   \qquad {\rm and}  \qquad |\psi'\>_{AB}   = \sqrt {1-x^2}  \,   \sum_{n=0}^{\infty}   \,  x^n  \,  |n\>_A\otimes |n+1\>_B\, , 
\end{align}
for some $x\in  [0,1)$.   

For every fixed $x\not  = 0$,  there is one and   only one operator $V_B$ satisfying the condition  $|\psi'\>_{AB}  = (I_A\otimes  V_B )  |\psi\>_{AB}$, namely the  shift operator 
$V_B   =   \sum_{n=0}^\infty   |n+1\>\<n|$.  However,  $V_B$ is only an isometry, but  not a unitary.   This means that, if we define the states of system   $S_A$ as equivalence classes of pure state under local unitary equivalence, the two states $  |\psi\>\<\psi|$ and $|\psi'\>\<\psi'|$ would end up into  two different equivalence classes.

  One way to address the problem is to relax the requirement of reversibility and to consider the monoid of isometries, defining 
 \begin{align}
 \Transf (S)    :  =  \{   V\cdot V^\dag\, :   \quad V  \in  \Lin  (S) \, ,~  V^\dag V  = I    \} \, . 
 \end{align}
 Given two purifications of the same state, say $|\psi\>$ and $|\psi'\>$, it is possible to show that at least one of the following possibilities holds: 
\begin{enumerate}
\item $|\psi'\>   =  ( I_A\otimes V_B )\,  |\psi\>$ for some isometry  $V_B$ acting on system $S_B$
\item $|\psi\>   =  ( I_A\otimes V_B )\,  |\psi'\>$ for some isometry  $V_B$ acting on system $S_B$. 
\end{enumerate}

Unfortunately, this uniqueness property is not automatically valid in every system satisfying the Logical Conservation of Information. 
Still, we will now show a regularity condition, under which the uniqueness property is satisfied: 
\begin{defi}\label{def:reg}
Let $S$ be a system satisfying the Logical Conservation of Information, let $M \subseteq \Transf (S)$ be a monoid,  and let $\Deg_{\sf M} (\psi)$ be the set  defined by 
\begin{align}
\Deg_{\sf M} (\psi)    =    \Big\{  \map V  \,\psi\, : \quad \map V \in  \sf M  \Big\}  \, .  
\end{align}    
We say that the monoid  $\sf M  \subseteq \Transf(S)$ is {\em regular}  iff 
\begin{enumerate}
\item for every pair of states $\psi, \psi'  \in\St (S)$, the condition $\Deg_{\sf M}  (\psi)  \cap  \Deg_{\sf M}  (\psi')\not  =  \emptyset$ implies that there exists a transformation $\map U  \in\sf M$ such that  $\psi'  =    \map U  \psi$ or $\psi  =   \map U  \psi'$, 
\item for every pair of transformations $\map V, \map V'  \in  \sf M$,  there exists a transformation  $\map W  \in \sf M$ such that  $\map V  =  \map W  \circ \map V'$ or $\map V' =  \map W  \circ \map V$. 
\end{enumerate}
\end{defi}

The regularity conditions are satisfied in quantum theory by the monoid of isometries. 
\begin{eg}[Isometric channels in quantum theory]
Let   $S$ be a quantum system with separable Hilbert space $\spc H$, of dimension $d\le \infty$.  Let $\St (S)$ the set of all pure quantum states, and let  $\Transf (S)$ be the monoid of all isometric channels.  

We now show that the monoid $\sf M  = \Transf (S)$ is regular.  The first regularity condition is immediate, because for every pair of unit vectors $|\psi\>$ and $|\psi'\>$ there exists an isometry (in fact, a unitary) $V$ such that $|\psi'\>   =  U |\psi\>$.  Trivially, this implies the relation $|\psi'\>\<\psi'|  =    U  |\psi\>\<\psi| U^\dag$ at the level of quantum states and isometric channels.  

 Let us see that the second regularity condition holds.  Let $V,  V' \in \Lin (\spc H)$ be two isometries on $\spc H$, and let $\{|i\>\}_{i=1}^d$ be the standard basis for $\spc H$.  Then,   the isometries $V$ and $V'$ can be written as 
\begin{align}
V  =  \sum_{i=1}^{d}      |\phi_i\>\< i|     \qquad {\rm and} \qquad V'  =  \sum_i   |\phi_i'\>\<i| \, ,
\end{align}
where $\{  |\phi_i\>\}_{i=1}^d$ and $\{  |\phi_i'\>\}_{i=1}^d$ are  orthonormal vectors (not necessarily forming bases for the whole Hilbert space $\spc H$).    
Define the subspaces  $S  = \Span \{  |\phi_i\>\}_{i=1}^d$ and   $S'  = \Span \{  |\phi'_i\>\}_{i=1}^d$, and let $\{|\psi_j\>\}_{j=1}^r$   and $\{|\psi'_j\>\}_{j=1}^{r'}$   be orthonormal bases for the orthogonal complements $S^\perp$ and  $S^{\prime \perp}$, respectively.  If $r\le r'$, we  define the isometry 
\begin{align}
W  =   \left ( \sum_{i=1}^d  |\phi_i'\>\<\phi_i|    \right)  +  \left(  \sum_{j=1}^r  |\psi_j'\>\<\psi_j|  \right)  \, ,  
\end{align}
and we obtain   the condition $V'=  WV$.   Alternatively, if $r'\le r$, we can define the isometry
\begin{align}
W =   \left ( \sum_{i=1}^d  |\phi_i\>\<\phi_i'|    \right)  +  \left(  \sum_{j=1}^r  |\psi_j\>\<\psi_j'|  \right)  \, ,  
\end{align}
and we obtain the condition $V=  W V'$.   At the level of isometric channels, we obtained the condition $\map V '  =  \map W \circ \map V$ or the condition $\map V  =  \map W \circ \map V'$, with $\map V  (\cdot)  =  V\cdot V^\dag$,  $\map V' (\cdot)=  V'\cdot V^{\prime \dag}$, and $\map W (\cdot)   =  W \cdot  W^\dag$.
 
The fact that the monoid of all isometric channels is regular implies that other monoids of isometric channels are also regular.  For example,  if the Hilbert space $\spc H$ has the tensor product structure $\spc H =  \spc H_A\otimes \spc H_B$, then the monoid of local isometric channels, defined by isometries  of the form  $I_A\otimes V_B$,  is regular. 
More generally, if the Hilbert space is decomposed as  
\begin{align}
\spc H  =  \bigoplus_k   \,  \left(  \spc H_{A,k} \otimes \spc H_{B,k}  \right)  \, ,
\end{align}
then the monoid of isometric channels generated by isometries of the form
\begin{align}
V  =  \bigoplus_k  \,  \left(   I_{A,k}  \otimes  V_{B,k} \right) 
\end{align}
is regular. 
\end{eg} 

We are now in position to derive the purification property for general closed systems: 
\begin{prop}\label{prop:regularity}
Let $S$ be a closed system.   Let $A$ be an agent and let $B=  A'$ be its maximal adversary.  If $\Act (B;S)$ is a regular monoid,  the condition $\Tr_B  [\psi]  = \Tr_B [\psi']$ implies that there exists  some invertible transformation $\map V_B  \in \Transf (B;S)$ such that   the relation  $\psi'  =  \map V_B \psi$ or the relation $\psi  =  \map V_B  \psi'$ holds. 
\end{prop} 

The proof is provided in Appendix \ref{app:regularity}.   In conclusion, we obtained the following  
\begin{cor}[Purification]
Let $S$ be a closed system, let $A$ be an agent in $S$, and let $B=  A'$ be its maximal adversary. If the monoid $\Act (B;S)$ is regular, then  every state $\rho \in \St (S_A)$ has  a  purification   $\psi\in\St (S)$, {\em i.e.} a state a such that  $\rho   =  \Tr_B  [\psi]$.  
Moreover, the purification is essentially unique: if $\psi'\in\St (S)$ is another  state with $\Tr_B  [\psi]  =  \rho$, then there exists a reversible  transformation $\map V_B\in\Act (B;S)$ such that  the relation $\psi'   =   \map V_B  \psi$ or the relation $\psi   =  \map V_B  \psi'$ holds.  
\end{cor}

\section{Example: group representations on quantum state spaces}\label{sec:groups}

We conclude the paper with a macro-example, involving group representations in closed-system quantum theory.  The point of this example is to illustrate the general notion of purification introduced in this paper and to characterize the sets of mixed states associated to different agents.

As system $S$, we consider  a quantum system with Hilbert space $\spc H_S$, possibly of infinite dimension.  We let $\St (S)$ be the set of pure quantum states, and let $\grp G (S)$ be the group of all unitary channels. With this choice,  the total system is closed and satisfies the Physical Conservation of Information. 


Suppose that agent $A$ is able to perform a group of transformations, such as {\em e.g.} the group of phase shifts on a harmonic oscillator, or the group of  rotations of a spin $j$ particle.  Mathematically, we focus our attention on unitary channels arising from some    representation of a given compact group $\grp G$.  Denoting the representation as $U: \grp G  \to  \Lin  (\spc H_S) \, ,  g\mapsto U_g$,   the group  of Alice's actions is 
\begin{align}
G_A   =   \Big\{ \map U_g (\cdot) =  U_g \cdot U_g^\dag  :  \quad    g\in\grp G   \Big\}
 \, .
 \end{align}     
  The maximal  adversary of $A$ is the agent $B= A'$ who is able to perform all unitary channels $\map V$ that commute with those  in $\grp G_A$, namely, the unitary channels in the group 
\begin{align}
\grp G_B :  =  \Big\{    \map V  \in  \grp G (S)  \, :  \quad    \map V \circ \map U_g  =  \map U_g\circ \map V  \quad \forall g\in\grp G  \Big\} \, .
\end{align}  
  Specifically,  the channels $\map V$ correspond to unitary operators $V$  satisfying the relation 
  \begin{align}\label{cocycle}
  V  U_g     =   \omega (V,g)  ~    U_g  V  \qquad \forall g\in \grp  G\,,
  \end{align}
  where, for every fixed $V$,  the function  $\omega (V,  \cdot):  \grp G  \to  \CC$ is a multiplicative character, {\em i.e.}  a one-dimensional representation of the group $\grp G$.    
  
 Note that,    if two unitaries $V$ and $W$ satisfy Eq. (\ref{cocycle}) with multiplicative characters  $\omega (V,\cdot) $ and $\omega (W, \cdot)$, respectively, then their product $VW$ satisfies Eq. (\ref{cocycle}) with multiplicative character $\omega (VW, \cdot)  =  \omega (V,\cdot)  \, \omega (W, \cdot)$.     This means that  the function $\omega:  \grp G_B  \times  \grp G  \to \CC$ is a multiplicative {\em bicharacter:} $\omega (V, \cdot)$ is a multiplicative character for $\grp G$ for every fixed $V\in \grp G_B$, and, at the same time,  $\omega (\cdot,  g) $ is a multiplicative character for $\grp G_B$ for every fixed $g\in\grp G$.

  The adversarial group $\grp G_B$ contains the commutant of the representation $U: g\mapsto  U_g$, consisting of all the unitaries $V$ such that 
   \begin{align}
  V  U_g     =     U_g  V  \qquad \forall g\in \grp  G\,. 
  \end{align}   
 The unitaries in the commutant satisfy Eq. (\ref{cocycle}) with  the trivial multiplicative character $\omega (V,g)  =1$,  $\forall g\in\grp G$.    In general, the adversarial group may contain other unitary operators, corresponding to non-trivial multiplicative characters.  The full characterization of the adversarial group  is provided by the following theorem:  
 
\begin{theo}\label{theo:semidirect}
Let $\grp G$ be a compact group,  let $U: \grp G \to \Lin  (\spc H)$ be a projective representation of $\grp G$, and let $\grp G_A$ be the group of channels $\grp G_A : =  \{  U_g \cdot U_g^\dag \,  \quad g\in\grp G\}$.   Then,  the adversarial group $\grp G_B$ is isomorphic to the semidirect product   $\grp A\ltimes  U'$, where  $U'$ is the commutant of the  representation $U: \grp G\to \Lin (\spc H_S)$, and $\grp A$ is  an Abelian subgroup  of the group of permutations of $\Irr  (U)$, the set of irreducible representations contained in the decomposition of $U$. 
\end{theo}
The proof is provided in Appendix \ref{app:adversarial}, and a simple example is presented in Appendix \ref{app:example}. 
 
In the following, we will illustrate the construction of the state space $S_A$ in a the prototypical example where the group $\grp G$ is a compact connected Lie group.   
 
\subsection{Connected Lie groups} 
 
When  $\grp G$ is a compact connected Lie group, the characterization of the adversarial group is simplified by the following theorem: 
\begin{theo}\label{theo:connectedLie}
If $\grp G$ is a compact connected Lie group, then the  Abelian subgroup $\grp A$ of Theorem \ref{theo:semidirect} is trivial, and   all the solutions of Eq. (\ref{cocycle}) have  $\omega (V, g)  =1 ~ \forall g\in\grp G$. 
\end{theo}   
The proof is provided in Appendix \ref{app:connectedLie}.  

For compact connected Lie groups,  the   the adversarial group coincides exactly  with the commutant of the representation $U: \grp G \to \Lin (\spc H_S)$.  
An  explicit expression can be obtained in terms of the isotypic decomposition  \cite{fulton2013representation} 
\begin{align}\label{isotypic}
U_g   =  \bigoplus_{j\in  \set{Irr}  (U)}  \,   \Big(     U_g^{(j)}\otimes  I_{\spc M_j} \Big) \, ,    
\end{align}
where $\set{Irr}  (U)$ is the set of irreducible representations (irreps)  of $\grp G$  contained in the decomposition of $U$,  $U^{(j)}:  g\mapsto U_g^{(j)}$ is the irreducible representation of $\grp G$ acting on the representation space $\spc  R_j$, and  $I_{\spc M_j}$ is the identity acting on the multiplicity space $\spc M_j$.     From this expression, it is clear that the adversarial group $\grp G_B$  consists of unitary gates $V$ of the form 
\begin{align}\label{Vp}
V  =  \bigoplus_{j\in \set{Irr}  (U)}   \Big (   I_{\spc R_j}\otimes V_j \Big) \, ,
\end{align} 
where $I_{\spc R_j}$ is the identity operator on the representation space $\spc R_j$, and $V_j$ is a generic unitary operator on the multiplicity space $\spc M_j$.  

In general, the agents $A$ and $B=A'$ do not form a dual pair.  Indeed, it is not hard to see that the maximal adversary of $B$ is the agent $C= A^{\prime \prime}$ that can perform every unitary channel $\map U  (\cdot)  =  U \cdot U^\dag$ where $U$ is a unitary operator of the form 
\begin{align}\label{Upp}
U  =  \bigoplus_{j\in\set{Irr}  (U)} \,   \Big(     U_j  \otimes  I_{\spc M_j} \Big) \, ,   
\end{align}
where $U_j$ is a generic unitary operator on the representation space $\spc R_j$.  When $A$ and $B$ form a dual par, the groups $G_A$ and $G_B$ are sometimes called {\em gauge groups} \cite{marvian2014generalization}.  

    It is now easy to characterize the subsystem $S_A$.  Its states are  equivalence classes of pure states under the relation $|\psi\>\<\psi|  \simeq_{A} |\psi'\>\<\psi'|$ iff
\begin{align}
\exists  U_B  \in  \grp G_B  \qquad {\rm such~that}   \qquad |\psi'\>   =    U_B  |\psi\>  \, .   
\end{align}
It is easy to see that two states in the same equivalence class must satisfy the condition 
\begin{align}
\Tr_B  (|\psi'\>\<\psi'|)   =  \Tr_B  (|\psi\>\<\psi|) \, ,
\end{align}
where the ``partial trace over agent $B$'' is $\Tr_B$ is the  map  
\begin{align}
\Tr_B (\rho)   : =  \bigoplus_{j\in \set{Irr}  (U)}\,    \Tr_{\spc M_j}  [  \Pi_j \, \rho  \,  \Pi_j]   \, ,  
\end{align}
$\Pi_j$ being the projector on the subspace $\spc R_j\otimes \spc M_j$.  

 Conversely, it is possible to show that the state  $\Tr_B  (|\psi\>\<\psi|) $ completely identifies the equivalence class  $[  |\psi\>\<\psi| ]_A$. 
 \begin{prop}\label{prop:LieEquivalence}
Let $|\psi\> ,|\psi'\>  \in  \spc H_S$ be two unit vectors  such that $\Tr_B   (|\psi\>\<\psi|)   =  \Tr_B  (|\psi'\>\<\psi'|)$.  Then, there exists a unitary operator $U_B  \in  \grp G_B$ such that $|\psi'\>  =  U_B|\psi\>$. 
\end{prop}
  The proof is provided in Appendix \ref{app:LieEquivalence}. 

\medskip 

We have seen that the states of system $S_A$ are in one-to-one correspondence with the density matrices of the form $\map T_B (|\psi\>\<\psi|)$, where $|\psi\>  \in  \spc H_S$ is a generic pure state.  Note that the rank of the density matrices $\rho_j$ in  Eq. (\ref{rank})  cannot be larger than the dimensions of the spaces  $\spc R_j$  and $\map M_j$, denoted as $d_{\spc R_j}$ and $d_{\spc M_j}$, respectively.  Taking this fact into account, we can represent the states of $S_A$ as  
\begin{align}
\St (S_A)   \simeq \Big\{  \rho =  \bigoplus_{j\in \set{Irr} (U)}   \,  p_j \,  \rho_j   \,  : ~    \rho_j\in  \QSt (\spc R_j) \, ,  ~  \Rank(\rho_j)  \le \min \{d_{\spc R_j},  d_{\spc M_j} \}  \Big\} \, ,
 \end{align}
 where $\{p_j\}$ is a generic probability distribution.  The state space of system $S_A$ is {\em not convex}, unless the condition 
 \begin{align}
 d_{\spc M_j}  \ge d_{\spc R_j}  \qquad \forall j\in  \set{Irr} (U) \,  
 \end{align}   
 is satisfied.   Basically, in order to obtain a convex set of density matrices, we need the total system $S$ to be ``sufficiently large'' compared to its subsystem $S_A$.     This observation is a clue suggesting that the standard convex framework of quantum theory and general probabilistic theory could be considered as the effective description of subsystems of ``large'' closed systems.  
 
Finally, note that, in agreement with the general construction,  the pure states of system $S$ are {\em ``purifications"} of the states of the system $S_A$.  Every state of system $S_A$ can be obtained from a pure state of system $S$ by {\em ``tracing out"} system  $S_B$.      Moreover, every two purifications of the same state are connected by a unitary transformation in $\grp G_B$.


 \section{Conclusions and outlook}\label{sec:conclusions}  
 
 In this paper we adopted rather minimalistic framework, in which a { single} physical system  was described solely in terms of states and transformations, without introducing  measurements \footnote{Or at least, without assuming an {\em explicit} notion of measurement. Of course, one could always interpret certain transformations as ``measurement processes", but this interpretation is not necessary for the conclusions drawn  in this paper.  }.  This framework can be interpreted in two ways. One way is to think of it  as a fragment of the larger framework of operational-probabilistic theories \cite{chiribella2010probabilistic,hardy2013formalism,chiribella2014dilation,chiribella2016quantum}, in which systems can be freely composed and measurements are explicitly described.  
 The other way is to regard our framework as a dynamicist framework, meant to describe physical systems {\em per se}, independently of any observer.       Both approaches are potentially fruitful, and suggest extensions of the present work.  
  
 On the operational-probabilistic side, it is interesting to see how the definition of subsystem adopted  in this paper interacts with probabilities.  For example, we have seen in a few examples that the state space of a subsystem  is not always convex: convex combination of  allowed states are not necessarily  allowed states.   It is then natural to ask: under which condition is convexity retrieved?  
In a different context,  the non-trivial relation between  convexity and the  dynamical notion of system has been emerged  in a work of Galley and Masanes \cite{galley2018impossibility}. There, the authors studied alternatives to quantum theory where the closed systems have the same states and the same dynamics of  closed quantum systems, while the measurements are different from the quantum measurements.   Among these theories, they found that quantum theory is the only theory where subsystems have a convex state space. 
  These and similar clues are an indication that the interplay between dynamical notions and probabilistic notions plays an important role in determining the structure of  physical theories.  Studying this interplay is a promising avenue of future research. 
 
 On the opposite end of the spectrum, it is interesting to explore how far the measurement-free approach can reach.  
    An interesting research project is to analyze the notions of subsystem, pure state, and purification, in the context of algebraic quantum field theory \cite{haag2012local} and quantum statistical mechanics \cite{bratteli1987operator}.     This is important because the   notion of pure state as an extreme point of the convex set breaks down for type III von Neumann  algebras \cite{yngvason2015localization}, whereas the notions  used in this paper (commutativity of operations, cyclicity of states) would still hold.    Another  promising  clue  is  the existence of  dual pairs of non-overlapping agents, which amounts to the requirement that the set of operations of each agent has trivial center and coincides with its double commutant.   A similar condition plays an important role in the algebraic framework, where the operator algebras with trivial center are known as  {\em factors}, and are at the basis of the theory of von Neumann algebras  \cite{murray1936rings,murray1937rings}.   
    
    Finally, another interesting direction is to enrich the structure of system with additional features, such as a metric, quantifying the proximity of states.    In particular, one may consider a strengthened formulation of the Conservation of Information, in which the physical transformations are required not only to be invertible, but also to preserve the distances.    It is then interesting to consider how the metric on the pure states of the whole system induces a metric on the subsystems, and to search for relations between global metric and local metric. Also in this case, there is a promising precedent, namely the  work of Uhlmann \cite{uhlmann1976transition}, which led to the notion of fidelity \cite{jozsa1994fidelity}.    All these potential avenues of future research suggest that the notions investigated in this work may find application in a variety of different contexts, and for a variety of interpretational standpoints.

 \section*{Acknowledgements}  
 
It is a pleasure to thank Gilles Brassard and Paul Paul Raymond-Robichaud for stimulating discussions on their recent work \cite{brassard2017equivalence},  Ad\'an Cabello, Markus M\"uller, and Matthias Kleinmann for providing motivation to the problem of deriving subsystems,  Mauro D'Ariano and Paolo Perinotti for the invitation to contribute to this Special Issue,  and  Christopher Timpson and Adam Coulton  for an invitation to present at the Oxford Philosophy of Physics Seminar Series, whose engaging atmosphere stimulated me to think about extensions of the Purification Principle.    I am also grateful to the three referees of this paper for useful suggestions, and to Robert Spekkens, Doreen Fraser, L\'idia del Rio, Thomas Galley, John Selby, Ryszard Kostecki, and David Schmidt for interesting discussions during the revision of the original manuscript.  
  
 This work is supported by the Foundational Questions Institute through grant FQXi-RFP3-1325, the National Natural Science Foundation of China through grant 11675136, the Croucher Foundation, the Canadian Institute for Advanced Research (CIFAR), and the Hong Research Grant Council through grant 17326616.  This publication was made possible through the support of a grant from the John Templeton Foundation. The opinions expressed in this publication are those of the authors and do not necessarily reflect the views of the John Templeton Foundation. The authors also acknowledge the hospitality of Perimeter Institute for Theoretical Physics.  Research at Perimeter Institute is supported by
the Government of Canada through the Department of
Innovation, Science and Economic Development Canada
and by the Province of Ontario through the Ministry of
Research, Innovation and Science.

\appendix 
\section{Proof that definitions   (\ref{def:composition}) and  (\ref{def:action}) are well-posed}\label{app:wellposed}

We give only the proof for  definition (\ref{def:composition}), as the other proof follows the same argument.  

\begin{prop}
If the transformations $\map S,  \widetilde {\map S}  ,\map T , \widetilde {\map T}  \in \Act(A;S)^{\prime \prime}$  are such that $[\map S]_{A}  =[ \widetilde {\map S}]_{A}$ and $[\map T]_{A} =  [\widetilde {\map T}]_{A}$, then $[\map S \circ \map T]_{A}  =  [\widetilde{\map S} \circ \widetilde{\map T}]_{A}$. 
\end{prop}
\Proof
Let $(\map S_1,  \map S_2,\dots, \map S_m) \subset  \Act(A;S)^{\prime \prime} $ and $(\map T_1,\map T_2,\dots, \map T_n) \subset \Act(A;S)^{\prime \prime}$ be two finite sequences such that 
\begin{align}
\nonumber &\map S_1   = \map S    \, , \qquad \map S_m  =  \widetilde {\map S} \, ,  \qquad  \Deg_{A'}  (\map S_i) \cap \Deg_{A'} (\map S_{i+1}) \not =  \emptyset    \, \quad \forall i \in \{1,\dots,  m-1\} \\ 
& \map T_1   = \map T    \, , \qquad \map T_n  =  \widetilde {\map T} \, ,  \qquad  \Deg_{A'}  (\map T_j) \cap \Deg_{A'} (\map T_{j+1}) \not =  \emptyset    \, \quad  \forall j \in \{1,\dots,  n-1\}    
\label{conditions}
\end{align} 
Without loss of generality, we assume that the two finite sequences have the same length $m=n$.  When this is not the case, one can always add dummy entries and ensure that the two sequences have the same length: for example, if $m<n$, one can always define $\map S_i : =  \map S_m$ for all $i  \in   \{m+1, \dots,  n\}$.     

Equations (\ref{conditions}) mean that for every $i$ and $j$ there exist  transformations $\map B_i,\widetilde{\map B_i},  \map C_j, \widetilde{\map C_j} \in \Act(A;S)'$ such that  
\begin{align}
\nonumber 
\map B_i \circ \map S_i     &=  \widetilde{\map B_i} \circ {\map S_{i+1}}  \\
\map C_j \circ \map T_j     &=  \widetilde{\map C_j} \circ {\map T_{j+1}}  
\end{align}
Using the above equalities for $i=j$, and using the fact that transformations in $\Act (A;S)'$ commute with transformations in $\Act (A;S)^{\prime \prime}$, we obtain  
\begin{align}
 \nonumber 
  \big(  \map B_i \circ \map C_i\big)  \circ \big(\map S_i  \circ \map T_i\big)  &   =   \big(\map B_i  \circ \map S_i\big)  \circ  \big(\map C_i \circ \map T_i \big) \\
 \nonumber
 &   =   \big(\widetilde{\map B_i } \circ \map S_{i+1}\big)  \circ  \big(\widetilde{\map C_i} \circ \map T_{i+1} \big) \\
   & =  \big(  \widetilde{\map B_i} \circ \widetilde{\map C_i}\big)  \circ \big(\map S_{i+1}  \circ \map T_{i+1}\big) \, . 
 \end{align}
 In short, we proved that 
 \begin{align}\label{productdeg}
 \Deg_{A'}  (  \map S_i  \circ \map T_i)\cap     \Deg_{A'}  (  \map S_{i+i}  \circ \map T_{i+1})  \not  =  \emptyset \, \qquad \forall i\in  \{1,\dots, n-1\} \, .
 \end{align}
 To conclude, observe that the sequence $(\map S_1\circ\map T_1, \map S_2\circ \map T_2,\dots, \map S_n\circ \map T_n)$ satisfies $\map S_1\circ \map T_1  =\map S  \circ \map T$, $\map S_n\circ\map T_n =  \widetilde{ \map S} \circ \widetilde {\map T}$, and Eq. (\ref{productdeg}).  By definition, this means that the transformations $\map S\circ \map T$ and $\widetilde {\map S} \circ \widetilde{\map T}$ are in the same equivalence class.   
\qed

\section{ The commutant of the local channels}\label{app:commutant}  

Here we show that the commutant of the quantum channels of the form $\map A\otimes \map I_B$ consists of quantum channels of the form $\map I_A\otimes  \map B$.     

Let $\map C \in  \CChan (S)$ be a quantum channel that commutes with all channels of the form $\map A\otimes \map I_B$, with $\map A  \in  \CChan  (A)$.  For a fixed unit vector $|\alpha\> \in  \spc H_A$, consider the  erasure channel $\map A_{\alpha} \in \CChan (A)$ defined by  
\begin{align}
\map A_{\alpha}  (\rho)  =  |\alpha\>\<\alpha| \, \Tr [\rho] \qquad \forall \rho  \in  \Lin (A) \,.
\end{align}   
 Then, the commutation condition $\map C  \circ (\map A_{\alpha}  \otimes \map I_B)  =  (\map A_{\alpha} \otimes \map I_B) \circ \map C$ implies 
\begin{align}\label{back}
  \map C  \Big( |\alpha\>\<\alpha|  \otimes  |\beta\>\<\beta|    \Big)    &  = \map C      \Big [  \Big ( \map  A_{\alpha} \otimes \map I_B\Big)     \Big(   |\alpha\>\<\alpha|  \otimes  |\beta\>\<\beta| \Big) \Big] \nonumber   \\
    &  =    \Big( \map  A_{\alpha} \otimes \map I_B\Big)   \Big[   \map C  \Big( |\alpha\>\<\alpha|  \otimes  |\beta\>\<\beta|\Big)\Big] \nonumber  \\
    &  =     |\alpha\>\<\alpha|  \otimes  \Tr_A  \Big[\map C   \Big( |\alpha\>\<\alpha|  \otimes  |\beta\>\<\beta|  \Big) \Big]       \qquad  \forall |\beta \>  \in \spc H_B \, .
\end{align}
Tracing over $B$ on both sides of Eq. (\ref{back}), we obtain  
\begin{align}
 \Tr_B\Big[\map C  \Big( |\alpha\>\<\alpha|  \otimes  |\beta\>\<\beta|   \Big)\Big]   =  |\alpha\>\<\alpha|  \,  . 
 \end{align}
The above relation  implies that the state $\map C  \Big( |\alpha\>\<\alpha|  \otimes  |\beta\>\<\beta|   \Big)$ is of the form  
\begin{align}
\map C  \Big( |\alpha\>\<\alpha|  \otimes  |\beta\>\<\beta|   \Big) &  =  |\alpha\>\<\alpha|  \otimes \map B   (|\beta\>\<\beta|) \, ,
\end{align}
for some suitable channel $\map B  \in  \CChan (B)$. 
Since $|\alpha\>$ and $|\beta\>$ are arbitrary, we obtained    $\map C      =  \map I_A \otimes \map B$.

\section{Subsystems associated to finite dimensional von Neumann algebras}\label{app:algebras}  
Here we prove the statements made in the main text about quantum channels with Kraus operators in a given algebra.

\subsection{The commutant of  $\CChan (\sf A)$.}
The purpose of this subsection is to prove the following theorem: 
\begin{theo}\label{theo:commutant}
Let $\sf A$ be a von Neumann subalgebra of $M_d (\CC)$, $d<\infty$, and let $\CChan (\sf A)$ be  the set of quantum channels   with Kraus operators in $\sf A$.   Then, the commutant of $\CChan (\sf A)$ is the set of channels with Kraus operators in the algebra $\sf A'$. In formula,
\begin{align}
\CChan  (\sf A)'   =    \CChan  (\sf A')  \, .  
\end{align}
\end{theo}  
The proof consists of a few lemmas, provided in the following.   

\begin{Lemma}\label{lem:diagonal}
Every channel $\map D  \in  \CChan  (A)'$ must satisfy the condition 
\begin{align}\label{diagonal}
\map P_l  \circ \map D \circ \map P_k  =  0  \qquad \forall  l
\not  = k \, ,
\end{align}
where $\map P_k$ is the CP map $\map P_k  (\cdot)  :  =   \Pi_k  \cdot \Pi_k$,  and $\Pi_k$ is the  projector on the subspace $\spc H_{A_k}\otimes \spc H_{B_k}$ in Eq. (\ref{wedderburn}). 
\end{Lemma}

\Proof 
Consider the quantum channel $\map C  \in  \CChan  (\sf A)$ defined as 
\begin{align}
\map C   : =  \bigoplus_k  \,   \Big(   |\alpha_k\>\<\alpha_k|  \, \Tr_{A_k}    \otimes   \map I_{B_k}   \Big) \circ  \map P_k   \, ,
\end{align} 
where each $|\alpha_k\>$ is a  generic (but otherwise fixed) unit vector in $\spc H_{A_k}$ and $  \map I_{B_k}$ is the identity map on $\Lin (  \spc H_{B_k})$.   By definition, every channel $\map D \in  \CChan (\sf A)'$ must satisfy the condition $\map C \circ \map D  =  \map D \circ \map C$.  In particular, we must have 
\begin{align}
\map D  (  |\alpha_k\>\<\alpha_k|  \otimes |\beta_k\>\<\beta_k|)  &  =   (\map D \circ \map C)  (  |\alpha_k\>\<\alpha_k|  \otimes |\beta_k\>\<\beta_k|)  \nonumber  \\
 &  =   (\map C \circ \map D)  (  |\alpha_k\>\<\alpha_k|  \otimes |\beta_k\>\<\beta_k|)  \nonumber  \\
 &  =  \bigoplus_l  \,   \Big(   |\alpha_l\>\<\alpha_l|  \otimes      \Tr_{A_l}  \Big[   ( \map P_l  \circ \,  \map D)  (  |\alpha_k\>\<\alpha_k|  \otimes |\beta_k\>\<\beta_k| )  \, \Big]   \Big)
\end{align}
  Applying the CP map  $\map P_l$  on both sides of the above equality, we obtain the relation
\begin{align}\label{key}
(\map P_l  \circ \map D)  (  |\alpha_k\>\<\alpha_k|  \otimes |\beta_k\>\<\beta_k|)     =      |\alpha_l\>\<\alpha_l|  \otimes    \map M_l   (  |\alpha_k\>\<\alpha_k|  \otimes |\beta_k\>\<\beta_k| )   \, ,
\end{align}
where $\map M_l $ is the map from $M_{d}  (\CC)$ to $\Lin  (\spc H_{A_l})$ defined as $\map M_l :  =  \Tr_{A_l}   \circ  \map P_l  \circ \map D$.    

Note that the right hand side  of Eq. (\ref{key}) depends on the choice of vector $|\alpha_l\>$,   which is arbitrary.  On the other hand,  the left hand side does not depend on $|\alpha_l\>$.  Hence,   the only way that the two sides of Eq. (\ref{key}) can be equal for $k\not  =  l$ is that they are both equal to 0.  
Moreover, since  $|\alpha_k\>$ and $|\beta_k\>$ are arbitrary vectors in $\spc H_{A_k}$ and $\spc H_{B_k}$, respectively, Eq. (\ref{key}) implies the relation  
\begin{align}
(\map P_l  \circ \map D  )    (\rho)  =  0   \qquad \forall \rho  \in \Lin  (\spc H_{A_k} \otimes \spc H_{B_k})    \, , \quad \forall  l\not =  k \, .
\end{align}
Since $\rho$ is an arbitrary operator in $ \Lin  (\spc H_{A_k} \otimes \spc H_{B_k})$, we conclude that the relation $\map P_l\circ \map D \circ \map P_k  =  0$ holds for every $l\not  = k$.  
 \qed

\medskip 

\begin{Lemma}\label{lem:endo}
Every channel $\map D  \in  \CChan  (A)'$ must satisfy the conditions 
\begin{align}\label{pkc=pkcpk}
 \map D \circ \map P_k    =  \map P_k  \circ \map D \circ \map P_k   \qquad \forall  k \, .
\end{align}
and 
\begin{align}\label{cpk=pkcpk}
 \map P_k  \circ  \map D     =  \map P_k  \circ \map D \circ \map P_k   \qquad \forall  k \, .
\end{align}
In short:  $\map D \circ \map P_k  = \map P_k \circ \map D$ for every $k$. 
\end{Lemma}

\Proof  Define $\map D_k  :  =  \map D\circ \map P_k$. Then,  the Cauchy-Schwarz inequality yields 
\begin{align}
\Big|\< \phi  |     \Pi_i  \,   \map D_k    (\rho) \,  \Pi_j  \,  |\phi\>\Big|  &    \le   \sqrt{ \< \phi  |     \Pi_i  \,   \map D_k   (\rho) \,  \Pi_i  \,  |\phi\>   ~  \< \phi  |     \Pi_j  \,   \map D_k   (\rho) \,  \Pi_j  \,  |\phi\>  }\nonumber \\
&  \le   \sqrt{ \< \phi  |     (\map P_i \circ   \map D \circ \map P_k)   (\rho) \,    |\phi\> ~  \< \phi  |    (\map P_j \circ   \map D\circ \map P_k )  (\rho) \,    |\phi\>  } \, .
\end{align}
Thanks to  Lemma \ref{lem:diagonal}, we know  the right hand side is 0 unless $i=j=k$.  Since the vector $|\phi\>$ is  are arbitrary, the condition $|\< \phi  |     \Pi_i  \,   \map D_k   (\rho) \,  \Pi_j  \,  |\phi\>|=0$  implies the 
relation  
$\Pi_i  \,   \map D_k   (\rho) \,  \Pi_j     =     0$.    
Using this fact, we obtain  the relation  
\begin{align}
\nonumber 
(\map D \circ \map P_k)  (\rho)  & =  \map D_k (\rho)   \\    
  &   =     \sum_{i,j}   \,  \Pi_i  \,    \map D_k (\rho)   \, \Pi_j \nonumber  \\
  &   =      \,  \Pi_k  \,    \map D_k (\rho)   \, \Pi_k  \nonumber \\
&  =      (\map P_k\circ \map D \circ \map P_k )(\rho)   \, , 
\end{align}
valid for arbitrary density matrices $\rho$, and therefore for arbitrary matrices in $M_d (\CC)$. In conclusion, Eq. (\ref{cpk=pkcpk}) holds.  

The proof of Eq. (\ref{pkc=pkcpk}) is analogous to that of Eq. (\ref{cpk=pkcpk}), with the only difference that it uses  the {\em adjoint  map}, which for a generic linear map $\map L  :  \Lin (\spc H_S)  \to \Lin (\spc H_S)$  is defined  by the relation 
\begin{align}
\Tr [  \map L^\dag (O) \,  \rho  ]   :  =  \Tr [ O  \,  \map L  (\rho)]   \qquad \forall    O  \in  M_d (\CC) \, ,   ~ \forall  \rho  \in  M_d (\CC) \, .
\end{align}
 Specifically, we define the map $\widetilde{\map D}_k  :  =   \map P_k  \circ \map D$. Then, we obtain the relation
 \begin{align}
\nonumber \Big|\, \<  \phi|   \, \widetilde{\map D}_k ( \Pi_i \rho  \Pi_j) \, |\phi\>   \Big|  &  = \Big|  \Tr  \big[ \widetilde {\map D}_k^\dag (|\phi\>\<\phi|)      \Pi_i \rho  \Pi_j \big]\Big| \nonumber \\
 &  = \left|  
 \Tr  \left[  
   \left( 
    \sqrt{ \widetilde {\map D}_k^\dag (|\phi\>\<\phi|)   }   \Pi_i \sqrt \rho
    \right)\,  
    \left(  \sqrt \rho    \Pi_j    \sqrt{ \widetilde {\map D}_k^\dag (|\phi\>\<\phi|   } 
    \right) 
    \right]
    \right|  \nonumber \\  
 & \le \sqrt{    \Tr \left[  \map D_k^\dag (|\phi\>\<\phi| ) \,  \Pi_i  \rho  \Pi_i \right] \,     \Tr \left[  \map D_k^\dag (|\phi\>\<\phi| ) \,  \Pi_j  \rho  \Pi_j \right]}  \nonumber \\
&=\sqrt{  \<  \phi|   \, \widetilde{\map D}_k ( \Pi_i \rho  \Pi_i) \, |\phi\> ~   \<  \psi|   \, \widetilde{\map D}_k ( \Pi_j \rho  \Pi_j) \, |\psi\>} \nonumber    \\
& =  \sqrt{  \<  \phi|   \,( \map  P_k \circ \map D\circ  \map P_i) ( \rho  ) \, |\phi\> ~   \<  \psi|   \,  (\map P_k  \circ \map D \circ \map P_j) ( \rho) \, |\psi\>}    \, ,
\end{align}  
where the right hand side is 0 unless $i=j=k$  (cf. Lemma \ref{lem:endo}).  Since the condition $|\, \<  \phi|   \, \widetilde{\map D}_k ( \Pi_i \rho  \Pi_j) \, |\phi\> |=0, \, \forall |\phi\> \in\spc H_S$ implies the condition $\widetilde{\map D}_k ( \Pi_i \rho  \Pi_j)=0$,  we obtained the relation    
\begin{align}
\widetilde{\map D}_k  (\Pi_i \rho   \Pi_j)    =     0   \qquad {\rm unless} \quad  i=j=k \, .
\end{align}
Using this fact, we obtain the equality
\begin{align}
( \map P_k\circ \map D)  (\rho)  &  =   \widetilde{\map D}_k  (\rho)  \nonumber \\
 &  =   \sum_{i,j}  \, \widetilde{\map D}_k   ( \Pi_i  \rho  \Pi_j)   \nonumber \\
 & =  (\widetilde{\map D}_k  \circ \map P_k  )    (\rho)  \nonumber \\
 &  =  (\map P_k \circ \map D \circ \map P_k)  (\rho)  \, .    
\end{align}
Since the equality holds for every $\rho$,  this proves Eq. (\ref{cpk=pkcpk}).  
  \qed 

\medskip  

Lemma \ref{lem:endo} guarantees that the linear map $ \map D  \circ \map P_k$ sends $\Lin (\spc R_k \otimes \spc M_k)$ into itself.  
It is also easy to see that the map $\map D \circ \map P_k$  has a simple form:  
\begin{Lemma}\label{lem:product}
For every channel $\map D  \in  \CChan  (A)'$, one has 
\begin{align}
\map D \circ \map P_k     =   (\map I_{A_k}  \otimes \map B_k)  \circ \map P_k  \qquad \forall k
\end{align}
where  $\map I_{A_k}$ is the identity map from $\Lin (\spc H_{A_k})$ to itself, and  $\map B_k$ is a quantum channel from $\Lin  (\spc H_{A_k})$ to itself.  
\end{Lemma}

\Proof Straightforward extension of the proof in Appendix \ref{app:commutant}. \qed 

\medskip 

Using the notion of adjoint, we can now prove the following
\begin{Lemma}\label{lem:preserve}
For every channel $\map D  \in  \CChan  (A)'$, the adjoint $\map D^\dag$ preserves the elements of the algebra $\sf A$, namely $\map D^\dag (C)  =  C$ for all $C\in\sf A$.   
\end{Lemma}

\Proof    Let $C $ be a generic element of $\sf A$.     By Eq. (\ref{wedderburn}), one has the equality   
\begin{align}\label{C}  C  =  \bigoplus_k   (  C_k\otimes I_{B_k})  =\bigoplus_k  \map P_k  (C)  \end{align}  
Using Lemma \ref{lem:product} and the definition of adjoint, we obtain  
\begin{align}
\Tr  [  \map D^\dag  (C) \, \rho]  & =  \Tr[   C\,   \map D (\rho)]   \nonumber \\
  &   = \sum_k  \,    \Tr [  \map P_k  (C) \,  \map D (\rho)  ]  \nonumber \\
  &   = \sum_k  \,    \Tr [ C \,     (\map P_k \circ \map D) (\rho)  ] \nonumber  \\
&   = \sum_k  \,    \Tr [ C \,     (\map P_k \circ \map D \circ \map P_k) (\rho)  ] \nonumber  \\
&   = \sum_k  \,    \Tr \Big  [\map P_k (C) ~     (\map D   \circ \map P_k)   (\rho)  \,  \Big]  \nonumber  \\
&  =  \sum_k  \, \Tr  \Big\{     ( C_k\otimes I_{B_k} ) ~  [ (\map I_{A_k}  \otimes \map B_k)  \circ \map  P_k ] (\rho)  \Big\}  \, ,  
\end{align}
having used Lemma \ref{lem:product} in the last equality.   Then, we  use the fact that the channel $\map B_k$ is trace-preserving, and therefore its adjoint $\map B_k^\dag$ preserves the identity. Using this fact, we can continue the chain of equalities as   
\begin{align}
\Tr  [\map D^\dag  (C)]&  =  \sum_k  \, {\Tr}  \Big\{   [ C_k\otimes   \map B_k^\dag  (I_{B_k}) ] ~ \map  P_k (\rho) \Big\}\nonumber \\
&  =  \sum_k  \, \Tr  \Big[  (C_k\otimes   I_{B_k} ) ~ \map  P_k (\rho) \Big]\nonumber \\
&  =  \sum_k  \, \Tr  \Big[  \map P_k(C_k\otimes   I_{B_k} ) ~ \rho \Big]\nonumber \\
&  =   \, \Tr  \left[  \left(\bigoplus_k C_k\otimes   I_{B_k} \right) ~ \rho \right]\nonumber \\
&= \Tr [C\rho] \, ,
\end{align}
having used Eq. (\ref{C}) in the last equality. 
Since the equality holds for every density matrix $\rho$, we proved the equality $\map D^\dag (C)  =  C$. \qed 

\medskip  

We are now in position to prove Theorem \ref{theo:commutant}.  

{\bf Proof of Theorem \ref{theo:commutant}.}       Let $\map D $ be a quantum channel  in  $\CChan (\sf A)'$.  Then,  Lemma \ref{lem:preserve} guarantees that the adjoint  $\map D^\dag$ preserves all operators in the algebra $\sf A$.   Then,  a result due to  Lindblad \cite{lindblad1999general} guarantees that all the Kraus operators of $\map D$ belong to the algebra $\sf A'$.   This proves the inclusion $\CChan  (\sf A)'  \subseteq \CChan (\sf A')$.  

The converse inclusion is immediate:  if a channel $\map D$ belongs to $\CChan (A')$, it  commutes with all channels in $\CChan (\sf A)$ thanks to the block diagonal form of the Kraus operators [cf. equations  (\ref{blockA}) and (\ref{blockA'})]. \qed 

\subsection{States of  subsystems associated to  finite dimensional von Neumann algebras }

Here we provide the proof of Proposition \ref{prop:algebrastates}, adopting the notation ${\sf B}  :  = \sf A'$. 

   The proof uses the following lemma:

\begin{Lemma}[No signalling condition]\label{lem:nosigalg}
For every channel $\map D  \in  \CChan  (\sf B)$, one  has  $\Tr_{\sf B}  \circ \map D  =  \Tr_{\sf B}$.  
\end{Lemma}  
\Proof   By definition, the partial trace channel $\Tr_{\sf B}$ can be written as 
\begin{align}
\Tr_{\sf B}   =  \bigoplus_k  \,   (\map I_{A_k}  \otimes \Tr_{B_k}) \circ \map P_k
\end{align}
For every channel $\map D \in \CChan (\sf B)$, we have 
\begin{align}
\Tr_{\sf B}  \circ \map    D 
&  =     \bigoplus_k  \,    (\map I_{A_k}  \otimes \Tr_{B_k}) \circ \map P_k     \circ \map D  \nonumber \\   
&  =     \bigoplus_k  \,    (\map I_{A_k}  \otimes \Tr_{B_k})      \circ  (\map I_{A_k}  \otimes \map B_k)  \circ \map P_k  \nonumber \\   
&  =     \bigoplus_k  \,    \Big [  \map I_{A_k}  \otimes  ( \Tr_{B_k}  \circ \map B_k  ) \Big]  \circ \map P_k  \nonumber \\   
&  =     \bigoplus_k  \,    (\map I_{A_k}  \otimes \Tr_{B_k}   ) \circ \map P_k  \nonumber \\   
&  =  \Tr_{\sf B}
\end{align}
where  the second equality follows  from Lemma \ref{lem:product}, and the third equality follows from the fact that $\map B_k$ is trace-preserving.     \qed

\medskip

{\bf Proof  of proposition \ref{prop:algebrastates}.}    Suppose that $\rho$ and $\sigma$ are equivalent for $A$.    By definition, this means that there exists  a finite sequence $(\rho_1,\rho_2,\dots,  \rho_n)$ such that 
\begin{align}
\rho_1 =\rho  \, ,  \qquad \rho_n  =  \sigma  \, ,\qquad    {\rm and} \qquad  \Deg_{B} (\rho_i)  \cap  \Deg_{B} (\rho_{i+1})\not  =  \emptyset   \quad \forall  i\in  \{1,2,\dots,  n-1\} \, .   
\end{align} 
The condition of non-trivial intersection implies that, for every $i \in  \{1,2,\dots, n-1\}$,  one has 
\begin{align}
 \map D_i  \, (\rho_i)     =    \widetilde{\map D}_i \,  (\rho_{i+1})  \, ,
\end{align}
where  $\map D_i$ and  $\widetilde{ \map D}_i$ are two quantum channels in  $\CChan (\sf B)$.    Tracing over $\sf B$ on both sides we obtain the relation 
\begin{align}
 (\Tr_{\sf B}  \circ\map D_i   ) \, (\rho_i)     =  (\Tr_{\sf B} \circ  \widetilde{\map D}_i) \,  (\rho_{i+1}) \, , 
\end{align} 
and, thanks to Lemma \ref{lem:nosigalg},  $\Tr_{\sf B}  [\rho_i]  =  \Tr_{\sf B}  [\rho_{i+1}]$. 
Since the equality holds for every $i\in  \{1,\dots,  n-1\}$, we obtained the condition $\Tr_{\sf B}  [\rho]  =  \Tr_{\sf B}  [\sigma]$.  
In summary, if two states $\rho$ and $\sigma$ are equivalent for $A$, then $\Tr_{\sf B}  [\rho]  =  \Tr_{\sf B}  [\sigma]$.  

To prove the converse, it is enough to define the channel  $\map D_0  \in  \CChan (\sf B)$ as 
\begin{align}\label{d0}
\map D_0 (\rho)   :  = \bigoplus_{k}  \,      \Tr_{B_k}  [ \map P_k (\rho)  ] \otimes \beta_k \, ,
\end{align}
where each $\beta_k$ is a fixed (but otherwise generic) density matrix in $\Lin  (\spc H_{B_k})$. Now,   if the equality $\Tr_{\sf B}  [  \rho]  =  \Tr_{\sf B} [\sigma]$ holds, then  also the equality  $\map D_0  (\rho)   =  \map D_0  (\sigma)$ holds.  This proves that the intersection between $\Deg_{B} (\rho)$ and $\Deg_{B} (\sigma)$  is non-empty, and therefore $\rho$ and $\sigma$ are equivalent for $A$. \qed

\subsection{Transformations of subsystems associated to finite dimensional von Neumann algebras}

Here we prove that all transformations of system $S_A$ are of the form $\map A   =  \bigoplus_k  \map A_k$,  where each   $\map A_k$ is a quantum channel from $\Lin (\spc H_{A_k})$ to itself.  The proof is based on the following lemmas: 

\begin{Lemma}\label{lem:simplestructure}
For every channel $\map C  \in  \CChan (\sf A)$, one has the relation  
\begin{align}
\map P_k  \circ \map C    =    ( \map A_k  \otimes \map I_{B_k})  \circ \map P_k  \, ,
\end{align}
where $\map A_k$ is a quantum channel from $\Lin (\spc H_{A_k})$ to itself.  
\end{Lemma}
\Proof   Let
\begin{align}
\map C (\rho) =  \sum_i \,   C_i \,  \rho \, C_i^\dag \, , \qquad C_i  = \bigoplus_k  \,   (  C_{ik}  \otimes I_{B_k}) \, .  
\end{align}
be a Kraus representation of channel $\map C$.  The preservation of the trace amounts to the condition 
\begin{align}
I   &  =  \sum_i  C_i^\dag C_i    \nonumber  \\
&  = \bigoplus_k     \left (   \sum_i     C_{ik}^\dag C_{ik} \otimes I_{B_k}   \right)  \, ,
\end{align} 
which implies  
\begin{align}
 \sum_i     C_{ik}^\dag C_{I_k}   =  I_{A_k}  \qquad \forall k  \, .
\end{align}
Now,  we have 
\begin{align}   
(\map P_k  \circ \map C  ) (\rho)  
 &  =   \sum_{i}     (C_{ik} \otimes I_{B_k})  \, \map P_k (\rho) \,  (C_{ik}\otimes I_{B_k})^{\dag} \nonumber \\
 & =   (\map  A_k  \otimes \map I_{B_k}) \,[\map P_k (\rho)] \,,   \label{qwerty}   
\end{align}
where the channel $\map A_k$ is defined as   
\begin{align}
\map A_k (\sigma)  :  =   \sum_i  C_{ik} \, \sigma \, C_{ik}^\dag  \qquad \forall \sigma \in  \Lin (\spc H_{A_k})\, .  
\end{align} 
Since the density matrix $\rho$ in Eq. (\ref{qwerty}) is arbitrary, we proved the relation $\map P_k\circ \map C  = (  \map A_k\otimes \map I_{B_k})\circ \map P_k$. \qed 

\begin{Lemma}\label{lem:Ak}
For  two channels  $\map C, \map C' \in \CChan (\sf A)$,  let $\map A_k$ and $\map A_k'$ be the quantum channels defined in Lemma \ref{lem:simplestructure}.  Then, the following are equivalent 
\begin{enumerate}
\item   $\Tr_{\sf B}  \circ \, \map C  =  \Tr_{\sf B}  \circ\, \map C'$  
\item $\map A_k  =  \map A_k'~ $   for every $k$. 
\end{enumerate}
\end{Lemma}
\Proof   $2  \Longrightarrow 1$.   For channel $\map C$ we  have 
\begin{align}
\Tr_{\sf B}  \circ  \,\map C  
&  =    \bigoplus_k  \,   (\map I_{A_k} \otimes \Tr_{B_k})   \circ  \map P_k  \circ \map C  \nonumber  \\
&  =    \bigoplus_k  \,   (\map I_{A_k} \otimes \Tr_{B_k})   \circ    (\map A_k  \otimes  \map I_{B_k}) \circ \map P_k  \nonumber  \\
&  =    \bigoplus_k  \,   (\map A_k  \otimes \Tr_{B_k})   \circ    \map P_k   \, . \label{16aprile}
\end{align}
Similarly, for channel $\map C'$ we have  
\begin{align}\label{16aprile2}
\Tr_{\sf B}  \circ  \,\map C'  =   \bigoplus_k  \,   (\map A'_k  \otimes \Tr_{B_k})   \circ    \map P_k  \,.
\end{align} 
 Clearly, if $\map A_k $ and $\map A_k'$ are equal for every $k$, then the partial traces $\Tr_{\sf B}  \circ \, \map C$ and $\Tr_{\sf B} \circ  \, \map C'$ are equal. 
 
  $1  \Longrightarrow 2$.    Suppose that     partial traces $\Tr_{\sf B}  \circ\, \map C$ and $\Tr_{\sf B} \circ \,\map C'$ are equal. Then, Equations (\ref{16aprile}) and (\ref{16aprile2}) imply the equality  
  \begin{align}\label{fame}
  (\map A_k  \otimes \Tr_{B_k})   \circ    \map P_k   =    (\map A'_k  \otimes \Tr_{B_k})   \circ    \map P_k  \qquad \forall k \, .
  \end{align}
In turn, the above equality implies $\map A_k   =  \map A_k'$, $\forall k$,  as one can easily verify by applying both sides of Eq. (\ref{fame}) to a generic product operator $X_k\otimes Y_k$,  with $X_k \in \Lin (  \spc H_{A_k})$ and $Y_k  \in  \Lin  (\spc H_{B_k})$. \qed 

         \begin{Lemma}\label{lem:trb}
Two channels $\map C , \map C' \in  \CChan (\sf A)$   are equivalent for $A$ if and only if $\Tr_{\sf B}  \circ \,\map C  =  \Tr_{\sf B}  \circ\, \map C'$.    
\end{Lemma}  
\Proof    Suppose that $\map C$ and $\map C'$ are equivalent for $A$.  By definition, this means that there exists a finite sequence $(  \map C_1,\map C_2,\dots, \map C_n)  \subset  \CChan  (\sf A)$ such that 
\begin{align}
\map C_1  =  \map C  \, , \qquad \map C_n  =  \map C'   \, , \qquad  \Deg_B (\map C_i)  \cap \Deg_B (\map  C_{i+1}) \not  = \emptyset   \,\quad  \forall  i\in \{1,\dots,  n-1\} \, .
\end{align}
This means that, for every $i$, there exist two channels $\map D_i  ,  \widetilde{\map D}_i  \in  \CChan (\sf B)$ such that 
\begin{align}
\map D_i\circ\map  C_i  =  \widetilde {\map D}_i    \circ \map C_{i+1}   \, .
\end{align}
Tracing over $\sf B$ on both sides,  we obtain 
\begin{align}
\Tr_{\sf B}\circ\,  \map D_i\circ\map  C_i  =  \Tr_{\sf B}\circ\,  \widetilde {\map D}_i    \circ \map C_{i+1}   \, , 
\end{align}
and, using the no signalling condition of Lemma \ref{lem:nosigalg},
\begin{align}
\Tr_{\sf B}\circ  \, \map  C_i  =  \Tr_{\sf B}\circ \, \map C_{i+1}   \, , 
\end{align}
Since the above relation holds for every $i$, we obtained the equality $\Tr_{\sf  B} \circ\,  \map C  = \Tr_{\sf B}  \circ \,\map C'$. 

Conversely, suppose that $\Tr_{\sf  B} \circ \,  \map C  = \Tr_{\sf B}  \circ \,\map C'$. Then, Lemma \ref{lem:Ak} implies the equality 
\begin{align}
\map A_k  =  \map A_k'  \qquad \forall k \,,
\end{align}
where $\map A_k$ and $\map A_k'$ are the quantum channels defined in Lemma \ref{lem:simplestructure}.

Now, let $\map D_0$ be the channel in $\CChan (\sf B)$ defined in Eq. (\ref{d0}).  
By definition, we have 
\begin{align}\label{bastissima}
\map D_0  \circ \map C   & =   \bigoplus_k \,      (\map I_{A_k}  \otimes \beta_k  \,  \Tr_{B_k})  \circ \map P_k \circ \map C 
\nonumber  \\  
& =   \bigoplus_k \,      (\map I_{A_k}  \otimes \beta_k  \,  \Tr_{B_k})  \circ   (\map A_k  \otimes \map I_{B_k})   \circ \map P_k 
\nonumber  \\  
& =   \bigoplus_k \,      (\map A_k  \otimes \beta_k  \,  \Tr_{B_k})  \circ   \map P_k  \, .
\end{align}
Similarly, we have 
\begin{align}
\map D_0  \circ \map C  =     \bigoplus_k \,      (\map A'_k  \otimes \beta_k  \,  \Tr_{B_k})  \circ   \map P_k  \, .
\end{align}
Since $\map A_k$ and $\map A_k'$ are equal for every $k$, we conclude that $\map D_0  \circ \map C$ is equal to $\map D_0 \circ \map C'$.   This means that the intersection between $\Deg (\map C)$ and $\Deg (\map C')$ is non-empty, and, therefore $\map C$ is equivalent to $\map C'$ modulo $B$. \qed 
\medskip  

Combining Lemmas \ref{lem:Ak} and \ref{lem:trb}, we obtain the following corollary: 
\begin{cor}
For  two channels  $\map C, \map C' \in \CChan (\sf A)$,  let $\map A_k$ and $\map A_k'$ be the quantum channels defined in Lemma \ref{lem:simplestructure}.   Then, the following are equivalent: 
\begin{enumerate}
\item $\map C$ and $\map C'$ are equivalent for $A$
\item $   \bigoplus_k \map A_k  = \bigoplus_k \map A_k'$. 
\end{enumerate}
\end{cor}
\Proof By Lemma \ref{lem:trb}, $\map C$ and $\map C'$ are equivalent for $A$ if and only if the condition $\Tr_{\sf B}  \circ \map C  =  \Tr_{\sf B}  \circ \map C'$ holds.   By Lemma \ref{lem:Ak}, the condition  $\Tr_{\sf B}  \circ \map C  =  \Tr_{\sf B}  \circ \map C'$ holds if and only if one has $\map A_k =\map A_k'$ for every $k$.  In turn, the latter condition holds if and only if  the equality $   \bigoplus_k \map A_k  = \bigoplus_k \map A_k'$ holds. \qed 
\medskip  

In summary, the transformations of system $S_A$ are characterized as 
\begin{align}
\Transf (S_A)  =  \bigoplus_k  \, \CChan  (A_k) \, ,
\end{align}
where $\CChan(A_k)$ is the set of all quantum channels from $\Lin (\spc H_{A_k})$ to itself.  

To conclude, we observe that the transformations of $S_A$ act in the expected way.  To this purpose, we consider the {\em restriction map}  
\begin{align}\pi_{\sf A}  :   \CChan   (\sf A)  \to   \bigoplus_k  \, \CChan  (A_k)  \, , \quad     \map C  \mapsto    \bigoplus_k \, \map A_k \, ,
\end{align}  
where $\map A_k$ is defined as in Lemma \ref{lem:simplestructure}. 

Using the restriction map, we can prove the following propositions: 
\begin{prop}
For every channel $\map C \in\CChan  (\sf A)$ we have the relation 
\begin{align}
\Tr_{\sf B}   \circ  \,  \map C    =      \pi_{\sf A}  (\map C)  \circ \Tr_{\sf B} 
\end{align}
In words,  evolving system $S$ with $\map C$ and then computing the local  state of system $S_A$ is the same as computing the local state  of system $S_A$ and then evolving it with $\pi_{\sf A}  (\map C)$.
\end{prop}
\Proof Using Lemma \ref{lem:simplestructure}, the proof is straightforward:  
\begin{align}
\Tr_{\sf B}  \circ \, \map C  & =  \bigoplus_k  \,   (\map I_{A_k} \otimes \Tr_{B_k}) \circ \map P_k \circ \map  C  \nonumber  \\
  &  =    \bigoplus_k  \,   (\map I_{A_k} \otimes \Tr_{B_k}) \circ   (\map  A_k\otimes \map I_{B_k}) \circ \map P_k   \nonumber \\
   &  =    \bigoplus_k  \,      \map  A_k   \circ  (\map I_{A_k} \otimes \Tr_{B_k}) \circ \map P_k   \nonumber \\
 &  =     \left(  \bigoplus_k  \,      \map  A_k\right)   \circ  \left[ \bigoplus_l   (\map I_{A_l} \otimes \Tr_{B_l}) \circ \map P_l  \right]    \nonumber \\
&=  \pi_{\sf A}  (\map C)  \circ \Tr_{\sf B} \, .\end{align}  \qed 
\medskip  

\begin{prop}
For every pair of channels $\map C_1, \map C_2  \in \CChan (\sf A)$, we have  the  homomorphism relation 
\begin{align}
\pi_{\sf A}   (\map C_1  \circ \map C_2)   =    \pi_{\sf A}  (\map C_1)  \circ \pi_{\sf A}   (\map C_2) \, .
\end{align}
\end{prop}
\Proof   Let us write the channels $\pi_{\sf A}  (\map C_1)$, $\pi_{\sf A}  (\map C_2)$, and  $\pi_{\sf A}   (\map C_1  \circ \map C_2)$ as 
\begin{align}
\pi_{\sf A}  (\map C_1)  = \bigoplus_k  \, \map A_{1k}   \, , \qquad \pi_{\sf A}  (\map C_2)  = \bigoplus_k  \, \map A_{2k} \, ,\qquad  {\rm and}  \qquad  \pi_{\sf A}  (\map C_1 \circ \map C_2)  = \bigoplus_k  \, \map A_{12k} \,.  
\end{align} 
With this notation, we have  
\begin{align}
(   \map A_{12k}\otimes \map I_{B_k}  ) \circ \map P_k   &   =     \map P_k  \circ \map C_1  \circ \map C_2   \nonumber \\ 
&    =     ( \map A_{1k}\otimes \map I_{B_k}  )  \circ \map P_k   \circ \map C_2  \nonumber \\
&    =  ( \map A_{1k} \otimes \map I_{B_k}  )    \circ (\map A_{2k}  \otimes I_{B_k}  )  \circ \map P_k   \nonumber \\
&    =  \Big[  ( \map A_{1k} \circ \map A_{2k}  ) \otimes \map I_{B_k}  \Big]     \circ \map P_k    \qquad \forall k  \, .
\end{align}
From the above equation, we obtain the equality  $\map A_{12k}  =  \map A_{1k}  \circ \map A_{2k}$ for all $k$.  In turn, this equality implies the desired result: 
\begin{align}
  \pi_{\sf A}  (\map C_1)  \circ \pi_{\sf A}  (\map C_2)  &  = \left (   \bigoplus_k  \map A_{1k}  \right)  \circ \left( \bigoplus_l   \map A_{2l}  \right)  \nonumber  \\
    &  =  \bigoplus_k       \map A_{1k}  \circ \map A_{2k}  \nonumber \\
    &   =  \bigoplus_k  \map A_{12 k}  \nonumber \\
    &   =   \pi_{\sf A}  (  \map C_1  \circ \map C_2) \, .
\end{align}
\qed

\section{Basis-preserving and multiphase-covariant channels}\label{app:multiphase}      

\subsection{Proof of Theorem \ref{theo:dualmultiphase}}\label{app:multi-bpres}

Here we prove that the monoid of multiphase covariant channels on $S$   (denoted as ${\sf MultiPCov}(S)$) and the monoid of basis-preserving channels on $S$ (denoted as ${\sf BPres} (S)$) are one the commutant of the other.

The proof  uses a few lemmas, the first of which is fairly straightforward:  
\begin{Lemma}
${\sf BPres} (S)'  \subseteq {\sf MultiPCov}(S)$. 
\end{Lemma}
\Proof Every unitary channel  of the form $\map U_{\bs \theta}  =  U_{\bs \theta} \cdot U_{\bs\theta}^\dag$ is basis-preserving, and therefore every channel  $\map C$ in the commutant of ${\sf BPres} (S)$  must  commute with it.  By definition, this means that $\map C$ is  multiphase covariant.    \qed 

\medskip 

To prove  the converse inclusion, we use  the following characterization of multiphase covariant channels: 

\begin{Lemma}[Characterization of ${\sf MultiPCov}(S)$]\label{lem:krausmultiphasecov}
A channel $\map M  \in  \CChan (S)$ is multiphase covariant if and only if it has a Kraus representation of the form 
\begin{align}\label{multiphasecov}
\map M  (\rho)  =   \sum_{i=1}^r   M_i \rho   M_i^\dag  +  \sum_{k=1}^d \sum_{j\not  =  k}   \,  p (j|k) \,   |j\>\<k|  \,\rho  | k\>\<j|  \, ,
\end{align}
where each  operator $M_i$ is diagonal in the computational basis, and   each $p(j|k) $  is non-negative.   
\end{Lemma}

\Proof Let  $M  \in  \Lin  (\spc H_S\otimes \spc H_S)$ be the Choi operator of channel $\map M$.    For a multiphase covariant channel, the Choi operator must satisfy the commutation relation  \cite{d2001optimal,chiribella2005extremal} 
\begin{align}
[M ,  U_{\bs \theta}  \otimes  \overline U_{\bs \theta}]  = 0  \qquad \forall \bs \theta \in  [0,2\pi)^{\otimes d} \, .
\end{align} 
This condition implies that $M$ must have the form 
\begin{align}
M = \sum_{s, t}    M_{ss,tt}  ~ |s\>\<t| \otimes |s\>\<t|      +  \sum_k \sum_{j\not  =  k} \,    M_{jk,jk}\,  |j\>\<j| \otimes |k\>\<k | \, ,
\end{align}
where the $d\times d$ matrix $[\Gamma_{s,t}]:=  [  M_{ss,tt}]_{s,t\in  \{  1,\dots,  d\}}$ is positive semidefinite and each coefficient $M_{st,st}$ is non-negative.  Then, Eq. (\ref{multiphasecov}) follows from  diagonalizing the matrix $\Gamma$ and using the relation $\map M (\rho)   =  \Tr  [  M  \,  (I  \otimes \rho^T)]$, where $\rho^T$ is the transpose of $\rho$ in the computational basis.  \qed

\medskip 

From Eq. (\ref{multiphasecov}) one can show every multiphase covariant channel commutes with every basis-preserving channel:  

\begin{Lemma}
${\sf MultiPCov}(S)  \subseteq  {\sf BPres}  (S)'$.  
\end{Lemma}
 \Proof  Let $\map B \in {\sf BPres}  (S)$ be a generic basis-preserving channel, and let $\map M \in{\sf MultiPCov}(S)$ be a generic multiphase covariant channel.  Using the characterization of Eq. (\ref{multiphasecov}), we obtain
\begin{align}
\nonumber \map M \circ \map B (\rho)    & =  \sum_i  \,  M_i  \map B (\rho )  M_i^\dag   +\sum_k   \sum_{j\not  =  k } \, p(j|k)    |j\>\<k|  \map B (\rho)   |k\>\<j|  \\  
 \nonumber &  =
     \sum_i  \,    \map B (M_i  \rho  M_i^\dag  )    +  \sum_k \sum_{j\not  =  k } \, p(j|k)    |j\>\<k|  \map B (\rho)   |k\>\<j|   \\ 
 \nonumber &  =
     \sum_i  \,    \map B (M_i  \rho  M_i^\dag  )    + \sum_k  \sum_{j\not  =  k } \, p(j|k)    |j\>  \,  \<k| \rho  |k\>  \,  \<j|  \\ 
     \nonumber &  =
     \sum_i  \,    \map B (M_i  \rho  M_i^\dag  )    + \sum_k  \sum_{j\not  =  k } \, p(j|k) \,  \map B( |j\>\<j|)    \,  \<k| \rho  |k\>  \,   \\ 
 \nonumber &  =           \map B \left (\sum_i    M_i  \rho  M_i^\dag      +   \sum_k   \sum_{j\not  =  k } \, p(j|k)  |j\>  \,  \<k| \rho  |k\>  \,  \<j| \right)  \\ 
 &=  \map B \circ \map M  (\rho) \qquad \forall \rho  \in \Lin (S) \, .
\end{align}
The second equality used the fact that the Kraus operators of $\map B$ are diagonal in the computational basis  \cite{buscemi2005inverting,buscemi2007quantum} and therefore commute with each operator $M_i$.    The third equality uses the relation $\<k|  \map B (\rho)  |k\>  =  \<  k|\rho  |k\>$, following from the fact that $\map B$ preserves the computational basis  \cite{buscemi2005inverting,buscemi2007quantum}. 
\qed

\medskip  

Summarizing, we have shown that the multiphase covariant channels are the commutant of the basis-preserving channels:  

\begin{cor}\label{cor:multi=bpres'}
${\sf MultiPCov}(S)  = {\sf BPres}  (S)'$.  
\end{cor}

Note that Corollary \ref{cor:multi=bpres'} implies the relation  
\begin{align}\label{acab}
{\sf MultiPCov}(S)'  = {\sf BPres}  (S)^{\prime \prime}    \supseteq   {\sf BPres}  (S) \, . 
\end{align}  To conclude the proof of Theorem \ref{theo:dualmultiphase}, we prove the converse inclusion:  
\begin{Lemma}\label{lem:multi'<bpres}
$ {\sf MultiPCov}(S)'  \subseteq  {\sf BPres}  (S)$.  
\end{Lemma} 
\Proof  A special case of multiphase covariant channel is the erasure channel $\map M_k $ defined by $\map M_k (\rho)   =  |k\>\<k|$ for every $\rho  \in  \Lin (S)$.   For a generic channel $\map C\in  {\sf MultiPCov}(S)'$, one must have 
\begin{align}
\map C  (|k\>\<k|)     =  \map C \circ  \map M_k   ( |k\>\<k|)  =   \map M_k \circ \map C   (|k\>\<k|)    =      |k\>\<k|  \, .     
\end{align}
Since the above condition must hold for every $k$,  the channel $\map C$ must be   basis-preserving.  \qed 

\medskip  

Combining Lemma \ref{lem:multi'<bpres} and Eq. (\ref{acab}) we obtain:
\begin{cor}\label{cor:bpres=multi'}
$ {\sf MultiPCov}(S)'   =  {\sf BPres}  (S)$.  
\end{cor}
  
\medskip  
Putting Corollaries \ref{cor:multi=bpres'} and \ref{cor:bpres=multi'} together, we  have an immediate proof of Theorem \ref{theo:dualmultiphase}.

\subsection{Proof of Eq. (\ref{classicalchan})}\label{app:classical}

Here we show that the transformations on system $S_A$ are classical channels.    To construct the transformations of $S_A$,  we have to partition the double commutant of $\Act (A;S)=  {\sf MultiPCov}(S)$ into equivalence classes.  

First, recall that  $  {\sf MultiPCov}(S)^{\prime \prime}  =   {\sf MultiPCov}(S)$  (by Theorem \ref{theo:dualmultiphase}). Then, note the following property: 
\begin{Lemma}\label{lem:samediag>sameclass}
If two channels $\map M, \widetilde{\map M}  \in  {\sf MultiPCov}(S)$ satisfy the condition  
\begin{align}\label{samediag}
\<k| \,\map M (|j\>\<j|)\, |k\>  =  \<k| \,\widetilde{\map M} (|j\>\<j|)\, |k\> \, ,
\end{align}
then  $[\map M]_{A'}  =  [\widetilde {\map M}]_{A'}$. 
\end{Lemma}
\Proof  Define the completely dephasing channel  $\map D  =  \sum_k    \,   |k\>\< k|  \cdot  |k\>\<k|$.  Clearly, $\map D$ is basis-preserving. Using the idempotence relation $\map D \circ \map D =  \map D$, we obtain 
  \begin{align}
 \nonumber \Big ( \map D  \circ  \map M \Big)  \, (\rho)     &=   \Big ( \map D \circ \map D  \circ  \map M \Big)     \, (\rho)  \\
 &=   \Big ( \map D  \circ  \map M \circ \map D  \Big)     \, (\rho)  \nonumber \\
   &  =   \Big ( \map D  \circ  \map M \Big )   \,  \Big ( \sum_{j}\,  |j\>\<j|  \,  \<j|\rho |j\>  \Big)  \nonumber \\
   &  =   \sum_{j}\,  \<j|\rho |j\> \,   \map D      \,  \Big (     \map M (  |j\>\<j| )     \Big)  \nonumber \\
   &  =   \sum_{j,k}\,  \<j|\rho |j\> ~   \< k|   \map M (  |j\>\<j| )  |k\>   ~   |k\>\<k|  \, .
    \end{align}
Likewise, we have
 \begin{align}
 \Big ( \map D  \circ  \widetilde{\map M} \Big)  \, (\rho)     =   \sum_{j,k}\,  \<j|\rho |j\> ~   \< k|   \widetilde{\map M} (  |j\>\<j| )  |k\>   ~   |k\>\<k |  \, .
    \end{align}
If condition (\ref{samediag}) holds, then the equality $\map D  \circ \map M  =  \map D \circ \widetilde {\map M}$ holds, meaning that  $\Deg  (\map M)$ and $\Deg(\widetilde{\map M})$ have non-empty intersection.   Hence, $\map M$ and $\widetilde{\map M}$ must be in the same equivalence class. \qed

\medskip 

The converse of Lemma \ref{lem:samediag>sameclass} holds:  
\begin{Lemma}\label{lem:sameclass>samediag}
If two channels $\map M, \widetilde{\map M}  \in  {\sf MultiPCov}(S)$ are in the same equivalence class, then they must satisfy condition  (\ref{samediag}). 
\end{Lemma}

\Proof 
If $\map M$ and $\widetilde{\map M}$ are in the same equivalence class, then there exists a finite sequence $(\map M_1, \map M_2,\dots,\map M_n)$ such that 
\begin{align*}
\map M_1  =  \map M \, ,\qquad \map M_n  = \widetilde{\map M}   \, ,\qquad  \forall i \in \{1,\dots, n-1\}\,  \exists \map B_i, \widetilde{\map B_i} \in {\sf BPres}  (S) :     \quad  \map B_i\circ \map M_i  =  \widetilde{\map  B_i}  \circ \map M_{i+1} \, .
\end{align*} 
The above condition implies  
  \begin{align}
  \< k|  \,\map M_i   (\rho) \, |k\>     =  \Tr  [   \map M_i   (\rho) \,  |k\>\<k|]    =    \< k|  \,\map B_i \circ \map M_i   (\rho) \, |k\>    =   \< k|\,  \widetilde{\map B_i} \circ \map M_{i+1}    (\rho) \, |k\>  =     \< k| \, \map M_{i+1}   (\rho)\,  |k\>  \,, 
  \end{align}
  for all $i  \in  \{  1,\dots,  n-1\}$ and for all $\rho  \in  \Lin (\rho)$.  In particular, choosing $\rho=  |j\>\<j|$ we obtain  
  \begin{align}
    \< k| \, \map M_i   (|j\>\<j|)\,  |k\>     =   \< k|  \,\map M_{i+1}   (|j\>\<j|) \, |k\> \qquad \forall i\in \{1,\dots, n-1\}  \, , \forall j,k  \in  \{1,\dots, d\}\, . 
  \end{align}
    Hence, Eq. (\ref{samediag}) follows. \qed  
    
    \medskip

\section{Classical systems and the resource theory of coherence}  

Here we consider  agents who have access to various types of free operations in the resource theory of coherence. 
We start from the types of operations that give rise to classical systems, and then show two examples that do not  have this property.

\subsection{Operations that lead to classical subsystems}\label{app:classicalsi}  

Consider the following monoids of operations
\begin{enumerate}
\item {\em Strictly incoherent operations \cite{yadin2016quantum}, i.e.}  quantum channels $\map T$ with the property that, for every Kraus operator $T_i$, the map $\map T_i  (\cdot)  =  T_i \cdot T_i  $ satisfies the condition $\map D  \circ \map T_i  = \map T_i  \circ \map D$, where $\map D$ is the completely dephasing channel.
\item {\em Dephasing covariant  operations \cite{chitambar2016critical,chitambar2016comparison,marvian2016quantify}, i.e.}  quantum channels $\map T$ satisfying the condition  $\map D  \circ \map T  = \map T  \circ \map D$.
\item {\em Phase covariant channels  \cite{marvian2016quantify}, i.e.} quantum channels   $\map T$ satisfying the condition $\map T\circ \map U_\varphi  =  \map U_\varphi  \circ \map T$, $\forall  \varphi  \in  [0,2\pi)$, where $\map U_\varphi $ is the unitary channel associated to the unitary matrix $U_\varphi  =   \sum_k    \,   e^{i  k\varphi}\,  |k\>\<k|$.
\item {\em Physically incoherent  operations \cite{chitambar2016critical,chitambar2016comparison}, i.e.}  quantum channels that are convex combinations of channels $\map T$ admitting a Kraus representation where each Kraus operator   $T_i$ is of the form   
\begin{align}\label{physinc}
T_i   =    U_{\pi_i}   \,  U_{{\bs \theta}_i} \,  P_i \,,
\end{align} where $U_{\pi_i}$ is a unitary that permutes the elements of the computational basis,  $U_{{\bs \theta}_i}$ is a diagonal unitary, and $P_i$ is a projector on a subspace spanned by a subset of vectors in the computational basis.  
\item {\em Classical channels, i.e.}  channels satisfying $\map T  =   \map D \circ \map T \circ \map D$. 
\end{enumerate} 

We now show that all the above operations define classical subsystems according to our construction.  

The first ingredient in the proof is the observation that  each of the monoids 1-5  contains the monoid of classical channels. 
Then, we can apply the following lemma: 
\begin{Lemma}\label{lem:basispres}
Let $\sf M  \subseteq  \CChan (S)$ be a monoid of quantum channels, and let $\sf M'$ be its commutant.  If $\sf M$ contains the monoid of classical channels, then $\sf M'$ is contained in the set of basis-preserving channels.  
\end{Lemma}
\Proof  Consider the erasure channel $\map C_k$ defined by $\map C_k(\rho)  :  =|k\>\<k| \, \Tr [\rho]$, $\forall \rho \in \Lin (\spc H_S)$.  Clearly, the erasure channel is a classical channel.  Then, every channel $\map B  \in  \sf M'$,must satisfy the condition 
  \begin{align}
  \map B  (|k\>\<k|)    =  \map B  \circ \map C_k  (|k\>\<k|) =  \map C_k  \circ \map B (|k\>\<k|)   =  |k\>\<k| \, .
  \end{align}  
Since  $k$ is generic, this implies that $\map B$ must be basis-preserving. \qed 

\medskip

Furthermore, we have the following
\begin{Lemma}\label{lem:if}
Let $\Act (A;S) \subseteq \CChan (S)$ be a set of quantum channels that contains the monoid of classical channels. If  two quantum states $\rho, \sigma \in \St (S)$ are equivalent for $A$, then they must have the same diagonal entries. Equivalently, they must satisfy $\map D (\rho) =  \map D(\sigma)$.  
\end{Lemma}
\Proof Same as the first part of the proof of Proposition \ref{prop:diagrho}.   Suppose that Condition 1 holds, meaning that  there exists a sequence $(\rho_1, \rho_2,\dots, \rho_n)$ such that  
\begin{align}
\rho_1 = \rho  \, , \qquad \rho_n=  \sigma \, , \qquad  \forall i \in  \{1,\dots, n-1\} \,  \exists   \map B_i  \, ,  \widetilde {\map B_i} \in  \Act(B;S)  :  ~  \map B_i  (\rho_i)  =  \widetilde {\map B_i}  (\rho_{i+1}) \, ,
\end{align} 
where  $\map B_i$ and $\widetilde {\map B_i}$ are channels in the commutant $\Act(A;S)'$.   The above equation  implies 
\begin{align}
 \< k|  \map B_i  (\rho_i)  |k\>  =   \<  k|  \widetilde{\map B_i}  (\rho_{i+1})  |k\> \,. 
\end{align}
Now, we know that the commutant $\Act(A;S)'$ consists of basis-preserving channels (Lemma \ref{lem:basispres}). Since every basis-preserving channel satisfies  the relation $  \< k|  \map B  (\rho)  |k\> =  \<  k|   \rho  |k\> $ \cite{buscemi2005inverting,buscemi2007quantum}, we obtain that all the density matrices $(\rho_1,\rho_2,\dots ,\rho_n)$ must have the same diagonal entries, namely $\map D (\rho_1)  = \map D(\rho_2)  =  \dots = \map D(\rho_n)$. \qed  

\medskip  

Now, we observe that the completely dephasing channel $\map D$ is contained in the commutant of all the monoids 1-5. This fact is evident for the monoids 1,2 and 5, where the commutation with $\map D$ holds by  definition. For the monoid 3, the commutation with $\map D$ has been proven in \cite{chitambar2016critical,chitambar2016comparison}, and for the monoid 4 it has been proven in \cite{marvian2016quantify}.

Since $\map D$ is contained in the commutant of all the monoids 1-5, we can use the following obvious fact:  
\begin{Lemma}\label{lem:onlyif}
Let $\Act (A;S) \subseteq \CChan(S)$  be a monoid of quantum channels and suppose that its commutant $\Act(A;S)'$ contains the dephasing channel $\map D$.   If two  quantum states $\rho, \sigma \in\St (S)$ satisfy $\map D (\rho)  =  \map D(\sigma)$, then they are equivalent for $A$. 
\end{Lemma}  
\Proof Trivial consequence of the definition.  \qed 
\medskip 

Combining Lemmas \ref{lem:if} and \ref{lem:onlyif}, we obtain the following 
\begin{prop}\label{prop:classicalstates}
Let $\Act(A;S)  \subseteq \CChan (S)$ be a monoid of quantum channels on system $S$. If $\sf Act(A;S)$ contains the monoid of classical channels,  and if the the commutant $\Act (A;S)'$ contains the completely dephasing channel $\map D$, then two  states $\rho, \sigma \in  \St (S)$ are equivalent for $A$ if and only if 
 $\map D(\rho)  =  \map D (\sigma)$.  
\end{prop}
\Proof   Same as the proof of Proposition \ref{prop:diagrho}. \qed 

\medskip  

Proposition \ref{prop:classicalstates} implies that the states of the subsystem $S_A$ are in one-to-one correspondence with diagonal density matrices. Since the conditions of the proposition are satisfied by  all the monoids 1-5, each of these monoids defines the same state space. 

The same result holds for the transformations:  
\begin{prop}\label{prop:classicalchannels}
Let $\Act(A;S)  \subseteq \CChan (S)$ be a monoid of quantum channels. If $\sf Act(A;S)$ contains the monoid of classical channels,  and if the the commutant $\Act (A;S)'$ contains the completely dephasing channel $\map D$, then two  transformations $\map S, \map T  \in  \Transf (S)$ are equivalent for $A$ if and only if 
 $\map D \circ \map T\circ \map D  =  \map D \circ \map T \circ \map D$.  
\end{prop}

\Proof Same as the proofs of Lemmas  \ref{lem:samediag>sameclass} and \ref{lem:sameclass>samediag}. \qed 
\medskip  

Proposition \ref{prop:classicalchannels} implies that the transformations of subsystem $S_A$ can be identified with classical channels.  Hence, system $S_A$ is exactly the $d$-dimensional classical subsystem of the quantum system $S$.  In summary, each of the monoids 1-5 defines  the same $d$-dimensional classical subsystem. 

\subsection{Operations that do not lead to classical subsystems}\label{app:classicalno}  

Here we show that our construction does not associate classical subsystems to the monoids of incoherent and maximally incoherent operations.  To start with, we recall the definitions of these two subsets:   
\begin{enumerate}
\item The {\em maximally incoherent operations} are  the  quantum channels $\map T$ that map diagonal density matrices to diagonal density matrices, namely $\map T  \circ \map D   =   \map D \circ \map T \circ \map D$, where $\map D$ is the completely dephasing channel. 
\item The {\em Incoherent operations} are   the quantum channels $\map T$ with the property that, for every Kraus operator $T_i$, the map $\map T_i  (\cdot)  =  T_i \cdot T_i  $ sends diagonal matrices to diagonal matrices, namely  $\map T_i  \circ \map D   =   \map D \circ \map T_i \circ \map D$. 
\end{enumerate}
Note that each set of operations contains the set of classical channels.  Hence, the commutant of each set of operation consists of (some subset of) basis-preserving channels (by Lemma \ref{lem:basispres}).  

Moreover, both sets of operations 1 and 2 contain the set of quantum channels $\map C_\psi$ defined by the relation 
\begin{align}
\map C_\psi  (\rho) =     |1\>\<1|    \,  \<\psi|  \rho  |\psi\>     +        \frac{ I-  |1\>\<1|}{d-1}    \,      \Tr [  (I-  |\psi\>\<\psi|) \,  \rho] \qquad\forall \rho\in \Lin (\spc H_S) \, ,    
\end{align}
where $|\psi\> \in  \spc H_S$ is a fixed (but otherwise arbitrary) unit vector. The fact that both monoids contain the channels $\map C_\psi$ implies a strong constraint on their commutants:  
\begin{Lemma}
The only basis-preserving quantum quantum channel $\map B  \in  {\sf BPres} (S)$ satisfying the property  $\map B\circ \map C_{\psi}  =  \map C_{\psi}  \circ \map B$ for every $|\psi\>\in\spc H_S$ is the identity channel. 
\end{Lemma}
\Proof   The commutation property implies the relation 
\begin{align}
(  \map C_{\psi}  \circ \map B )\,  (|\psi\>\<\psi|)    & =    (  \map B  \circ \map C_\psi )\, (|\psi\>\<\psi|)\nonumber \\
&  =  \map B   (  |1\>\<1|)  \nonumber \\
& =  |1\>\<1|    \, ,
\end{align}
where we  used the fact that $\map B$ is basis-preserving.   Tracing both sides of the equality with the projector $|1\>\<1|$, we obtain the relation 
\begin{align}\label{polar}
1   &  =  \<1  |   (  \map C_{\psi}  \circ \map B )\,  (|\psi\>\<\psi|)  |1\> \nonumber \\
&  =     \<\psi|   \, \map B   (|\psi\>\<\psi|)  \,  |\psi\>  \, ,
\end{align}
the second equality following from the definition of channel $\map C_\psi$.  In turn, Eq. (\ref{polar}) implies the relation $\map B  (|\psi\>\<\psi|)  =   |\psi\>\<\psi|$. Since $|\psi\>$ is arbitrary, this means that $\map B$ must be the identity channel.  \qed 

\medskip 
In summary, the commutant of the set of incoherent channels consists only of the identity channel, and so is the  the commutant of the set of maximally incoherent channels.  Since the commutant is trivial, the equivalence classes are trivial, meaning that the subsystem $S_A$ has exactly the same states and the same transformations of the original system $S$.   In short, the subsystem associated to the incoherent (or maximally incoherent) channels is the full quantum system.

\section{Enriching the sets of transformations}\label{app:category}  

 Here we provide a mathematical construction that enlarges the sets of transformations in the ``baby category'' with objects $S, S_A,$ and $S_B$.   This construction provides a realization of a catagorical structure known as splitting of idempotents  \cite{selinger2008idempotents,coecke2018two}.

We as have seen in the main text, our basic construction does not provide transformations from the subsystem $S_A$ to the global system $S$.  One could  introduce such transformations  by hand, by defining an {\em embedding} \cite{del2015resource}: 
\begin{defi}
An {\em embedding} of $S_A$ into $S$ is a map $\map E_{A}:   \St (S_A) \to  \St (S)$ satisfying the property 
\begin{align}
  \Tr_B  \circ \map E_{A} =  \map I_{S_A} \, .
\end{align} 
In other words, $\map E_A$ associates a representative to every equivalence class  $\rho  \in  \St (S_A) $.
\end{defi}

{\em A priori},  embeddings need not be physical processes. Consider the example of a classical system, viewed as a subsystem of a closed quantum system as in Subsection \ref{subsec:coherence}.    An embedding would  map each classical probability distribution $(p_1, p_2, \dots, p_d)$ into a pure quantum state $|\psi\>  =  \sum_k  \, c_k  \,  |k\>$ satisfying the condition $|c_k|^2  =  p_k$ for all $k\in \{  1,\dots, d\}$.  If the embedding were a physical transformation, there would be a way to physically transform every classical probability distributions into a corresponding  pure quantum state, a fact that is impossible in standard quantum theory.  

When building a new physical theory, one could {\em postulate} that there exist an embedding $\map E_A$ that is physically realizable.  In that case, the transformations from $S_A$ to $S$ would be those in  the set 
  \begin{align}
  \Transf(S_A  \to  S)   =  \Big\{  \map T\circ \map E_A \, : \quad \map T\in\Transf (S)  \Big\}    \, ,
  \end{align}
and similarly for the transformations from $S_B$ to $S$.  
The transformations from $S_A$ to $S_B$ would be those in the set 
 \begin{align}
  \Transf(S_A  \to  S_B)   =  \Big\{  \Tr_{A} \circ \map T\circ \map E_A \, : \quad \map T\in\Transf (S)  \Big\}    \, ,
  \end{align}
and similarly for the transformations from $S_B$ to $S_A$.  
In that new theory, the  old set of transformations from $S_A$  should be replaced by the new set 
\begin{align}
\widetilde{\Transf} (S_A)   =  \Big\{  \Tr_B  \circ \map T  \circ \map E_A  \, :  \quad \map T\in \Transf (S) \Big\} \, ,
\end{align}
so that the structure of category is preserved. 
Similarly, the old set of transformations from $S_B$ to $S_B$ should be replaced by the new set .  
\begin{align}
\widetilde{\Transf} (S_B)   =  \Big\{  \Tr_A  \circ \map T  \circ \map E_B  \, :  \quad \map T\in \Transf (S) \Big\} \, .
\end{align}

When this is done, the embeddings define two idempotent morphisms $\map P_A     : = \map E_A  \circ \Tr_B$ and $\map P_B  : =  \map E_B  \circ \Tr_A$, {\em i.e.}  two morphisms satisfying the conditions 
\begin{align}
\map P_A\circ \map P_A  =  \map P_A \qquad {\rm and} \qquad \map P_B\circ \map P_B  =  \map P_B \,. 
\end{align}     
The partial trace and the embedding define a   {\em splitting of idempotents}, in the sense of Refs.   \cite{selinger2008idempotents,coecke2018two}. The splitting of idempotents was considered in the categorical framework as a way to define general decoherence maps, and, more specifically, decoherence maps to classical subsystems  \cite{coecke2018two,gogioso2018categorical}.

\section{The total system as a subsystem}\label{app:SaS}
For every system satisfying the Non-Overlapping Agents Requirement, the system $S$ can be regarded as a subsystem: 
\begin{prop}
Let $S$ be a system satisfying the Non-Overlapping Agents Requirement,  let $A_{\max}$ be the maximal agent, and $S_{A_{\max}}$ be the associated subsystem. Then, one has $S_{A_{\max}}\simeq S$, meaning that there exist two isomorphisms $\gamma: \St (S)  \to \St \left(S_{A_{\max}}\right)$ and $\delta:  \Transf (S)  \to \Transf \left(S_{A_{\max}}\right)$ satisfying the condition 
\begin{align}\label{functorial}
\gamma  (  \map T   \psi)   =  \delta (\map T) \, \gamma (\psi) \, , \qquad \forall \psi \in\St (S) \, ,\forall \map T\in\Transf (S) \, .
\end{align}
\end{prop}
\Proof 
The Non-Overlapping Agents Requirement guarantees that the commutant $\Act (A_{\max};S)'$ contains only the identity transformation. Hence, the equivalence class $[\psi]_{A_{\max}}$ contains only the state $\psi$.  Hence, the partial trace  $\Tr_{A_{\max}'}:   \psi   \mapsto  [\psi]_{A_{\max}} $ is a bijection from $\St (S)$ to $\St \left(S_{A_{\max}}\right)$.  
Similarly, the equivalence class $[\map T]_{A_{\max}}$ contains only    the transformation $\map T$.   Hence, the restriction $\pi_{A_{\max}} :  \map T \mapsto  [\map T]_{A_{\max}}$ is a bijective function between $\Transf (S)$ and $\Transf \left(S_{A_{\max}}\right)$.  Such a function is an homomorphism of monoids, by equation  (\ref{def:composition}).  Setting $\delta  : =  \pi_{A_{\max}}$ and $\gamma:  =  \Tr_{A_{\max}'}$, the condition (\ref{functorial}) is guaranteed by equation (\ref{def:action}). \qed

\section{Proof of Proposition \ref{prop:regularity}}\label{app:regularity}

By definition, the condition $\Tr_B  [\psi]  = \Tr_B [\psi']$ holds if and only if there exists a finite sequence $(\psi_1,\psi_2,\dots, \psi_n)$ such that 
\begin{align}\label{chainagain}
\psi_1  =  \psi   \, ,\qquad \psi_n =  \psi'   \, ,\qquad  \forall i\in\{1,\dots, n-1\}  \quad  \exists \map V_i \, ,\widetilde{\map V_i}  \in  \Act (B;S) : ~   \map V_i \psi_i   = \widetilde{\map V}_i  \psi_{i+1}  \, . \
\end{align} 
Our goal is to prove  that there exists an adversarial action  $\map V_B\in\Act (B;S)$ such that the relation  $\psi'  =  \map V_B \psi$ or $\psi  = \map V_B  \psi'$ holds.   

We will proceed by induction on $n$, starting from the base case $n=2$.  In this case,  we have  $\Deg_B (\psi)  \cap \Deg_B  (\psi')\not = \emptyset$.   Then, the first regularity condition implies that  there exists a transformation $\map V_B  \in   \Act (B;S)$ such that at least one of  the relations  $\map V_B    \psi   =  \psi' $ and  $\psi =  \map V_B  \psi'$ holds.  This proves the validity of the base case.

   Now, suppose that the induction hypothesis holds for all sequences of length $n$, and suppose that $ \psi$ and $\psi'$ are equivalent through a sequence of length $n+1$, say $(\psi_1, \psi_2, \dots, \psi_n, \psi_{n+1})$.       Applying the induction hypothesis to the sequence $(\psi_1, \psi_2, \dots, \psi_n)$, we obtain that there exists a transformation $\map V \in\Act (B;S)$ such that at least one of the relations $\psi_n  =  \map V   \psi$ and $\psi  =  \map V  \psi_n$ holds. Moreover, applying the induction hypothesis to the pair $(\psi_n,  \psi_{n+1})$ we obtain  that there exists a transformation  $\map V'  \in  \Act (B;S)$ such that $\psi_{n+1}  =  \map V'   \psi_n$, or $\psi_n  =  \map V'  \psi_{n+1}$.  Hence, there are four possible cases:  
\begin{enumerate}
\item   
$\psi_n  =  \map V   \psi$ and $\psi_{n+1}  =  \map V'   \psi_n$.  In this case, we have $\psi_{n+1}=  (\map V'\circ \map V)  \psi$, which proves the desired statement. 
\item 
$\psi_n  =  \map V   \psi$ and $\psi_n  =  \map V'  \psi_{n+1}$. In this case, we have $\map V   \psi  = \map V'  \psi_{n+1}$, or equivalently $\Deg_B (\psi)  \cap \Deg_B  (\psi_{n+1})   \not  = \emptyset$.    Applying the induction hypothesis to the sequence $(\psi,\psi_{n+1})$, we obtain the desired statement. 
\item 
$\psi  =  \map V  \psi_n$ and $\psi_{n+1}  =  \map V'   \psi_n$.   Using the second regularity condition, we obtain that there exists a transformation $\map  W\in\Act (B;S)$ such that at least one of the relations $\map V  =  \map W  \circ \map V'$ and $\map V'  =  \map W \circ \map V$ holds.   Suppose that  $\map V  =  \map W  \circ \map V'$. In this case, we have 
\begin{align}
\psi  =  \map V \psi_n  = (\map W  \circ \map V' ) \psi_n   =   \map W  \psi_{n+1} \, .
\end{align} 
Alternatively, suppose  that $\map V'  =  \map W  \circ \map V$. In this case, we have  
\begin{align}
\psi_{n+1}  =  \map V' \psi_{n}  = (\map W  \circ \map V ) \psi_n   =   \map W  \psi \, .
\end{align} 
In both  cases, we proved the desired statement.  
\item
$\psi  =  \map V  \psi_n$ and $\psi_n  =  \map V'  \psi_{n+1}$. In this case, we have $\psi  =  (\map V\circ \map V' ) \psi_{n+1}$, which proves the desired statement. 
\end{enumerate}
\qed

\section{Characterization of the adversarial group}\label{app:adversarial}
  
 Here we provide the proof of Theorem \ref{theo:semidirect},  proving a canonical decomposition of the elements of the adversarial group.  The proof proceeds in a few steps:

 \begin{Lemma}[Canonical form of the elements of the adversarial group]\label{theo:permute}
 Let $U:  g\mapsto U_g $ be a projective representation of the group $\grp G$, let $\Irr (U)$ be the set of irreducible representations contained in the isotypic decomposition of $U$, and let $\omega:  \grp G \to \CC$ be a multiplicative character of $\grp G$. Then, the commutation relation 
\begin{align}\label{eqa}
V  U_g  =  \omega (g)~  U_g V  \qquad \forall g\in\grp G 
\end{align} 
holds iff 
\begin{enumerate}
\item  The map $U^{(j)}  \mapsto  \omega  \,   U^{(j)}$ is a permutation of the set $\Irr (U)$, denoted as $\pi: \Irr (U) \to \Irr (U)$. In other words, for every irrep $U^{(j)}$ with $j\in\Irr (U)$, the irrep  $\omega\,  U^{(j)}$ is equivalent to an irrep $k \in\Irr (U)$, and the correspondence between $j$ and $k$ is bijective. 
\item The multiplicity spaces $\spc M_j$ and $\spc M_{\pi(j)}$ have the same dimension.  
\item  The unitary operator $V$ has the canonical form    $V  =   U_{\pi}   V_0$  where $V_0$ is an unitary operator in the commutant $U'$ and $U_\pi$ is a permutation operator satisfying  
\begin{align}
U_\pi  \Big ( \spc R_j \otimes \spc M_j \Big)  =  \Big ( \spc R_{\pi(j)} \otimes \spc M_{\pi(j)}   \Big) \qquad \forall  j\in\Irr (U) \, .
\end{align}
\end{enumerate}
 \end{Lemma}
 
\Proof  Let us use the isotypic decomposition of $U$, as in  Eq. (\ref{isotypic}).   We define  
 \begin{align}
 V_{j,k}   : =  \Pi_j\,  V \,  \Pi_k \, ,
 \end{align}
 where $\Pi_j$  ($\Pi_k$) is the projector onto $\spc R_j\otimes \spc M_j$  ($\spc R_k\otimes \spc M_k$).  
 Then,  Eq. (\ref{eqa}) is equivalent to the condition 
 \begin{align}
 V_{j,k}   \, \Big(U_g^{(k)}  \otimes I_{\spc M_k}  \Big)    =   \omega (g) \,   \Big(U_g^{(j)}  \otimes I_{\spc M_j}  \Big)  \,    V_{jk} \, , \qquad \forall  g\in\grp G \, , \forall j,k \, , 
 \end{align} 
 which in turn is equivalent to the condition 
 \begin{align}\label{basta}
 \<\alpha |V_{j,k} |\beta\>  \,    \,   U_g^{(k)}      =   \omega (g) \,   U_g^{(j)}   \,   \<\alpha| V_{j,k}|\beta\> \, , \qquad \forall  g\in\grp G \, , \forall j,k \, , \forall |\alpha\>  \in  \spc M_j\, , \forall |\beta\>  \in  \spc M_k \,  , 
 \end{align} 
where  $\<\alpha| V_{j,k}|\beta\>$ is a shorthand for the partial matrix element $ (I_{\spc R_j} \otimes \<\alpha|     ) \,  V_{j,k} \, ( I_{\spc R_k} \otimes |\beta\>  )$. 
 
Eq. (\ref{basta}) means that each operator $\<\alpha| V_{j,k}|\beta\>$ intertwines the two  representations $U^{(k)}$ and $ \omega  \,  U^{(j)}$.  Recall  that each representation is irreducible.  Hence,  the second Schur's lemma \cite{fulton2013representation} implies that $\<\alpha| \, V_{j,k}\,  |\beta\>$ is zero if the two representations  are not equivalent.  
Note that there can be {\em at most}  one value of $j$ such that $U^{(k)}$ is equivalent to $\omega\,  U^{(j)}$.   If such a value exists, we denote it as $j =  \pi (k)$.      By construction,  the function $\pi:  \set{Irr} (U) \to \set {Irr}  (U) $ must be injective.

When $j=  \pi  (k)$, the first Schur's lemma \cite{fulton2013representation} guarantees that the operator $\<  \alpha|  \,  V_{\pi(k),k}|\beta\>$ is proportional to the partial isometry $T_{\pi(k), k}$
 that implements the equivalence of the two representations.  Let us write  
 \begin{align}
 \<\alpha|  \,  V_{\pi (k),k}\,  |\beta\>  =   M_{\alpha, \beta}  \,  T_{ \pi (k),  k }\, ,
 \end{align}
 for some $M^{(k)}_{\alpha,\beta}  \in  \CC$. Note also that, since the left hand side is sesquilinear in $|\alpha\>$ and $|\beta\>$, the right hand side should also be sesquilinear. Hence, we can find an operator $M_{ \pi (k),k}: \spc M_{k} \to \spc M_{\pi (k)} $  such that $M^{(k)}_{\alpha, \beta}  =  \<\alpha|  \,  M_{ \pi (k),k} \, |\beta\>$.  
 Putting everything together,  the operator $V$ can be written as 
\begin{align}
V  =  \bigoplus_{k\in \set{Irr}  (U)}  \,     \Big (   T_{ \pi  (k),k}  \otimes M_{\pi(k),k}  \Big)  \, .
\end{align}
Now, the operator $V$ must be unitary, and, in particular, it should satisfy the condition $V V^\dag  =  I$, which reads
\begin{align}
\bigoplus_{k\in \set {Irr}  (U)}    \Big (  I_{\spc R_{\pi(k)}}  \otimes M_{\pi(k),k}  M_{\pi  (k), k}^\dag \Big )   =  I \, . 
\end{align} 
The above condition implies that:  {\em i)} the function $\pi$ must be surjective, and {\em ii)} the operator  $M_{\pi(k), k}$ must be a co-isometry. From the relation $V^\dag V$ we also obtain that $M_{\pi(k), k}$ must be an isometry. Hence,  $M_{\pi(k)}$ is  unitary.

Summarizing, the condition (\ref{eqa}) can be satisfied only if there exists a permutation $\pi :  \set {Irr}  (U)  \to \set {Irr} (U)$ such that, for every $j$, 
\begin{enumerate}
\item the irreps $\omega \, U^{(k)}$ and $U^{\pi (k)}$   are equivalent
\item the multiplicity spaces   $\spc M_k$ and $\spc M_{\pi  (k)}$ are unitarily isomorphic.  
\end{enumerate}
 Fixing a unitary isomorphism $S_{\pi (k),k}:  \spc M_{k} \to \spc M_{\pi (k)}$, we can write every element of the adversarial group in the canonical form 
 $V   =       U_\pi  \,    V_0$, 
  where  $U_\pi$ is the permutation operator 
 \begin{align}\label{upi}
 U_{\pi}   =  \bigoplus_{k  \in \set {Irr} (U)}  \,    \Big  (   T_{\pi (k),  k}  \otimes S_{\pi(k),k}\Big)  \, ,
 \end{align} 
  and $V_0$ is an element of the commutant $U'$, i.e. 
 a generic unitary operator of the form 
 \begin{align}\label{vzero}
 V_0 =  \bigoplus_{k\in \set {Irr} (U)}  \,    \Big  (    I_j  \otimes   V_{0,k} \Big)  \, .
 \end{align}
 Conversely, if a permutation $\pi$ exists with the properties that for every $k\in  \Irr (U)$
 \begin{enumerate}
 \item $\omega \,  U^{(k)}$ and   $U^{(\pi(k))}$ are equivalent irreps
 \item $\spc M_k$ and $\spc M_{\pi (k)}$ are unitarily equivalent,
 \end{enumerate} 
 and if the  operator  $V$ has the form $V  =  U_\pi  V_0$, with $U_\pi$ and $V_0$ as in equations   (\ref{upi}) and (\ref{vzero}), then $V$ satisfies the commutation relation (\ref{eqa}). \qed

\medskip

We have seen that every element of the adversarial group can be decomposed into the product of a permutation operator, which permutes the irreps, and an operator in the commutant of the original group representation $U:  \grp G  \to  \Lin (\spc H)$.  We now observe that the allowed permutations have an additional structure: they must form an Abelian group, denoted as $\grp A$. 
\begin{Lemma}\label{theo:abelian}
The permutations $\pi$  arising from Eq. (\ref{eqa}) with a generic multiplicative character $\omega (V,\cdot)$  form an Abelian subgroup   $\grp A$ of the group of all permutations of $\set {Irr} (U)$. 
\end{Lemma}

\Proof Let $V$  and $W$ be two elements of the adversarial group $\grp G_B$, let   $\omega (V,\cdot)$ and $\omega (W,\cdot)$ be the corresponding characters, and let $\pi_V$ and $\pi_{W}$ be the permutations associated to  $\omega (V,\cdot)$ and $\omega (W,\cdot)$ as in Theorem \ref{theo:permute}, {\em i.e.} through the relation  
\begin{align}
\nonumber j=  \pi_V  (k) & \qquad \Longleftrightarrow    \qquad  U^{(j)}  \quad {\rm is~equivalent~to} \quad \omega (V,\cdot)  \,  U^{(k)}   \\
j=  \pi_W  (k)  &\qquad \Longleftrightarrow    \qquad  U^{(j)}  \quad {\rm is~equivalent~to} \quad \omega (W,\cdot)  \,  U^{(k)} \, .  
\end{align}
Now, the element $VW$ is associated to the permutation  $\pi_V \circ \pi_W$, while the element $WV$ is associated to the permutation $\pi_W\circ \pi_V$.    On the other hand,  the characters obey the equality 
\begin{align}
\omega (  VW, g)   =  \omega  (  V, g)  \omega (W, g)    = \omega (WV,  g)   \qquad \forall g\in\grp G\, .
\end{align}
Hence, we conclude that $\pi_V \circ \pi_W$ and $\pi_W\circ \pi_V$ are, in fact, the same permutation.   Hence, the elements of the adversarial group must correspond to an Abelian subgroup of the permutations of $\Irr (U)$.  \qed 

\medskip 
Combining Lemmas  \ref{theo:permute} and  \ref{theo:abelian}, we can now prove  Theorem \ref{theo:semidirect}.  

\medskip 

{\bf Proof  of Theorem \ref{theo:semidirect}.}  For different permutations in  $\grp A$, we can choose the isomorphisms  $S_{\pi (k),k}:  \spc M_k\to \spc M_{\pi(k)}$ such that the following property holds: 
\begin{align}
   S_{\pi_2\circ\pi_2  (k),  k}    =  S_{\pi_2 (\pi_1  (k)),    \pi_1 (k)} ~ S_{\pi_1 (k),k} \, ,  \qquad \forall \pi_1, \pi_2\,\in\grp {A} \, .
  \end{align}
  When this is done, the unitary operators $U_{\pi}$ defined in equation (\ref{upi}) form a faithful representation of the Abelian group $\grp A$.    Using the canonical decomposition of Theorem \ref{theo:permute}, every element of $  V\in  \grp G_B$ is decomposed {\em uniquely} as $V =  U_\pi  \,  V_0$, where $V_0$ is an element of the commutant $U'$. Note also  that the commutant $U'$ is a normal subgroup of the adversarial group: indeed, for every element $V \in  \grp G_B$ we have 
$  V    U'  V^\dag   =   U'$.   Since $U'$ is a normal subgroup and the decomposition $V=  U_\pi   V_0$ is unique for every $V\in\grp G_B$, it follows that the adversarial group $\grp G_B$ is the semidirect product   $ \grp A\ltimes  U'$. \qed

\section{Example: the phase flip group}\label{app:example}

Consider the Hilbert space $\spc H_S  =  \CC^2$, and suppose that agent $A$ can only perform the identity channel and the phase flip channel  $\map Z$, defined as 
\begin{align}
\map Z  (\cdot)     =  Z  \cdot Z  \, , \qquad Z  =  |0\>\<0  |  -  |1\>\<1|  \, . 
\end{align}  
Then, the actions of agent $A$ correspond to  the  unitary representation 
\begin{align}
U:   \Z_2  \to   \Lin  (S) \, , \qquad  k  \mapsto  U_k  =  Z^k \, .
\end{align} 
   The representation can be decomposed into two irreps, corresponding to the one-dimensional subspaces $\spc H_0  =  \Span  \{|0\>\}$ and $\spc H_1  =  \Span \{|1\>\}$.   
The corresponding irreps, denoted by 
\begin{align}
\nonumber \omega_0 : \quad &  \Z_2  \to \CC   \, , \quad  \omega (k)  =  1\\
\omega_1: \quad &  \Z_2 \to   \CC  \, , \quad \omega  (k)  =  (-1)^k   
\end{align}  are the only two irreps of the group and are multiplicative characters.

The condition $V  U_k=    U_k  V $ yields the solutions 
\begin{align}\label{1}
V  =   e^{i\theta_0} \,  |0\>\<0|  +  e^{i\theta_1}\,  |1\>\<1|  \, ,  \qquad \theta_0,\theta_1  \in  [0,2\pi) \, ,  
\end{align}
corresponding to the commutant $U'$.  The condition $V  U_k  =  (-1)^k \,     U_k  V$ yields the solutions 
\begin{align}\label{2}
V  =   e^{i\theta_0} \,  |0\>\<1|  +  e^{i\theta_1}\,  |1\>\<0|  \, ,  \qquad \theta_0,\theta_1  \in  [0,2\pi) \, .   
\end{align}
It is easy to see that the adversarial  group $\grp G_B$ acts irreducibly on $\spc H_S$.  

Let us consider now the  subsystem $S_A$.  The states of $S_A$ are equivalence classes  under the relation  
\begin{align}
|\psi\>  \simeq_{A}   |\psi'\> \qquad \exists  V\in\grp G_B   :   \quad  |\psi'\>  =  V  |\psi\>  \, . 
\end{align}   
It is not hard to see that the equivalence class of the state $|\psi\>$ is uniquely determined by the {\em unordered} pair  $\{  |\<0|\psi\> |  \, , \,  |\<1|  \psi\>|  \}$. In other words, the state space of system $S_A$ is 
\begin{align}
\St (S_A)  =  \Big\{~    \{ p,  1-p\}    \, , : \quad      p\in  [0,1]     \Big\} \,.
\end{align}
Note that in this case the state space  is not a convex set of density matrices. Instead,  it is the quotient of the set of diagonal density matrices, under the equivalence relation that two matrices with the same spectrum are equivalent.    

Finally, note that the transformations of system $S_A$ are trivial:   since the adversarial group $\grp G_B$ contains the group $\grp G_A$,  the group  $\grp G  (  S_A)= \pi_A (G_A)$ is trivial, namely 
\begin{align}
\grp G(S_A  )  =    \Big \{   \map I_{S_A} \Big\}    \, . 
\end{align}

\section{Proof of Theorem \ref{theo:connectedLie}}\label{app:connectedLie}  

Let $\grp G$ be a connected Lie group, and let $\mathfrak g$ be the Lie algebra.  Since $\grp G$ is connected, the exponential map reaches every element of the group, namely $\grp G  =  \exp[i  \mathfrak g]$.

Let $h\in\grp G$ be a generic element of the group, written as $h =  \exp[i    X]$ for some $X\in\mathfrak g$, and consider the  one-parameter subgroup $\grp H  =   \{   \exp[i\lambda  X]  \, , \lambda\in  \R  \}$.  
 For a generic element $g \in \grp H$, the corresponding unitary operator can be expressed as $U_g  =   \exp[i  \lambda K]$, where $K  \in  \Lin (S)$ is a suitable self-adjoint operator.  Similarly, the multiplicative character has the form $\omega (g)   =   \exp[i\lambda  \mu]$, for some real number $\mu  \in  \R$.  

Now, every element $V$ of the adversarial group must satisfy  the relation
\begin{align}
V \exp[i\lambda K]   =        \exp[i  \lambda ( K+  \mu \, I_S)] V  \qquad \forall \lambda\in \R  \, ,
\end{align} 
or equivalently, 
\begin{align}
\exp[i\lambda K]   =      V^\dag \,  \exp[i  \lambda ( K+  \mu \, I_S)] \,V  \qquad \forall \lambda\in \R  \, .
\end{align} 
Since the  operators $\exp[i\lambda K] $  and $\exp[i  \lambda ( K+  \mu \, I_S)]$   are unitarily equivalent, they must have the same spectrum  
This is only possible if the operators $K$ and $K+  \mu  \,  I_S$ have the same spectrum, which happens only if $\mu  =  0$.    

Now, recall that the one-parameter Abelian subgroup $\grp H$ is generic. Since every element of $\grp G$ is contained in some one-parameter Abelian subgroup $\grp H$,   we showed that $\omega (g)  =1$ for every  $g\in\grp G$.  

To conclude the proof, observe that the map $U^{(j)}  \mapsto  \omega \,  U^{(j)}$ is the identity, and therefore induces the trivial permutation on the set of irreps $\Irr (U)$.  Hence, the group of permutations $\grp A$ induced by multiplication by $\omega$   contains only the identity element.    \qed  

\section{Proof of Proposition \ref{prop:LieEquivalence}}\label{app:LieEquivalence}

\Proof It is enough to decompose the two states as 
\begin{align}
|\psi\>   =  \bigoplus_{j\in\set {Irr}  (U)} \,   \sqrt{p_j} \,    |\psi_j\>  \qquad {\rm and}  \qquad |\psi'\>   =  \bigoplus_{j\in\set {Irr} (U)} \,   \sqrt{p'_j} \,    |\psi'_j\> \, ,
\end{align}  
where $|\psi_j\>$ and $|\psi_j'\>$ are unit vectors in $\spc R_j\otimes \spc M_j$.   Using this decomposition, we  obtain 
\begin{align}\label{rank}
\map T_B  (  |\psi\>\<\psi|)   =  \bigoplus_{j\in\set {Irr} (U)}  \,   p_j  \,  \rho_j     \qquad {\rm and}  \qquad   \map T_B  (  |\psi\>\<\psi|)   =  \bigoplus_{j \in \set{Irr}  (U)}  \,   p'_j  \,  \rho'_j   \, ,
\end{align}
where $\rho_j$   ($\rho_j'$) is the marginal of $|\psi_j\>$   ($|\psi_j'\>$) on system $\spc R_j$.    It is then clear that the equality $\map T_B   (|\psi\>\<\psi|)   =  \map T_B  (|\psi'\>\<\psi'|)$ implies $p_j=p_j'$  and  $\rho_j  =  \rho_j'$ for every $j$.   Since the states $|\psi_j\>$ and $|\psi_j'\>$ have the same marginal on system $\spc R_j$, there must exist a unitary operator $U_j:  \spc M_j\to  \spc M_j$ such that   
\begin{align}|\psi_j'\>   =  (I_{\spc R_j} \otimes  U_j)\,  |\psi_j\> \, .
\end{align}  
We can then define the unitary gate 
\begin{align}
U_B  =  \bigoplus_{j\in \Irr  (U)}  \,   \Big(I_{\spc R_j}  \otimes  U_j\Big) \, ,
\end{align} 
which satisfies the property $U_B|\psi\>  =  |\psi'\>$.  By the characterization of Eq. (\ref{Vp}), $U_B$ is an element of $\grp G_B$.     \qed

\bibliographystyle{mdpi}
\makeatletter
\renewcommand\@biblabel[1]{#1. }
\makeatother
\bibliography{generic}

\end{document}